\documentclass[article,onecolumn,superscriptaddress,nofootinbib,floatfix,onecolumn]{revtex4}

\usepackage{amsmath, latexsym, amssymb, hyperref, graphicx, color, multirow, makecell}
\usepackage{tabularx}
\usepackage[T1]{fontenc}

\usepackage{dcolumn}
\usepackage{bm}
\usepackage{xspace}

\usepackage{listings}
\lstdefinestyle{mystyle}{mathescape,basicstyle=\small\ttfamily,frame=leftline,aboveskip=4mm,belowskip=4mm,xleftmargin=20pt,framexleftmargin=10pt,numbers=none,framerule=2pt,abovecaptionskip=0.0mm,belowcaptionskip=3.5mm}
\usepackage{float}
\usepackage{epsfig}
\usepackage{longtable}
\usepackage{rotating}
\usepackage{ulem}
\usepackage{amssymb}
\usepackage{amsmath}
\usepackage{txfonts}
\usepackage{upgreek}
\usepackage{mathtools}
\usepackage{tabularx}
\usepackage{booktabs}

\usepackage{soul} 

\usepackage[capitalize]{cleveref}
\definecolor{nicered}{rgb}{.7,.1,.1}
\definecolor{nicegreen}{rgb}{.1,.5,.1}
\definecolor{darkblue}{rgb}{0,0,.5}
\hypersetup{colorlinks, citecolor=nicegreen,linkcolor=nicered, urlcolor=darkblue}
\usepackage{multirow}
\usepackage{slashed}
\usepackage{verbatim}
\usepackage{listings}

\def\ttt#1{\texttt{\small #1}}

\providecommand{\qqbar}{q\overline{q}}
\providecommand{\ccbar}{c\overline{c}}
\providecommand{\bbbar}{b\overline{b}}

\newcommand{\pp}{p-p}
\newcommand{\ppbar}{p-$\overline{\mathrm{p}}$}
\providecommand{\pA}{p-A}
\providecommand{\pPb}{p-Pb}
\providecommand{\AaAa}{A-A}
\providecommand{\AaBa}{A-B}
\providecommand{\PbPb}{Pb-Pb}

\newcommand{\epem}{e^+e^-}
\newcommand{\mumu}{\mu^+\mu^-}
\newcommand{\tautau}{\tau^+\tau^-}
\newcommand{\tata}{\mathcal{T}_0}

\newcommand{\gaga}{\gamma\gamma}

\newcommand{\jpsi}{\mathrm{J}/\psi}

\providecommand{\ccbar}{c\overline{c}}
\providecommand{\bbbar}{b\overline{b}}
\providecommand{\ttbar}{t\overline{t}}

\newcommand{\etacOneS}{\mathrm{\eta_{c}(1\mathrm{S})}}
\newcommand{\etacTwoS}{\mathrm{\eta_{c}(2\mathrm{S})}}

\newcommand{\chicZero}{\mathrm{\chi_{c0}}}
\newcommand{\chicTwo}{\mathrm{\chi_{c2}}}

\newcommand{\etabOneS}{\mathrm{\eta_{b}(1\mathrm{S})}}
\newcommand{\etabTwoS}{\mathrm{\eta_{b}(2\mathrm{S})}}

\newcommand{\chibZero}{\mathrm{\chi_{b0}}}
\newcommand{\chibTwo}{\mathrm{\chi_{b2}}}

\providecommand{\mgg}{m_{\gamma\gamma}}

\providecommand{\sigmagagaX}{\sigma_{\gaga\to X}}

\newcommand{\sqrts}{\sqrt{s}}

\newcommand{\sqrtsgg}{\sqrt{s_{_{\gamma\,\gamma }}}}
\newcommand{\sqrtsnn}{\sqrt{s_{_\text{NN}}}}

\newcommand{\kt}{k_{\perp}}
\newcommand{\pT}{p_\mathrm{T}}
\newcommand{\ET}{E_\mathrm{T}}
\newcommand{\ABgagaX}{A \,B\,$\xrightarrow{\gaga}$ A\, $X$ \,B}

\newcommand{\Lumi}{\mathcal{L}}

\newcommand{\LumiInt}{\mathcal{L}_{\mathrm{\tiny{int}}}}

\newcommand{\gammaUPC}{\ttt{gamma-UPC}}
\newcommand{\helaconia}{\textsc{HELAC-Onia}}
\newcommand{\madgraph}{\textsc{MadGraph5\_aMC@NLO}}
\newcommand{\mgshort}{\textsc{MG5\_aMC}}
\newcommand{\starlight}{\textsc{Starlight}}
\newcommand{\superchic}{\textsc{Superchic}}

\usepackage{xspace}
\newcommand*{\eg}{e.g.\@\xspace}
\newcommand*{\ie}{i.e.,\@\xspace}
\newcommand*{\cm}{c.m.\@\xspace}

\begin{document}


\title{\gammaUPC: Automated generation of exclusive photon-photon processes in\\ ultraperipheral proton and nuclear collisions with varying form factors}

\author{Hua-Sheng Shao}\email{huasheng.shao@lpthe.jussieu.fr}
\affiliation{Laboratoire de Physique Th\'eorique et Hautes Energies (LPTHE), UMR 7589, Sorbonne Universit\'e et CNRS, 4 place Jussieu, 75252 Paris Cedex 05, France}

\author{David d'Enterria}\email{david.d'enterria@cern.ch}

\affiliation{CERN, EP Department, CH-1211 Geneva 23, Switzerland}

\begin{abstract}
\noindent
The automated generation of arbitrary exclusive final states produced via photon fusion in ultraperipheral high-energy collisions of protons and/or nuclei, \ABgagaX, is implemented in the \madgraph\ and \helaconia\ Monte Carlo codes. Cross sections are calculated in the equivalent photon approximation using $\gamma$ fluxes derived from electric dipole and charge form factors, and incorporating hadronic survival probabilities. Multiple examples of $\gaga$ cross sections computed with this setup, named \gammaUPC, are presented for proton-proton, proton-nucleus, and nucleus-nucleus ultraperipheral collisions (UPCs) at the Large Hadron Collider and Future Circular Collider. Total photon-fusion cross sections for the exclusive production of spin-0,\,2 resonances (quarkonia, ditauonium, and Higgs boson; as well as axions and gravitons), and for pairs of particles ($\jpsi\jpsi$, WW, ZZ, Z$\gamma$, $\ttbar$, HH) are presented. Differential cross sections for exclusive dileptons and  light-by-light scattering are compared to LHC data. This development paves the way for the upcoming automatic event generation of any UPC final state with electroweak corrections at next-to-leading-order accuracy and beyond.
\end{abstract}

\date{\today}

\maketitle


\section{Introduction}

The electromagnetic field of any charged particle accelerated at high energies can be identified in the equivalent photon approximation (EPA)~\cite{vonWeizsacker:1934nji,Williams:1934ad} as a flux of quasireal photons~\cite{Brodsky:1971ud,Budnev:1975poe} whose intensity is proportional to the square of its electric charge, $Z^2$. 
Although high-energy photon-photon processes have been studied in $\epem$ and $e$-p collisions since more than thirty years ago~\cite{Vermaseren:1982cz,Schuler:1997ex,Uehara:1996bgt}, as well as in the last twenty years with heavy ions at the Relativistic Heavy Ion Collider (RHIC)~\cite{Bertulani:2005ru}, this physics domain has received a particularly strong boost in the last ten years thanks to the greatly extended center-of-mass (\cm) energies and luminosities accessible in collisions with hadron beams at the Large Hadron Collider (LHC). The multi-TeV energies and high-luminosity beams available at the LHC, and the possibility of accelerating not just protons but heavy ions with charges up to $Z=82$ for lead (Pb) ions, has enabled a multitude of novel $\gaga$-collision measurements in ultraperipheral collisions (UPCs) of proton-proton (\pp), proton-nucleus (\pA), and nucleus-nucleus (\AaAa) as anticipated in~\cite{Baltz:2007kq,dEnterria:2008puz,deFavereaudeJeneret:2009db}. A nonexhaustive list of photon-fusion processes observed for the first time at the LHC includes 
light-by-light (LbL) scattering $\gaga\to\gaga$~\cite{ATLAS:2017fur,CMS:2018erd,ATLAS:2019azn,ATLAS:2020hii}, high-mass dileptons $\gaga\to\ell^+\ell^-$~\cite{CMS:2018erd,ATLAS:2015wnx,ATLAS:2017sfe,CMS:2018uvs,ATLAS:2020epq,ATLAS:2022ryk,CMS:2022arf}, and W-boson pair $\gaga\to\mathrm{W^+W^-}$~\cite{CMS:2013hdf,CMS:2016rtz,ATLAS:2016lse} production. 
\begin{figure}[htpb!]
\centering
\includegraphics[width=1.05\textwidth]{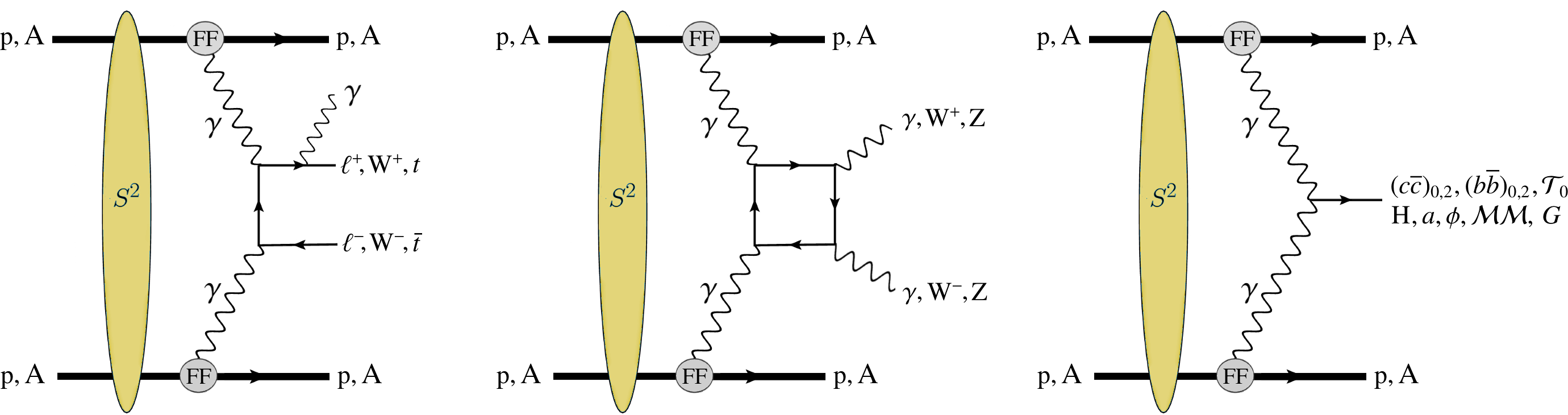}
\caption{Typical exclusive $\gaga$ collision processes in UPCs of proton and ions (with form factors FF and survival probabilities $S^2$) that can be automatically generated with the \gammaUPC\ code: $t$-channel charged particle pair production with final-state photon radiation (left), box diagrams for diboson production (center), and resonant production of SM and BSM spin-even states (right).
\label{fig:gaga_diags}}
\end{figure}
Competitive searches for anomalous quartic gauge couplings (aQGC)~\cite{dEnterria:2013zqi,Pierzchala:2008xc}, axion-like-particles (ALPs)~\cite{Knapen:2016moh}, Born--Infeld (BI) extensions of Quantum Electrodynamics (QED)~\cite{Ellis:2017edi}, or anomalous $\tau$ electromagnetic (e.m.) moments~\cite{delAguila:1991rm,Atag:2010ja,Beresford:2019gww,Dyndal:2020yen} have thereby been performed, and many more studies of the Standard Model (SM) and beyond (BSM) are open to study in the near future~\cite{Bruce:2018yzs,Klein:2020nvu,dEnterria:2022sut}. Multiple SM and BSM $\gaga$ processes accessible in UPCs at hadron colliders are displayed in Fig.~\ref{fig:gaga_diags} and listed in Table~\ref{tab:gagaPhys}.

\begin{table}[htpb!]
\tabcolsep=1.mm
\centering
\caption[]{Gold-plated SM and BSM processes accessible via photon-photon collisions in UPCs at hadron colliders.
\label{tab:gagaPhys}}
\vspace{0.2cm}
\begin{tabular}{lc} \hline
Process  & Physics motivation \\\hline
$\gaga\to\epem,\mumu$ & ``Standard candles'' for proton/nucleus $\gamma$ fluxes, EPA calculations, and higher-order QED corrections\\
$\gaga\to\tautau$ & Anomalous $\tau$ lepton e.m.\ moments~\cite{delAguila:1991rm,Atag:2010ja,Beresford:2019gww,Dyndal:2020yen}\\
$\gaga\to\gaga$ & aQGC~\cite{dEnterria:2013zqi}, ALPs~\cite{Knapen:2016moh}, BI QED~\cite{Ellis:2017edi}, noncommut.\ interactions~\cite{Horvat:2020ycy}, extra dims.~\cite{Atag:2010bh},...\\
$\gaga\to\tata$ & Ditauonium properties (heaviest QED bound state)~\cite{dEnterria:2022ysg,dEnterria:2022alo} \\
$\gaga\to(\ccbar)_{0,2},(\bbbar)_{0,2}$ & Properties of scalar and tensor charmonia and bottomonia ~\cite{Yu:2017rfi,Chapon:2020heu} \\
$\gaga\to\mathrm{XYZ}$ & Properties of spin-even XYZ heavy-quark exotic states~\cite{Goncalves:2021ytq}\\
$\gaga\to\mathrm{VM\,VM}$ & (with $\mathrm{VM} = \rho,\omega,\phi,\,\jpsi,\Upsilon$): BFKL-Pomeron dynamics~\cite{Kwiecinski:1998sa,Kwiecinski:1999hg,
Chernyak:2014wra,Goncalves:2015sfy}\\
$\gaga\to\mathrm{W}^+\mathrm{W}^-,\,\mathrm{Z}\mathrm{Z},\,\mathrm{Z}\gamma,\cdots$ & anomalous quartic gauge couplings~\cite{Pierzchala:2008xc,deFavereaudeJeneret:2009db,Chapon:2009hh,Baldenegro:2017aen} \\
$\gaga\to\mathrm{H}$ & Higgs-$\gamma$ coupling, total H width~\cite{dEnterria:2009cwl,dEnterria:2019jty} \\
$\gaga\to\mathrm{HH}$ & Higgs potential~\cite{Belusevic:2004pz}, quartic $\gamma\gamma \mathrm{HH}$ coupling \\
$\gaga\to\ttbar$ & anomalous top-quark e.m.\ couplings~\cite{deFavereaudeJeneret:2009db,dEnterria:2009cwl} \\
$\gaga \to \tilde{\ell}\tilde{\ell},\, \tilde{\chi}^+\tilde{\chi}^-,\, \mathrm{H^{++}H^{--}}$  & SUSY pairs: slepton~\cite{deFavereaudeJeneret:2009db,Beresford:2018pbt,Harland-Lang:2018hmi}, chargino~\cite{deFavereaudeJeneret:2009db,Godunov:2019jib}, doubly-charged Higgs bosons~\cite{deFavereaudeJeneret:2009db,Babu:2016rcr}.\\
$\gaga\to a,\phi, \mathcal{MM},\,G$ & ALPs~\cite{Knapen:2016moh,Goncalves:2021pdc}, radions~\cite{Lietti:2002rq}, monopoles~\cite{Kurochkin:2006jr,Dougall:2007tt,Baines:2018ltl,MoEDAL:2019ort}, gravitons~\cite{Atwood:1999zg,Zhou:2007wfr,Inan:2012zz},...\\
\hline
\end{tabular}
\end{table}

The photons coherently emitted from a charged hadron must have a wavelength larger than the size of the latter, such that they do not resolve the individual hadron constituents (partons or nucleons in the case of protons or nuclei, respectively) but see the coherent action of them. Such coherence emission condition forces the photons to be almost on-mass shell, limiting their virtuality $Q^2=-q^2$ to very low values\footnote{Natural units, $\hbar=c=1$, are used throughout the paper.} $Q^{2} < 1/R^{2}$, where $R$ is the charge radius: $Q^2\approx~0.08$~GeV$^2$ for protons with $R\approx~0.7$~fm, and $Q^2<$~4$\cdot10^{-3}$~GeV$^2$ for nuclei with $R_\mathrm{A}\approx 1.2\,A^{1/3}$~fm, for mass number $A>$~16. With the hadrons interacting only electromagnetically at large impact parameters without hadronic overlap, and surviving the emission of the quasireal photon, the $\gaga$ production processes are called exclusive or elastic (when only one hadron survives the UPC, the processes are called semiexclusive or semielastic). The photon spectra in the longitudinal direction have a typical $E_{\gamma}^{-1}$ bremsstrahlung-like spectrum up to energies of the order of $E_{\gamma}^\text{max}\approx\gamma_\mathrm{L}/R$, where $\gamma_\mathrm{L}=E_\text{beam}/m_\text{p,N}$ is the Lorentz relativistic factor of the proton (mass m$_\text{p}$~=~0.9383~GeV) or ion (nucleon mass m$_\text{N}$~=~0.9315~GeV), beyond which the $\gamma$ flux is further exponentially suppressed. The photon energies determine the rapidity of the produced system, $y = 0.5\, \ln(E_{\gamma_1}/E_{\gamma_2})$, and the \cm\ energy $W_{\gaga}=\mgg=\sqrt{4E_{\gamma_1}E_{\gamma_2}}$ which, for symmetric systems, is maximal at $y=0$ when $E_{\gamma_1}^\text{max}=E_{\gamma_2}^\text{max} \approx \gamma/b_\text{min}$ with $b_\text{min}$ the minimum impact parameter between the two charges of radius $R_\mathrm{A,B}$. Table~\ref{tab:1} summarizes the typical parameters for \pp, \pA, and \AaAa\ UPCs at the LHC and Future Circular Collider (FCC) energies, illustrating the impressive range of maximum photon-photon \cm\ energies $\sqrtsgg\approx 0.2$--30~TeV covered. The HL-LHC integrated luminosities for light-ion runs are taken from~\cite{Bruce:2018yzs,dEnterria:2022sut}, although there are intriguing proposals to significantly enhance them for Ca-Ca collisions~\cite{Krasny:2020wgx}.  Compared to the $\epem$ and \pp\ cases, the main advantage of studies of photon-fusion processes via \AaAa\ UPCs is the lack of pileup collisions and the huge $Z^2$ photon-flux boost that leads to $\gaga$ cross sections comparatively enhanced by factors of up to $Z^4\approx 50\cdot 10^6$ for \PbPb. On the other hand, proton beams at the LHC feature $\mathcal{O}(10^8)$ larger $\LumiInt$, have forward proton detectors available to tag such collisions at high masses~\cite{CMS:2021ncv,Tasevsky:2015xya}, and have harder $\gamma$ spectra compared to the heavy-ion case. All~such \pp\ differences eventually compensate for the \PbPb\ advantages above $W_{\gaga}\equiv\sqrtsgg\approx~100$--300~GeV (depending on single- or double-proton tagging)~\cite{CMS:2021ncv,Bruce:2018yzs}. Adding forward downstream proton spectrometers at 400~m in the LHC tunnel would cover collisions down to $W_{\gaga}\approx 50$~GeV~\cite{FP420RD:2008jqg}.\\

\begin{table}[htpb!]
\tabcolsep=2.5mm
\centering
\caption[]{Summary of the generic characteristics of photon-photon collisions in ultraperipheral proton and nuclear collisions at HL-LHC~\cite{Bruce:2018yzs,dEnterria:2022sut} and FCC~\cite{FCC:2018vvp,Dainese:2016gch} energies. For each colliding system, we quote its
(i) nucleon-nucleon (NN) \cm\ energy $\sqrtsnn$, (ii) integrated luminosity per typical run $\LumiInt$, (iii) beam energies $\rm E_{beam}$, (iv) Lorentz factor $\gamma_\mathrm{L}$, (v) effective charge radius $R_\mathrm{A}$, (vi) photon ``maximum'' energy $E_{\gamma}^\text{max}$ in the \cm\ frame, and
(vii) ``maximum'' photon-photon \cm\ energy $\sqrt{s_{\gaga}^\text{max}}$.}
\label{tab:1}
\vspace{0.2cm}
\begin{tabular}{lccccccc}
\hline
System  & $\sqrtsnn$ & $\LumiInt$ & $E_\text{beam1}+E_\text{beam2}$ & $\gamma_\mathrm{L}$ &  $R_\mathrm{A}$ & $E_{\gamma}^\text{max}$ & $\sqrt{s_{\gaga}^\text{max}}$
\\\hline
\PbPb\ & 5.52 ~TeV &  5~nb$^{-1}$ & 2.76 + 2.76~TeV & 2960 & 7.1 fm &  80 GeV & 160 GeV 
\\ 
Xe-Xe & 5.86~TeV &  30~nb$^{-1}$ & 2.93 + 2.93~TeV & 3150 & 6.1 fm & 100 GeV & 200 GeV 
\\ 
Kr-Kr & 6.46~TeV & 120~nb$^{-1}$ & 3.23 + 3.23~TeV & 3470 & 5.1 fm & 136 GeV & 272 GeV 
\\ 
Ar-Ar & 6.3 ~TeV & 1.1~pb$^{-1}$ & 3.15 + 3.15~TeV & 3390 & 4.1 fm & 165 GeV & 330 GeV 
\\ 
Ca-Ca   & 7.0 ~TeV & 0.8~pb$^{-1}$ & 3.5 + 3.5~TeV   & 3760 & 4.1 fm & 165 GeV & 330 GeV 
\\ 
O-O   & 7.0 ~TeV & 12.0~pb$^{-1}$ & 3.5 + 3.5~TeV   & 3760 & 3.1 fm & 240 GeV & 490 GeV 
\\ 
\pPb\  & 8.8~TeV  &   1~pb$^{-1}$ & 7.0 + 2.76~TeV  & 7450, 2960 & 0.7, 7.1 fm & 2.45~TeV, 130 GeV & 2.6~TeV 
\\ 
\pp\   & 14~TeV   &   150~fb$^{-1}$ & 7.0 + 7.0~TeV   & 7450 & 0.7 fm & 2.45~TeV& 4.5~TeV 
\\ \hline 
\PbPb\ &  39.4~TeV  & 110~nb$^{-1}$ & 19.7 + 19.7~TeV & 21\,100 & 7.1 fm & 600 GeV & 1.2~TeV 
\\ 
\pPb\  &  62.8~TeV  &  29~pb$^{-1}$ & 50. + 19.7~TeV  & 53\,300, 21\,100 & 0.7,7.1 fm & 15.2~TeV, 600 GeV & 15.8~TeV 
\\ 
\pp\   & 100~TeV  &   1~ab$^{-1}$ & 50. + 50.~TeV   & 53\,300 & 0.7 fm & 15.2~TeV &  30.5~TeV 
\\\hline
\end{tabular}
\end{table}

Studies of photon-photon physics in UPCs with hadron beams at RHIC, LHC, and FCC have been so far carried out mostly employing dedicated Monte Carlo (MC) event generators such as \starlight~\cite{Klein:2016yzr}, \superchic~\cite{Harland-Lang:2020veo}, or \textsc{fpmc} (for \pp\ UPCs only)~\cite{Boonekamp:2011ky}, where a subset of selectable physical processes has been previously coded at leading-order (LO) QED accuracy. There is an increasing experimental and phenomenological need to have at hand more versatile MC generators that can automatically produce any final state of interest, including new SM and BSM signals, as well as any potential backgrounds (including, \eg, the generation of additional photon and/or gluon emissions from the final state particles), and that can be extended to include next-to-leading (NLO) pure QED or full electroweak (EW) corrections. Standard MC tools to automatically generate any collider final state of interest are \madgraph\ (called \mgshort\ hereafter)~\cite{Alwall:2011uj,Alwall:2014hca} for generic SM/BSM studies, and \helaconia~\cite{Shao:2012iz,Shao:2015vga} for dedicated studies of charmonium and bottomonium physics. At variance with the UPC-only MC generators, \mgshort\ and \helaconia\ can not only produce any arbitrary final state but also generate events with additional higher-order real (photon and/or gluon) emissions, \mgshort\ is extendable to include also full NLO (real and virtual) EW corrections~\cite{Frederix:2018nkq}, and their full events are by default output in a convenient Les Houches Event (LHE) format~\cite{Alwall:2006yp} that can be automatically interfaced to external codes for the subsequent showering and hadronization (in the case of partonic final states) and/or decay of the produced particles.\\ 

In the case of \pp\ collisions, the \mgshort\  generator already contains the possibility to produce arbitrary photon-induced final states via two different setups. The first one uses the inclusive photon distribution function (PDF) of the proton~\cite{Frederix:2018nkq}, such as the LuxQED~\cite{Manohar:2016nzj}, NNPDF31luxQED~\cite{Bertone:2017bme}, MMHT2015qed~\cite{Harland-Lang:2019pla} or CT18lux~\cite{Xie:2021equ} ones, where the photon is mostly emitted from the individual partons of the proton, which does not survive the QED interaction.
The second setup, which is the main subject of this work, deals with the EPA case where only the coherent $\gamma$ emission by the proton is considered. 
The $\gamma$ flux currently implemented in \mgshort, dubbed ``improved Weizs\"acker-Williams'' (iWW) (following~\cite{Frixione:1993yw}), is obtained from the proton elastic electric ($E$) and magnetic ($M$) form factors in the dipole approximation\footnote{The \superchic\ MC generator uses the alternative fit from the A1 collaboration~\cite{A1:2013fsc}.}, $F_M=G_M^2$ and  $F_E=(4m_\mathrm{p}^2G_E^ 2+Q^2G_M^2)/(4m_\mathrm{p}^2+Q^2)$ where $G_E$ and $G_M$ are the ``Sachs'' form factors related by $G_E^2 = G_M^2 /7.78 = (1+Q^2/Q_0^2)^{-4}$, with $Q_0^2\approx0.71\,\text{GeV}^2$. The photon number density as a function of the fraction of the proton energy carried by the photon, $x = E_{\gamma}/E_\mathrm{p}$, reads~\cite{Budnev:1975poe}
\begin{eqnarray}
n_{\gamma/\mathrm{p}}^\mathrm{iWW}(x) & = & 
\frac{\alpha}{\pi}\,(1-x)\;\left[\varphi\left(x,Q^2_\text{max}/Q_0^2\right)-\varphi\left(x,Q^2_\text{min}/Q_0^2\right)\right]\;,
\mbox{with} \label{eq:flux_p}\\
 & & \varphi(x,Q)  = (1+c_1\,\vary)\,\left[-\ln\,\tfrac{1+Q}{Q}+\sum_{k=1}^3\tfrac{1}{k\,(1+Q)^k}\right] \, + \, \tfrac{(1-c_2)\,\vary}{4Q(1+Q)^3}\,
+ c_3\,\left(1+\tfrac{\vary}{4}\right)\,\left[\ln\,\tfrac{(1+Q)-c_2}{1+Q}+\sum_{k=1}^3\tfrac{c_2^k}{k(1+Q)^k}\right]
\end{eqnarray}
where $\alpha=1/137.036$ is the QED coupling, $\vary=x^2/(1-x)$, and 
$c_1=(1+7.78)/4+4\,m_\mathrm{p}^2/Q_0^2\approx 7.16$, $c_2=1-4m_\mathrm{p}^2/Q_0^2\approx -3.96$, and $c_3=(7.78-1)/c_2^4\approx 0.028$ are constants. The minimum momentum transfer squared is a function of $x$ and the proton mass, $Q^2_\text{min} \approx (x m_\mathrm{p})^2/(1 - x)$, and a value of $Q^2_\text{max} \approx 1$--2~GeV$^2$ is usually taken to warrant the ``onshellness'' of the photon\footnote{Older \mgshort\ versions~\cite{Alwall:2007st} used $Q^2_\text{max} = \mu^2_\mathrm{F}$ (factorization scale squared), which is not theoretically correct but not numerically important as the flux is almost negligible above $Q^2\approx 2$~GeV$^2$.}. However, as we discuss below, the current \mgshort\ implementation of \pp\ UPCs~\cite{deFavereaudeJeneret:2009db} does not explicitly consider the survival of the protons, a fact that does not warrant the exclusivity condition of the final state. Accounting for such effects has been usually done by introducing a correction factor to the cross section, called the ``survival probability'' $S^2$~\cite{Dokshitzer:1987nc}, which corresponds to the probability that both scattered protons do not dissociate due to secondary soft hadronic interactions (yellow ``blob'' in the Fig.~\ref{fig:gaga_diags} diagrams). Calculations of the survival factors are usually done in the impact parameter space, assuming factorization as in the EPA. Since the photon $Q$ is inversely proportional to the impact parameter of the \pp\ collision, which is usually much larger than the range of strong interactions, the proton survival probability in e.m.\ interactions has been so far \textit{de facto} taken as $S^2_{\gaga} = 1$ in \mgshort. However, since the average $Q^2$ increases with $\gamma$ energy, one expects a decreasing survival probability for processes with larger $W_{\gaga}$. Therefore, the current \mgshort\ EPA setup should be considered as just providing a reasonable upper value of the cross section for high-mass exclusive $\gaga$ processes in \pp\ UPCs.\\

This paper provides a description of the new ingredients that have been incorporated into the \mgshort\ and \helaconia\ MC codes in order to be able to generate any exclusive photon-photon final state of interest, not only with proton but also with nuclear beams, including two modelings of the underlying hadronic form factors and associated survival probabilities (represented, respectively, by the grey circle and the yellow ``blob'' in the diagrams of Fig.~\ref{fig:gaga_diags}).
The paper is organized as follows. Section~\ref{sec:sigma_gaga} provides a short reminder of the basic expressions to compute photon-fusion cross sections in the EPA framework. 
Section~\ref{sec:gaga_lumis} describes the new \gammaUPC\ proton and heavy-ion EPA photon fluxes incorporated into \mgshort/\helaconia\ based on the standard electric dipole form factor (EDFF) as well as on the charge form factor (ChFF), and associated survival factors for \pp, \pA, and \AaAa\ collisions. Results for a broad selection of exclusive $\gaga$ processes at hadron colliders are presented in Sections~\ref{sec:results} and~\ref{sec:diff_results}, including total cross sections for a large variety of resonances with even charge-conjugation ($C$) quantum number, BSM particles, as well as differential distributions for LbL and exclusive $\ell^+\ell^-$ production. Predictions for the latter are compared to the LHC data as well as to those of the \starlight\ and \superchic\ models. For all our calculations, the EDFF- and ChFF-based results are confronted and half the difference between their numerical cross sections is taken as indicative of the associated FF and $S^2$ uncertainties. Details on the \gammaUPC\ code output and ongoing developments of the framework to be implemented in upcoming releases are discussed in Section~\ref{sec:upcoming}. The paper is closed with a summary in Section~\ref{sec:summ}, and an appendix \ref{sec:app} with basic instructions to compile and run the code.

\section{Theoretical \texorpdfstring{$\gaga$}{gamma-gamma} cross sections}
\label{sec:sigma_gaga}

In the EPA framework, the exclusive production cross section of a final state $X$ via photon fusion in an UPC of hadrons A and B with charges $Z_{1,2}$, \ABgagaX, factorizes into the product of the elementary cross section at a given $\gaga$ \cm\ energy, $\sigmagagaX(W_{\gaga})$, convolved with the two-photon differential distribution of the colliding beams,
\begin{equation}
\sigma(\mathrm{A}\; \mathrm{B}\,\xrightarrow{\gaga} \mathrm{A} \; X \; \mathrm{B})=
\int \frac{dE_{\gamma_1}}{E_{\gamma_1}} \frac{dE_{\gamma_2}}{E_{\gamma_2}} \, \frac{\mathrm{d}^2N^{(\mathrm{AB})}_{\gamma_1/\mathrm{Z}_1,\gamma_2/\mathrm{Z}_2}}{\mathrm{d}E_{\gamma_1}\mathrm{d}E_{\gamma_2}} \sigmagagaX(W_{\gaga})\,.
\label{eq:two-photon}
\end{equation}
where 
\begin{equation}
\frac{\mathrm{d}^2N^{(\mathrm{AB})}_{\gamma_1/\mathrm{Z}_1,\gamma_2/\mathrm{Z}_2}}{\mathrm{d}E_{\gamma_1}\mathrm{d}E_{\gamma_2}} =  \int{\mathrm{d}^2\pmb{b}_1\mathrm{d}^2\pmb{b}_2\,P_\text{no\,inel}(\pmb{b}_1,\pmb{b}_2)\,N_{\gamma_1/\mathrm{Z}_1}(E_{\gamma_1},\pmb{b}_1)N_{\gamma_2/\mathrm{Z}_2}(E_{\gamma_2},\pmb{b}_2)}\,.\label{eq:2photonintegral}
\end{equation}
is derived from the convolution of the two photon number densities  $N_{\gamma_i/\mathrm{Z}_i}(E_{\gamma_i},\pmb{b}_i)$ with energies $E_{\gamma_{1,2}}$ at impact parameters $\pmb{b}_{1,2}$ from hadrons A and B, respectively\footnote{The vectors $\pmb{b}_{1}$ and $\pmb{b}_{2}$ have their origins at the center of each hadron, and, therefore, $|\,\pmb{b}_{1}-\pmb{b}_{2}|$ is the impact parameter between them.}; and $P_\text{no\,inel}(\pmb{b}_1,\pmb{b}_2)$ encodes the probability of hadrons A and B to remain intact after their interaction, which depends on their relative impact parameters. The $\gaga$ survival factor can then be written as
\begin{equation}
S^2_{\gaga} = \frac{
\int{\mathrm{d}^2\pmb{b}_1\mathrm{d}^2\pmb{b}_2\,P_\text{no\,inel}(\pmb{b}_1,\pmb{b}_2)\,N_{\gamma_1/\mathrm{Z}_1}(E_{\gamma_1},\pmb{b}_1)N_{\gamma_2/\mathrm{Z}_2}(E_{\gamma_2},\pmb{b}_2)}
}
{
\int{\mathrm{d}^2\pmb{b}_1\mathrm{d}^2\pmb{b}_2\,N_{\gamma_1/\mathrm{Z}_1}(E_{\gamma_1},\pmb{b}_1)N_{\gamma_2/\mathrm{Z}_2}(E_{\gamma_2},\pmb{b}_2)}
},
\label{eq:S2}
\end{equation}
where the numerator is the two-photon density accounting for finite-size effects, Eq.~(\ref{eq:2photonintegral}), and the denominator represents the integral of the two photon fluxes over all impact parameters without hadronic overlap constraint. The role of the modeling of $S^2_{\gaga}$ in \pp\ UPCs cross sections at the LHC has been discussed in~\cite{Dyndal:2014yea,Harland-Lang:2021ysd}.\\

In the case of \pp\ UPCs calculations that ignore the hadronic-nonoverlap condition, the $\gamma$ flux has no explicit dependence on the impact parameter, \ie\ $n_{\gamma}(E_{\gamma}) = \int{N_{\gamma/\mathrm{p}}(E_{\gamma},\pmb{b})\,\mathrm{d}^2\pmb{b}}$, the survival factor is unity, and the two-photon distribution just factorizes as the product of two PDF-like photon distributions,
\begin{equation}
\frac{\mathrm{d}^2N^{(\mathrm{pp,factorized})}_{\gamma_1/\mathrm{Z}_1,\gamma_2/\mathrm{Z}_2}}{\mathrm{d}E_{\gamma_1}\mathrm{d}E_{\gamma_2}} = n_{\gamma/\mathrm{p}}(x_1)\,n_{\gamma/\mathrm{p}}(x_2)\,,\label{eq:gagalumipoint}
\end{equation}
where $n_{\gamma/\mathrm{p}}(x)$ is given by Eq.~(\ref{eq:flux_p}) for the EPA case, or by LuxQED-type PDFs for inclusive $\gaga$ collisions, in the current \mgshort\ implementation.\\

A particular case of interest in two-photon physics is the production of
spin-0 and spin-2 resonances since, for real photons, the $\gaga \to$~vector process is forbidden by the Landau--Yang theorem~\cite{Landau:1948kw,Yang:1950rg}. The cross section for the exclusive production of a $C$-even resonance $X$ (with spin $J$, and $\Gamma_{\gaga}(X)$ two-photon width) through $\gaga$ fusion in an UPC of charged particles A and B, is given by~\cite{Budnev:1975poe}
\begin{equation}
\sigma(\mathrm{A}\; \mathrm{B}\,\xrightarrow{\gaga} \mathrm{A} \; X \; \mathrm{B}) = 
4\pi^2 (2 J+1)\frac{\Gamma_{\gaga}(X)}{m_X^2} 
    \left. \frac{\mathrm{d}{\Lumi}^{(\mathrm{A}\,\mathrm{B})}_{\gaga}}{\mathrm{d}W_{\gaga}} \right|_{W_{\gaga}=m_X},
\label{eq:sigma_AA_X}
\end{equation}
where $\frac{d{\Lumi}^{(\mathrm{A\,B})}_{\gaga}}{dW_{\gaga}}\big|_{W_{\gaga}=m_X}$ is the value of the effective two-photon luminosity at the resonance mass $m_X$, amounting to
\begin{eqnarray}
\frac{\mathrm{d}{\Lumi}^{(\mathrm{AB})}_{\gaga}}{\mathrm{d}W_{\gaga}}&=&\frac{2W_{\gaga}}{s_{_\mathrm{NN}}}\int{\frac{\mathrm{d}E_{\gamma_1}}{E_{\gamma_1}}\frac{\mathrm{d}E_{\gamma_2}}{E_{\gamma_2}}\delta\left(\frac{W_{\gaga}^2}{s_{_\mathrm{NN}}}-\frac{4E_{\gamma_1}E_{\gamma_2}}{s_{_\mathrm{NN}}}\right)\frac{\mathrm{d}^2N^{(\mathrm{AB})}_{\gamma_1/\mathrm{Z}_1,\gamma_2/\mathrm{Z}_2}}{\mathrm{d}E_{\gamma_1}\mathrm{d}E_{\gamma_2}}}\,.\label{eq:gagalumi}
\end{eqnarray}

The expressions above, Eqs.~(\ref{eq:two-photon})--(\ref{eq:2photonintegral}) and Eqs.~(\ref{eq:sigma_AA_X})--(\ref{eq:gagalumi}), are valid for any colliding system with the appropriate (charged lepton, proton, and/or heavy ion) photon fluxes and survival probabilities. For $\epem$ beams, the photon flux in the WW approximation~\cite{Kniehl:1996we} is commonly used (also cf.\ Eq.~(3) of~\cite{Flore:2020jau}) in Eq.~(\ref{eq:gagalumipoint}), with the maximum virtuality usually set to $Q^2_\mathrm{max} \approx 1$~GeV$^2$ when focusing on quasireal photon scatterings without the need to tag the $e^\pm$ transversely scattered at large angles. For proton beams one normally employs the $\gamma$ spectrum obtained from its elastic form factor, Eq.~(\ref{eq:flux_p}), whereas the impact-parameter-dependent expression from $b_\text{min}$ to infinity is used for the $\gamma$ spectrum of heavy ions~\cite{Bertulani:1987tz}. As aforementioned, in the case of proton and nuclear beams, an extra requirement needs however to be imposed to ensure that the collisions are truly exclusive, namely that they occur without hadronic interactions and subsequent breakup of the colliding particle beams. In the next section, we discuss the new photon fluxes and nonoverlap conditions incorporated into the \mgshort\ and \helaconia\ generators. 

\section{Effective photon-photon luminosities}
\label{sec:gaga_lumis}

\begin{figure}[htpb!]
\centering
\includegraphics[width=0.65\textwidth]{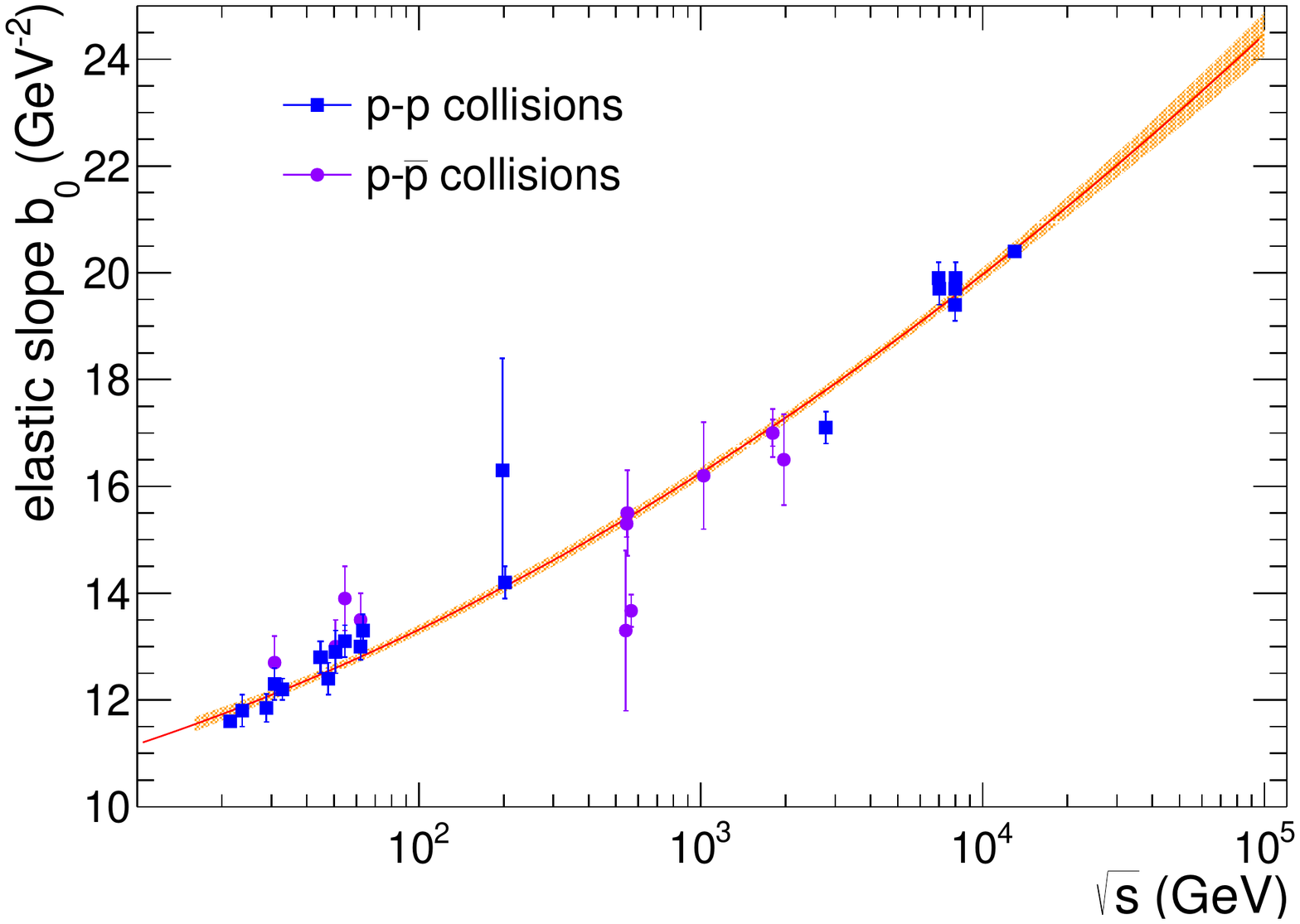}
\caption{Measurements of the low-$|t|$ exponential elastic slope $b_0$ in \pp~\cite{TOTEM:2012oyl,TOTEM:2013lle,TOTEM:2017asr,TOTEM:2018psk,STAR:2020phn,ATLAS:2014vxr,ATLAS:2016ygv} and \ppbar~\cite{ParticleDataGroup:2010dbb} collisions as a function of $\sqrts=\sqrt{s_{_\text{NN}}}$ (individual data points at the same $\sqrts$ have been slightly shifted to the left or right to improve visibility). The orange curve shows our fit to the data, $b_0 = A + B \,\ln(s_{_\text{NN}}) + C \ln^{2}(s_{_\text{NN}})$, with the parameters given in the text.
\label{fig:Bslope}}
\end{figure}

At variance with photon-photon processes from pointlike emitters, the effective $\gaga$ luminosity in UPCs with hadrons cannot be just simply factorized as a direct convolution of the product of the photon densities of the two beams, such as in Eq.~(\ref{eq:gagalumipoint}), because of their finite transverse profile and the consequent nonzero probability of concomitant hadronic interactions that can break the exclusivity condition. In past $\gaga$-fusion studies with \mgshort\ (see \eg~\cite{dEnterria:2009cwl,dEnterria:2013zqi,dEnterria:2019jty}), this effect has been often only partially accounted for either by imposing a maximum $Q^2_\mathrm{max} \approx 1$~GeV$^2$ value for the photon flux in \pp\ UPCs (a choice that de facto removes the most central $\gaga$ collisions with potential hadronic overlap), or by restricting the range of minimum impact parameters in the $\gamma$ fluxes to $b_\text{min}=R_\mathrm{A,B}$ plus an effective correction equivalent to the geometrical condition $|\pmb{b}_{1} - \pmb{b}_{2}| > R_\mathrm{A}+R_\mathrm{B}$~\cite{Cahn:1990jk} in the case of \pA\ and \AaAa\ UPCs. A more realistic approach is considered here, similar to the ones implemented in the \starlight\ and \superchic\ MC generators. 
The two-photon differential yield (\ref{eq:2photonintegral}), is now given by
\begin{eqnarray}
\frac{\mathrm{d}^2N^{(\mathrm{AB})}_{\gamma_1/\mathrm{Z}_1,\gamma_2/\mathrm{Z}_2}}{\mathrm{d}E_{\gamma_1}\mathrm{d}E_{\gamma_2}}&=&\int{\mathrm{d}^2\pmb{b}_1\mathrm{d}^2\pmb{b}_2\, P_\text{no\,inel}\left(\left|\pmb{b}_1-\pmb{b}_2\right|\right)\,N_{\gamma_1/\mathrm{Z}_1}(E_{\gamma_1},\pmb{b}_1)N_{\gamma_2/\mathrm{Z}_2}(E_{\gamma_2},\pmb{b}_2)\,\theta(b_1-\epsilon R_\mathrm{A})\theta(b_2-\epsilon R_\mathrm{B})}\,.\label{eq:twophotonyield}
\end{eqnarray}
In this expression, $\theta(b_{1,2}-\epsilon R_\mathrm{A,B})$ is the Heaviside step function, and the $\epsilon>0$ parameter can be used to restrict the range of impact parameters depending on the concrete implementation of the photon EPA fluxes as explained below; and $P_\text{no\,inel}(b)$ is the probability to have no inelastic hadronic interaction at impact parameter $b$ given by standard opacity (optical density) or eikonal expressions~\cite{Glauber:1970jm}:
\begin{eqnarray}
P_\text{no\,inel}\left(b\right)&=&\left\{\begin{array}{ll} 
e^{-\,\sigma^{\mathrm{NN}}_{\text{inel}}\cdot T_{\mathrm{AB}}(b)}, & \text{for nucleus-nucleus UPCs}\\
e^{-\,\sigma^{\mathrm{NN}}_{\text{inel}}\cdot T_\mathrm{A}(b)}, & \text{for proton-nucleus UPCs}\\
\left|1-\Gamma(s_{_\text{NN}},b)\right|^2,\; \mbox{ with }\;\Upgamma(s_{_\mathrm{NN}},b)\propto e^{-b^2/(2b_0)} & \text{for \pp\ UPCs}\\
\end{array}\right..\label{eq:Psurv}
\end{eqnarray}
Here $T_\mathrm{A}(b)$ and $T_{\mathrm{AB}}(b)$ are the nuclear thickness and overlap functions respectively, commonly derived from the hadron transverse density profile via a Glauber MC model~\cite{Loizides:2017ack,dEnterria:2020dwq}, $\sigma^{\mathrm{NN}}_{\text{inel}} \equiv \sigma^{\mathrm{NN}}_{\text{inel}}(\!\sqrtsnn)$ is the inelastic NN scattering cross section parametrized as a function of $\sqrtsnn$ as in~\cite{dEnterria:2020dwq}, and $\Upgamma(s_{_\mathrm{NN}},b)$ is the Fourier transform of the \pp\ elastic scattering amplitude modelled by an exponential function~\cite{Frankfurt:2006jp}
with inverse slope $b_0 \equiv b_0(\!\sqrtsnn)$ dependent on the NN \cm\ energy. Figure~\ref{fig:Bslope} shows a compilation of all measurements of the $b_0$ slope extracted in elastic scattering measurements at low $-t\lesssim 0.3$~GeV$^{2}$ in \pp~\cite{TOTEM:2012oyl,TOTEM:2013lle,TOTEM:2017asr,TOTEM:2018psk,STAR:2020phn,ATLAS:2014vxr,ATLAS:2016ygv} and \ppbar~\cite{ParticleDataGroup:2010dbb} collisions as a function of $\sqrtsnn$. In principle, the elastic slope is defined at zero exchanged momenta ($t=0$), but the experimental determinations of $b_0$ depend on the actual chosen $|t|$-range used to extract it, and whether or not local deviations of the data from a pure exponential due to Coulomb-nuclear interference are taken into account. These facts explain some of the relative large dispersion of slopes measured at the same $\sqrts$ value, and uncertainties beyond the plotted experimental error bars should be expected in some cases. The experimental data have been fit here to the functional form $b_0(\sqrtsnn) = A + B \,\ln(s_{_\text{NN}}) + C \ln^{2}(s_{_\text{NN}})$, yielding $A=9.81$~GeV$^{-2}$, $B=0.211$~GeV$^{-2}$, and $C=0.0185$~GeV$^{-2}$ (for $s_{_\text{NN}}$ measured in GeV$^2$) with goodness-of-fit per degree-of-freedom of $\chi^2/N_\mathrm{dof}=2.3$. 
Whereas a simple logarithmic dependence $\ln(s_{_\text{NN}})$ is expected in the case of one-Pomeron exchange, the fit needs an extra $\ln^2(s_{_\text{NN}})$ term to reproduce the highest \cm\ energy data, a manifestation of the increasing role of multi-Pomeron exchanges at LHC energies and beyond~\cite{Schegelsky:2011aa}. Such a fit predicts $b_0 = 20.6, 24.5$~GeV$^{-2}$ for \pp\ collisions at LHC(14 TeV) and FCC(100 TeV), respectively. The photon number densities, $N_{\gamma/\mathrm{Z}}(E_\gamma,b)$, the key ingredient of Eq.~(\ref{eq:twophotonyield}), have been implemented as discussed next.\\

\begin{figure}[htbp!]
\centering
\includegraphics[width=0.49\textwidth,draft=false]{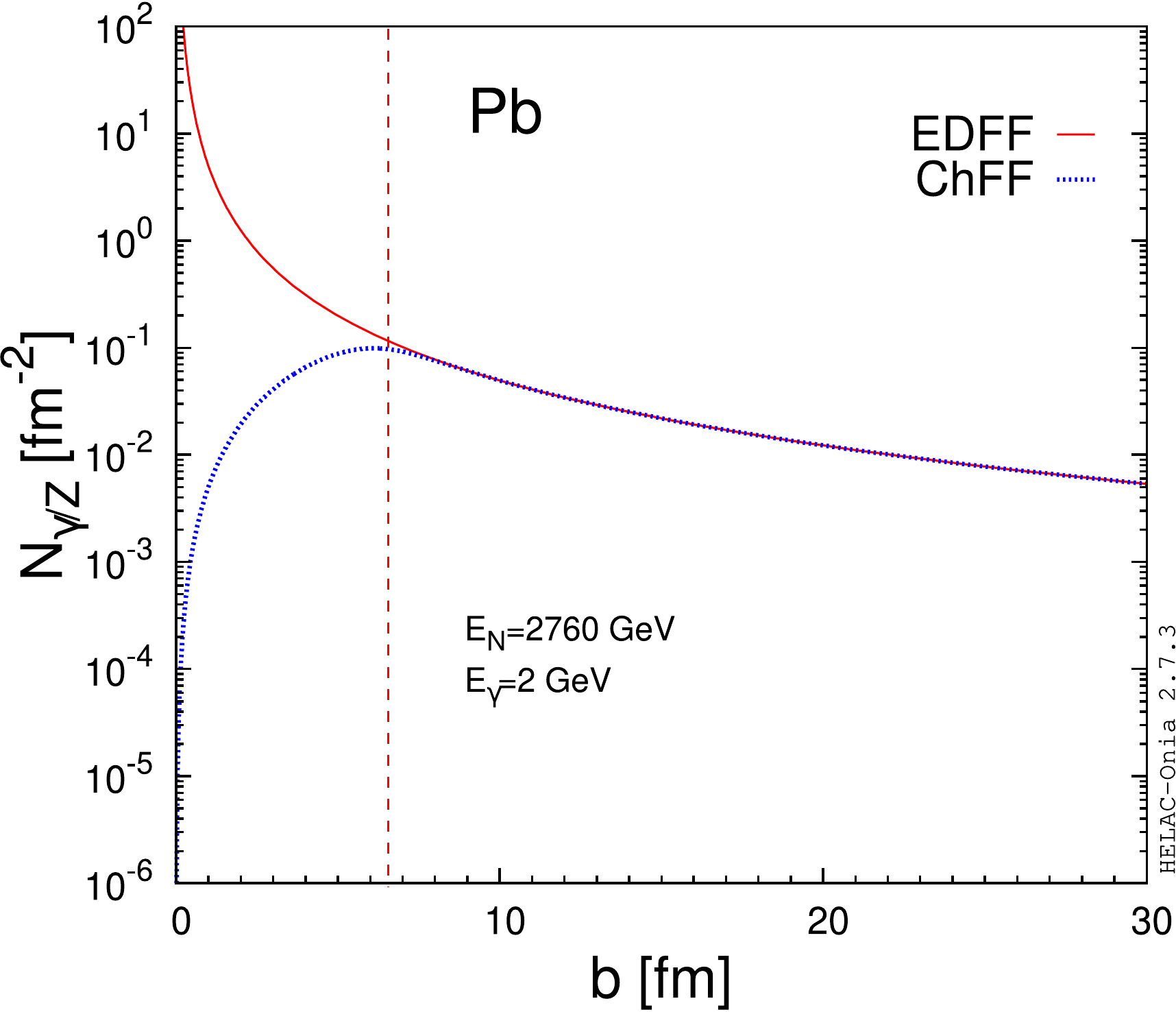}
\includegraphics[width=0.49\textwidth,draft=false]{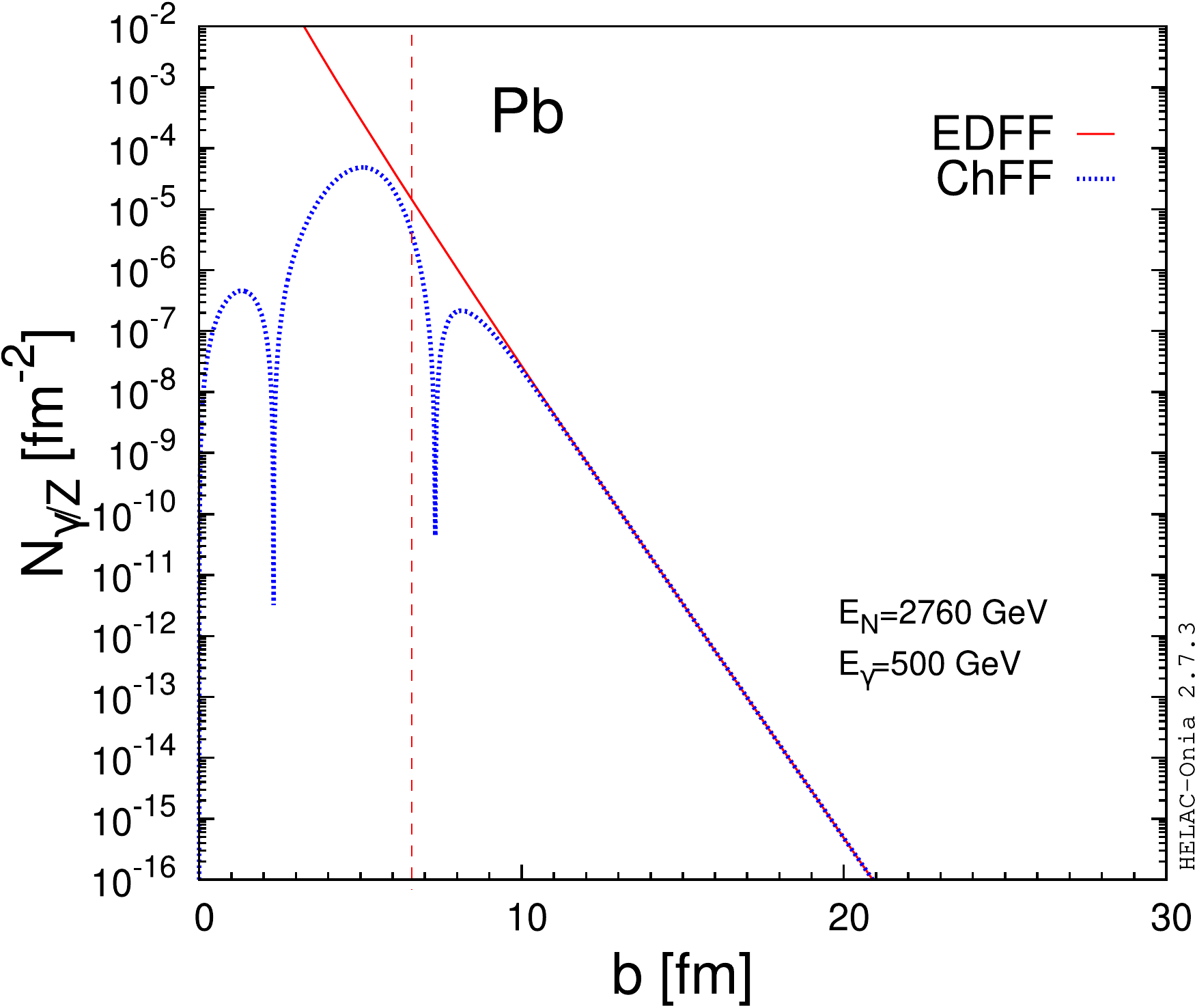}\\
\includegraphics[width=0.49\textwidth,draft=false]{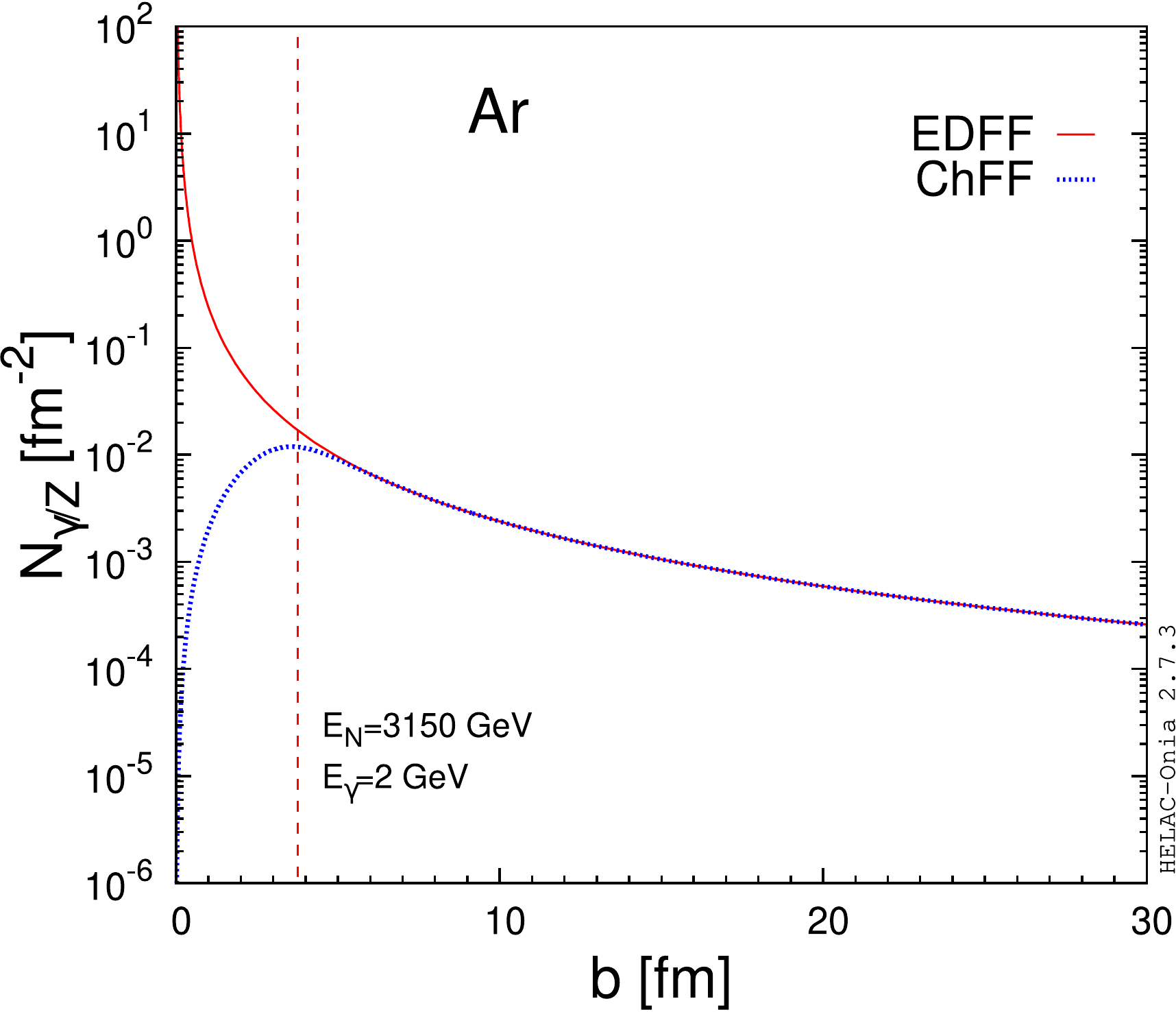}
\includegraphics[width=0.49\textwidth,draft=false]{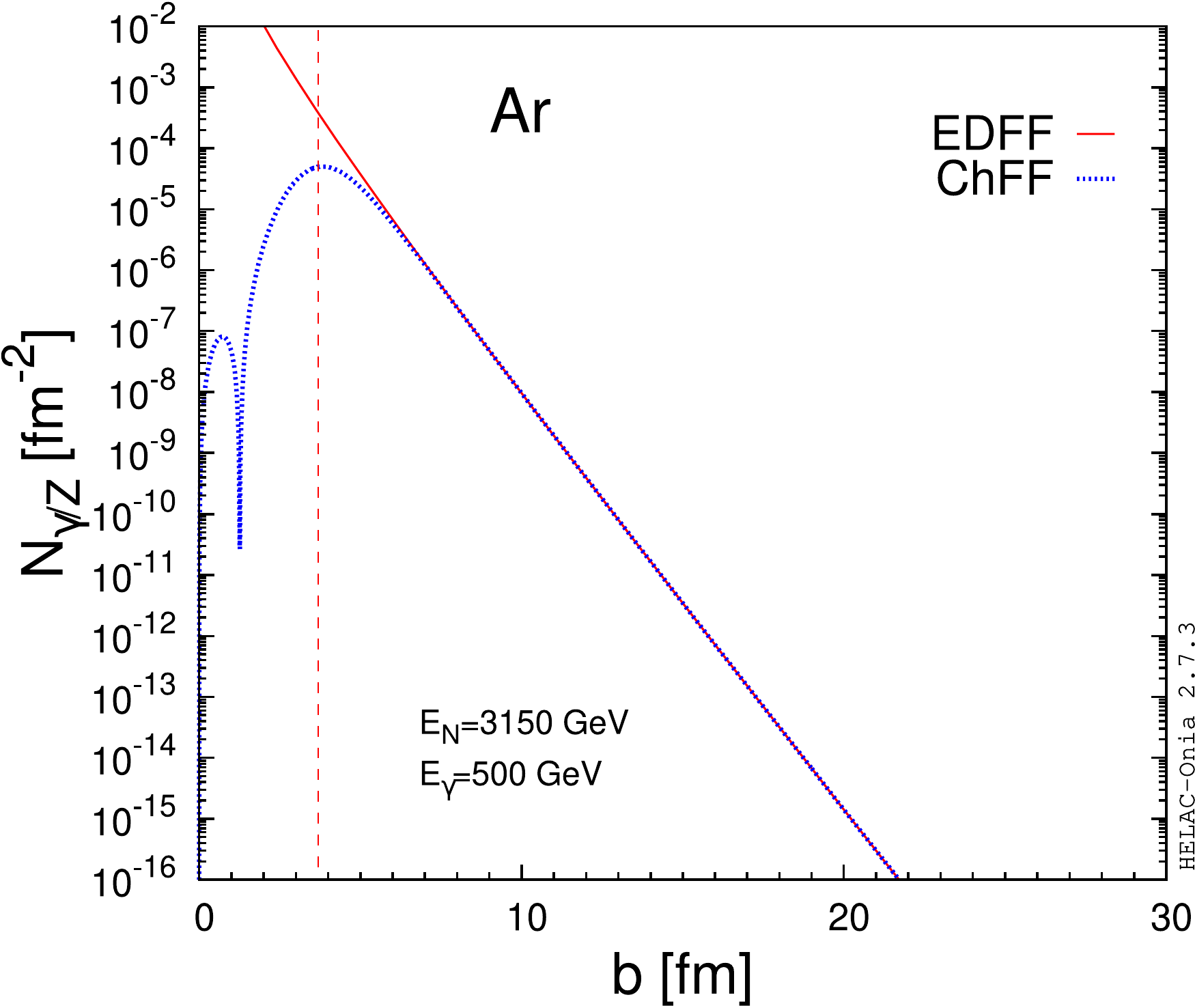}\\
\includegraphics[width=0.49\textwidth,draft=false]{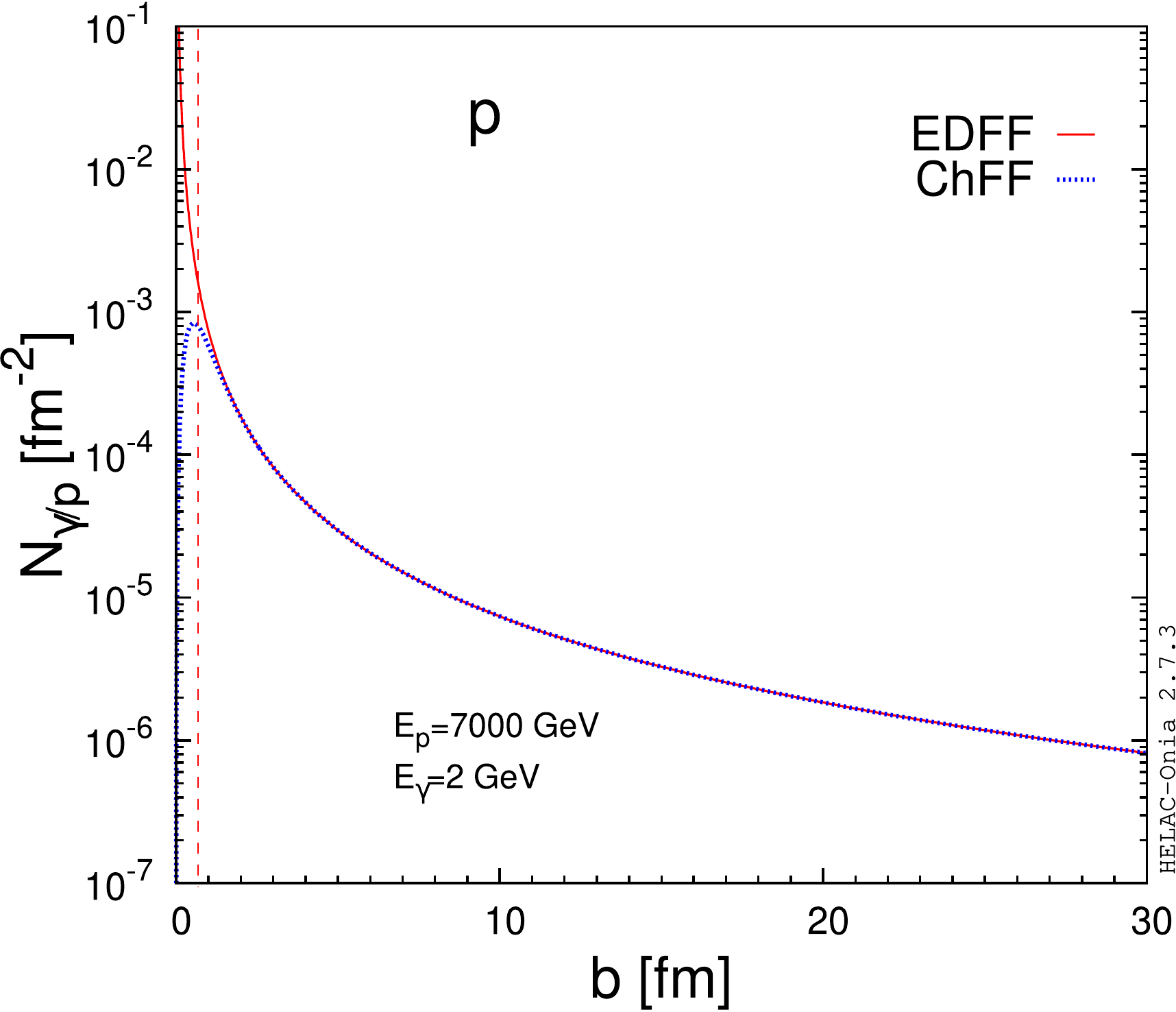}
\includegraphics[width=0.49\textwidth,draft=false]{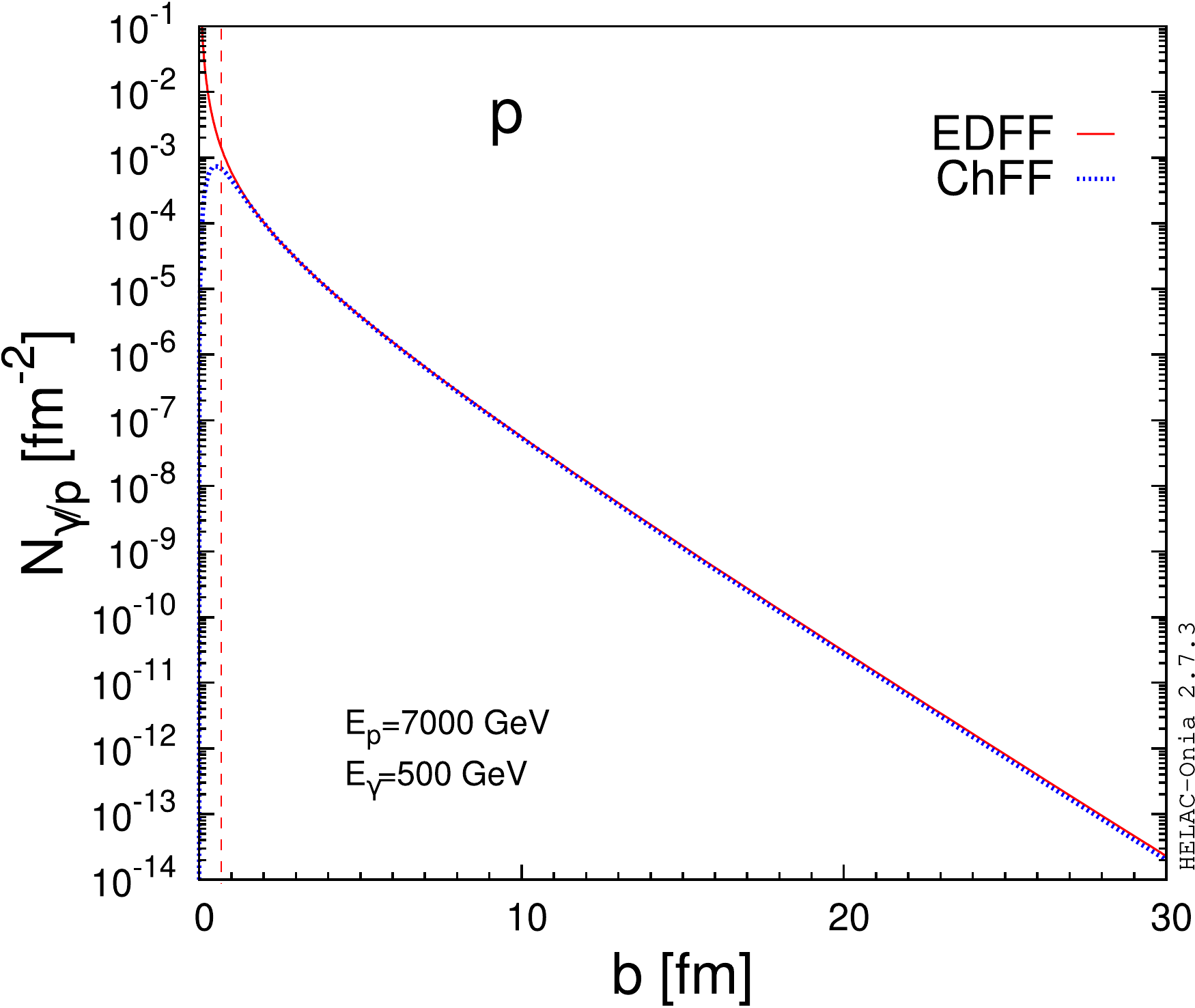}
\caption{Comparison of the photon number densities at low ($E_\gamma = 2$~GeV, left) and high ($E_\gamma = 500$~GeV, right) energies as functions of the impact parameter $b$, obtained with the two form factors considered here (EDFF, red solid, and ChFF, blue dashed) for Pb ions at $2.76$~TeV (top), Ar at $3.15$~TeV (middle), and proton at $7$~TeV (bottom). The vertical dashed red lines at $b\approx R_\mathrm{A}$ indicate the threshold lower-limit imposed on the integral of the EDFF fluxes.\label{fig:photonnumberdensity}}
\end{figure}

The first $\gamma$ flux considered in this work, and commonly used in the literature,
is derived from the electric dipole form factor (EDFF) of the emitting hadron. For ion beams with charge number $Z$ and Lorentz boost $\gamma_\mathrm{L}$, the photon number density at impact parameter $b$ obtained from its corresponding EDFF reads
\begin{eqnarray}
N^{\text{EDFF}}_{\gamma/\mathrm{Z}}(E_\gamma,b)&=&\frac{Z^2\alpha}{\pi^2}\frac{\xi^2}{b^2}\left[K_1^2(\xi)+\frac{1}{\gamma_\mathrm{L}^2}K_0^2(\xi)\right]\,,\label{eq:EDFF}
\end{eqnarray}
where $\xi=E_\gamma b/\gamma_\mathrm{L}$, and $K_i$'s are modified Bessel functions~\cite{Baltz:2007kq}. The first term inside the parentheses gives the flux of transversely polarized photons with respect to the ion direction, which dominates for relativistic nuclei, while the second one is the flux for longitudinally polarized photons. As aforementioned, the flux is exponentially suppressed for $E_\gamma \gtrsim \gamma_{\mathrm{L}}/b$ (corresponding to the $E_\gamma^\mathrm{max}$ values of Table~\ref{tab:1}). Since the EDFF photon number density is divergent when $b\to 0$ (Fig.~\ref{fig:photonnumberdensity}, blue dashed curves), the $\epsilon$ parameter in the integral Eq.~(\ref{eq:twophotonyield}) is usually taken as unity ($\epsilon^{\text{EDFF}}=1$), which is equivalent to restricting the integration to impact parameters $b_{1,2}>R_\mathrm{A,B}$ (vertical dashed lines in Fig.~\ref{fig:photonnumberdensity}, where we have taken the radius parameters as those of the corresponding Woods-Saxon nuclear profiles in Table~\ref{tab:NuclearProfile}).\\

For proton UPC fluxes, the same expression (\ref{eq:EDFF}) is applicable using $Z=1$. However, the EDFF flux for protons assuming 100\% survival probability (setting $P_\text{no\,inel}=1$ in Eq.~(\ref{eq:S2})) is not identical to the $b$-independent flux given by Eq.~(\ref{eq:flux_p}). Indeed, for $P_\text{no\,inel}=1$, one can analytically integrate (\ref{eq:EDFF}) over $b$, and obtain the effective photon PDF as
\begin{equation}
n_{\gamma/\mathrm{p}}^\mathrm{EDFF}(x)=n_{\gamma/\mathrm{p}}(x R_\mathrm{p} m_\mathrm{p}),\;
\mbox{ with }\;
n_{\gamma/\mathrm{p}}(\chi)=\frac{2\alpha}{\pi}\left[\chi K_0(\chi)K_1(\chi)-\left(1-\gamma_\mathrm{L}^{-2}\right)\frac{\chi^2}{2}\left(K_1^2(\chi)-K_0^2(\chi)\right)\right]\,,
\end{equation}
which is different than $n_{\gamma/\mathrm{p}}^\mathrm{iWW}(x)$ in Eq.~(\ref{eq:flux_p}) that keeps an explicit dependence on the photon (maximum and minimum) virtualities.\\

The second photon flux implemented in our code is that derived from the integral over the charge form factor (ChFF) of the nucleus~\cite{Vidovic:1992ik} [cf. Eq.~(43) there], \ie
\begin{eqnarray}
N^{\text{ChFF}}_{\gamma/\mathrm{Z}}(E_\gamma,b)&=&\frac{Z^2\alpha}{\pi^2}\left|\int_0^{+\infty}{\frac{d\kt \kt^2}{\kt^2+E_\gamma^2/\gamma_{\mathrm{L}}^2}F_\text{ch,A}\left(\sqrt{\kt^2+E_\gamma^2/\gamma_{\mathrm{L}}^2}\right)}\, J_1\left(b\kt\right)\right|^2\,,\label{eq:ChFF}
\end{eqnarray}
where $F_\text{ch,A}$ is the ChFF of the ion A emitting the photon, $\kt$ is the photon transverse momentum, related to its virtuality as $Q^2 =  \kt^2 + E_{\gamma}^2/\gamma_\mathrm{L}^2$, and $J_1$ is the Bessel function of the first kind. The ChFF can be related to the transverse density profile of the radiating ion A, via
\begin{eqnarray}
F_\text{ch,A}(q)&=&\int{\mathrm{d}^3\pmb{r} e^{i\pmb{q}\cdot \pmb{r}}\rho_\mathrm{A}(\pmb{r})}=\frac{4\pi}{q}\int_0^{+\infty}{\mathrm{d}r\rho_\mathrm{A}(r) r\sin{\left(qr\right)}}\,,\label{eq:FchA_ChFF}
\end{eqnarray}
with $q=\sqrt{k_{\perp}^2+m_\mathrm{N}^2 x^2}$, where the particle density $\rho_\mathrm{A}$ is normalized to unity
\begin{eqnarray}
\int{\mathrm{d}^3\pmb{r}\rho_\mathrm{A}(\pmb{r})}&=&1,\label{eq:rhonorm}
\end{eqnarray}
and the last equality of (\ref{eq:FchA_ChFF}) applies for isotropic $\rho_\mathrm{A}$ densities. A more generic density profile of nuclei is given by the $3$-parameter Woods-Saxon function~\cite{DeJager:1974liz,DeVries:1987atn}
\begin{eqnarray}
\rho_\mathrm{A}(r)&=&\rho_\mathrm{0,\mathrm{A}}\frac{1+w_\mathrm{A}\left(r/R_\mathrm{A}\right)^2}{1+\exp{\left(\frac{r-R_\mathrm{A}}{a_\mathrm{A}}\right)}},\label{eq:3param_WS}
\end{eqnarray}
with $\rho_\mathrm{0,\mathrm{A}}$ a normalization constant so that Eq.~(\ref{eq:rhonorm}) is fulfilled, and typical radial parameters ($R_\mathrm{A}$, $a_\mathrm{A}$, and $w_\mathrm{A}$) listed in Table~\ref{tab:NuclearProfile} for various nuclei.

\begin{table}[htpb!]
\tabcolsep=2.5mm
\centering
\caption[]{Parameters of the Woods-Saxon profile, Eq.~(\ref{eq:3param_WS}), for a variety of nuclei implemented in our code. For each ion we quote its mass number, charge, and radial parameters $R_\mathrm{A}$, $a_\mathrm{A}$, and $w_\mathrm{A}$ from Refs.~\cite{DeJager:1974liz,DeVries:1987atn,Loizides:2017ack}.
\label{tab:NuclearProfile}}
\vspace{0.2cm}
\begin{tabular}{lccccc} \hline
Nucleus  & $A$ & $Z$ & $R_\mathrm{A}$ [fm] & $a_\mathrm{A}$ [fm] &  $w_\mathrm{A}$ \\\hline
O & 16 & 8 & $2.608$ & $0.513$ & $-0.051$\\
Ar & 40 & 18 & $3.766$ & $0.586$ & $-0.161$\\
Ca & 40 & 20 & $3.766$ & $0.586$ & $-0.161$\\
Kr & 78 & 36 & $4.5$ & $0.5$ & $0$ \\
Xe & 129 & 54 & $5.36$ & $0.59$ & $0$ \\
Pb & 208 &  82 & $6.624$ & $0.549$ & $0$ \\
\hline
\end{tabular}
\end{table}
Plugging into Eq.~(\ref{eq:FchA_ChFF}) the 3-parameter Woods-Saxon function above, the following analytic ChFF formula can be derived:
\begin{eqnarray}
F_{\text{ch},A}(q)&=&\frac{4\pi^2\rho_{0,\mathrm{A}} a_\mathrm{A}^3 }{q^2 a_\mathrm{A}^2 \sinh^2{(\pi q a_\mathrm{A})}}\left\{\pi q a_\mathrm{A}\cosh{(\pi q a_\mathrm{A})}\sin{(qR_\mathrm{A})}\left[1-\frac{w_\mathrm{A} a_\mathrm{A}^2}{R_\mathrm{A}^2}\left(\frac{6\pi^2}{\sinh^2{(\pi q a_\mathrm{A})}}+\pi^2-3\frac{R_\mathrm{A}^2}{a_\mathrm{A}^2}\right)\right]\right.\nonumber\\
&&\left.-qR_\mathrm{A}\sinh{(\pi q a_\mathrm{A})}\cos{(qR_\mathrm{A})}\left[1-\frac{w_\mathrm{A} a_\mathrm{A}^2}{R_\mathrm{A}^2}\left(\frac{6\pi^2}{\sinh^2{(\pi q a_\mathrm{A})}}+3\pi^2-\frac{R_\mathrm{A}^2}{a_\mathrm{A}^2}\right)\right]\right\}\nonumber\\
&&+\underbrace{8\pi\hat{\rho}_{0,\mathrm{A}}a_\mathrm{A}^3\sum_{n=1}^{+\infty}{(-1)^{n-1}\frac{n e^{-n\frac{R_\mathrm{A}}{a_\mathrm{A}}}}{\left(n^2+q^2a_\mathrm{A}^2\right)^2}\left[1+12\frac{w_\mathrm{A} a_\mathrm{A}^2}{R_\mathrm{A}^2}\frac{n^2-q^2a_\mathrm{A}^2}{(n^2+q^2a_\mathrm{A}^2)^2}\right]}}_{F_{\text{ch},A,2}(q)}\nonumber\\
&=&F_{\text{ch},A,1}(q)+F_{\text{ch},A,2}(q),
\end{eqnarray}
which has been conveniently split into the last sum of two terms because the expression for $w_\mathrm{A}=0$ is already known from Ref.~\cite{Maximon:1966sqn} [cf. Eqs.~(1) and (20) there], and we are also able to analytically work out the integral in Eq.~(\ref{eq:ChFF}) for the second term $F_\text{ch,A,2}(q)$, as follows
\begin{eqnarray}
&&\hspace{-1.1cm}
\int_0^{+\infty}{\frac{d\kt \kt^2}{\kt^2+\frac{E_\gamma^2}{\gamma_{\mathrm{L}}^2}} F_{\text{ch},A,2}\left(\sqrt{k_{\perp}^2+\frac{E_\gamma^2}{\gamma_{\mathrm{L}}^2}}\right)J_1(b\kt)} = \nonumber\\
&=&\frac{E_\gamma}{\gamma_{\mathrm{L}}}8\pi \rho_{0,\mathrm{A}} a_\mathrm{A}^3\sum_{n=1}^{+\infty}{(-1)^{n-1}ne^{-n\frac{R_\mathrm{A}}{a_\mathrm{A}}}\left\{\left[\frac{K_1(\xi)}{n^4}-\sqrt{1+n^2 \tilde{a}_\mathrm{A}^{-2}}\frac{K_1\left(\tilde{B}_n\right)}{n^4}-\frac{\xi}{2n^2\tilde{a}_\mathrm{A}^2}K_0\left(\tilde{B}_n\right)\right]\right.}\nonumber\\
&&+12\frac{w_\mathrm{A} a_\mathrm{A}^2}{R_\mathrm{A}^2}\left[\frac{K_1(\xi)}{n^6}-\left(\frac{1}{n^6}+\frac{\xi^2(5n^2+3\tilde{a}_\mathrm{A}^2)}{24 n^2 (n^2+\tilde{a}_\mathrm{A}^2)^2 \tilde{a}_\mathrm{A}^2}\right)\sqrt{1+n^2\tilde{a}_\mathrm{A}^{-2}}K_1(\tilde{B}_n)\right.
\left.\left.-\left(\frac{\xi}{2n^4 \tilde{a}_\mathrm{A}^2}+\frac{\xi^3}{24(\tilde{a}_\mathrm{A}^2+n^2)\tilde{a}_\mathrm{A}^4}\right)K_0(\tilde{B}_{n})\right]\right\}\nonumber\\
&=&-\frac{E_\gamma}{\gamma_{\mathrm{L}}}8\pi \rho_{0,\mathrm{A}} a_\mathrm{A}^3K_1(\xi)\left[\text{Li}_3\left(-e^{-\frac{R_\mathrm{A}}{a_\mathrm{A}}}\right)+12\frac{w_\mathrm{A} a_\mathrm{A}^2}{R_\mathrm{A}^2}\text{Li}_5\left(-e^{-\frac{R_\mathrm{A}}{a_\mathrm{A}}}\right)\right]\nonumber\\
&&+\frac{E_\gamma}{\gamma_{\mathrm{L}}}8\pi \rho_{0,\mathrm{A}} a_\mathrm{A}^3\sum_{n=1}^{+\infty}{(-1)^{n-1}ne^{-n\frac{R_\mathrm{A}}{a_\mathrm{A}}}\left\{\left[-\sqrt{1+n^2 \tilde{a}_\mathrm{A}^{-2}}\frac{K_1\left(\tilde{B}_n\right)}{n^4}-\frac{\xi}{2n^2\tilde{a}_\mathrm{A}^2}K_0\left(\tilde{B}_n\right)\right]\right.}\nonumber\\
&&+12\frac{w_\mathrm{A} a_\mathrm{A}^2}{R_\mathrm{A}^2}\left[-\left(\frac{1}{n^6}+\frac{\xi^2(5n^2+3\tilde{a}_\mathrm{A}^2)}{24 n^2 (n^2+\tilde{a}_\mathrm{A}^2)^2 \tilde{a}_\mathrm{A}^2}\right)\sqrt{1+n^2\tilde{a}_\mathrm{A}^{-2}}K_1(\tilde{B}_n)\right.
\left.\left.-\left(\frac{\xi}{2n^4 \tilde{a}_\mathrm{A}^2}+\frac{\xi^3}{24(\tilde{a}_\mathrm{A}^2+n^2)\tilde{a}_\mathrm{A}^4}\right)K_0(\tilde{B}_{n})\right]\right\},
\end{eqnarray}
where we have used the notations $\tilde{a}_\mathrm{A}=a_\mathrm{A}\, E_\gamma/\gamma_{\mathrm{L}}$ and $\tilde{B}_n=\xi\sqrt{1+n^2 \tilde{a}_\mathrm{A}^{-2}}$, and $\text{Li}_m$'s are standard polylogarithms of order $m$. We opt for numerically integrating $F_{\text{ch},A,1}$ in Eq.~(\ref{eq:ChFF}), which is however nontrivial because the integrand involves highly oscillatory trigonometric functions and the $J_1$ Bessel function. Finally, we can solve $\rho_{0,\mathrm{A}}$ from the normalization condition Eq.~(\ref{eq:rhonorm}), yielding
\begin{eqnarray}
\rho_{0,\mathrm{A}}&=&\frac{1}{-8\pi a_\mathrm{A}^3\left[\text{Li}_3\left(-e^{\frac{R_\mathrm{A}}{a_\mathrm{A}}}\right)+12\frac{w_\mathrm{A} a_\mathrm{A}^2}{R_\mathrm{A}^2}\text{Li}_5\left(-e^{\frac{R_\mathrm{A}}{a_\mathrm{A}}}\right)\right]}\,.
\end{eqnarray}

\begin{figure}[htpb!]
\centering
\includegraphics[width=0.6\textwidth]{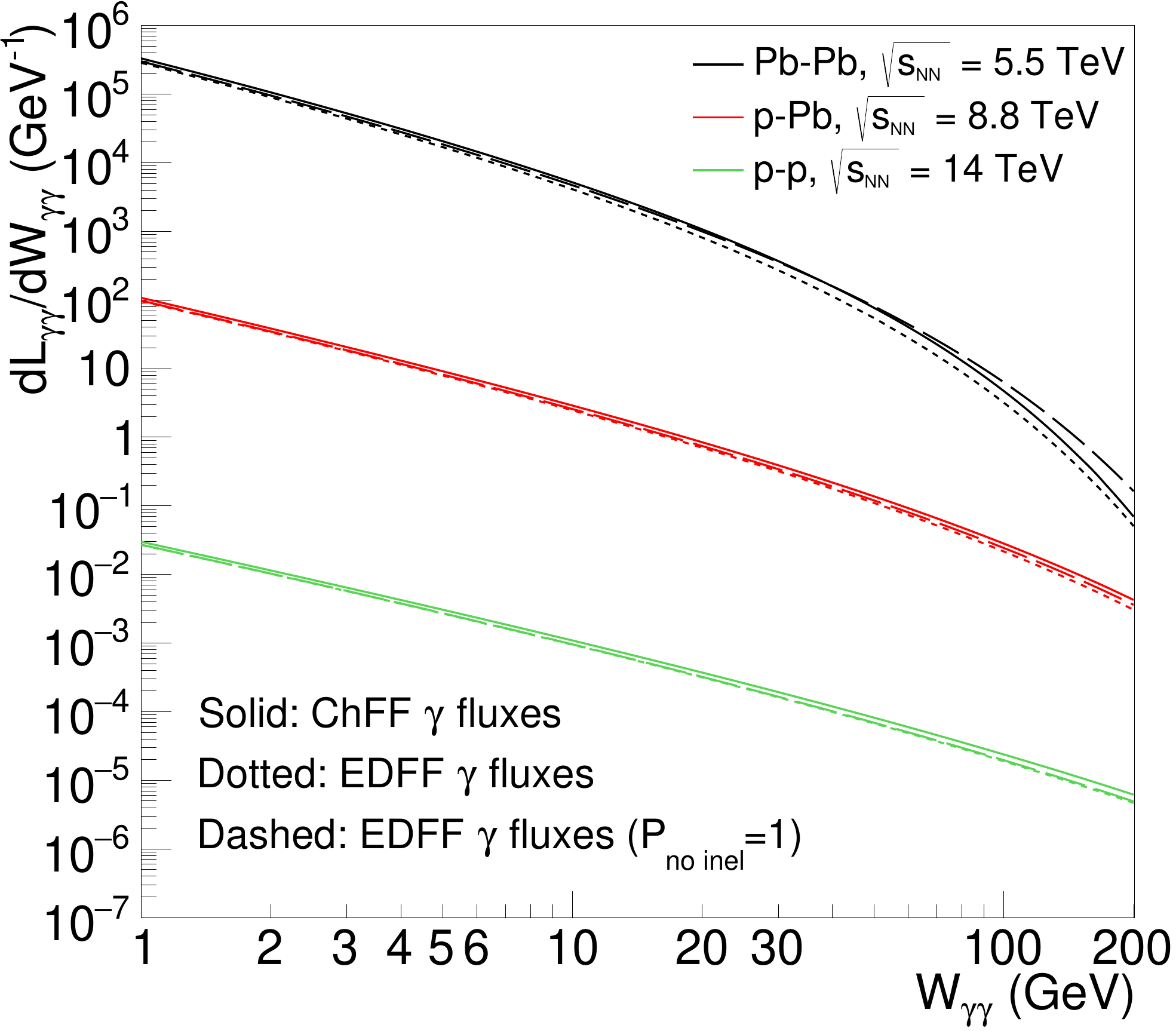}
\includegraphics[width=0.6\textwidth]{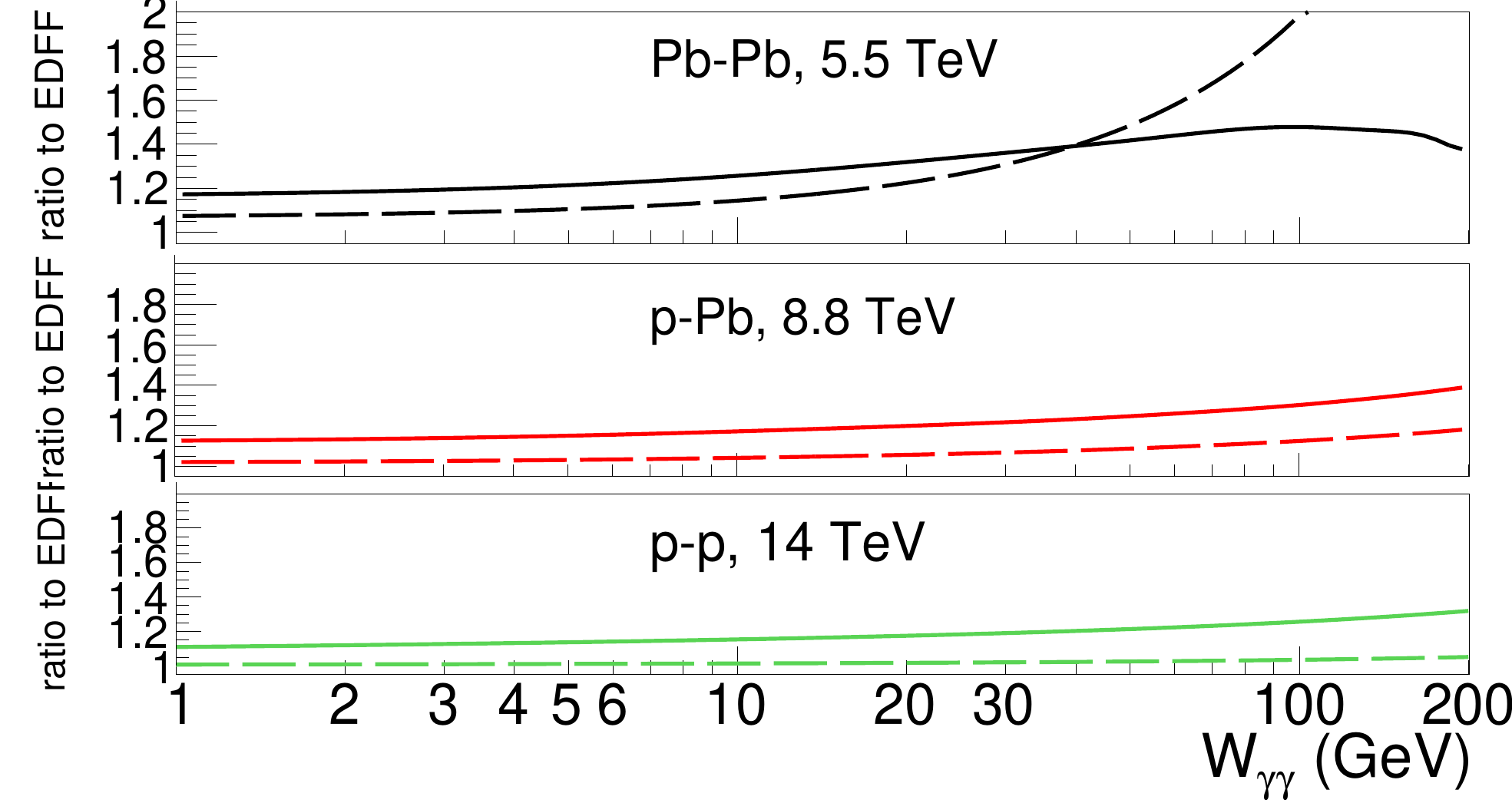}
\caption{Comparison of the effective photon-photon luminosities $\mathrm{d}{\Lumi}_{\gaga}/\mathrm{d}W_{\gaga}$ as a function of $W_{\gaga}$, Eq.~(\ref{eq:gagalumi}), for ultraperipheral \PbPb, \pPb, and \pp\ collisions at the LHC. The solid curves are obtained using ChFF, the dotted curves using EDFF, and the dashed curves using EDFF fluxes with $P_{\text{no\,inel}}=1$. The lower insets show the corresponding ratios over the EDFF-based luminosities.\label{fig:dLdWUPC}}
\end{figure}

For the proton case, we implement in Eq.~(\ref{eq:ChFF}) the dipole form factor~\cite{Klein:2003vd}
\begin{eqnarray}
F_{\text{ch},\mathrm{p}}(q)&=&\frac{1}{\left(1+q^2a_\mathrm{p}^2\right)^2}\label{eq:protonChFF}
\end{eqnarray}
with $a_\mathrm{p}^{-2}=Q_0^2=0.71~\text{GeV}^2$, resulting in the following ChFF $\gamma$ number density for the proton 
\begin{eqnarray}
N_{\gamma/\mathrm{p}}^{\text{ChFF}}(E_\gamma,b)&=&\frac{\alpha}{\pi^2}\frac{\xi^2}{b^2}\left\{\left[K_1(\xi)-\sqrt{1+\tilde{a}_\mathrm{p}^{-2}}K_1\left(\xi\sqrt{1+\tilde{a}_\mathrm{p}^{-2}}\right)\right]-
\frac{\xi}{2\tilde{a}_\mathrm{p}^2}K_0\left(\xi\sqrt{1+\tilde{a}_\mathrm{p}^{-2}}\right)
\right\}^2\,,
\end{eqnarray}
where $\tilde{a}_\mathrm{p}=a_\mathrm{p}\,E_\gamma/\gamma_{\mathrm{L}}$. In the limit $a_\mathrm{p}\to 0$, the ChFF flux reproduces the transversely polarized part of the EDFF flux, Eq.~(\ref{eq:EDFF}).\\

For the charge form factor, we can safely set the $\epsilon$ parameter to zero in Eq.~(\ref{eq:twophotonyield}), i.e., $\epsilon^{\text{ChFF}}=0$, because the photon number densities are well-behaved for $b\to 0$, as can be seen by the blue dashed lines in Fig.~\ref{fig:photonnumberdensity}. The ChFF is more realistic than the EDFF as it allows considering also the photon flux within the nuclei, namely for $b<R_\mathrm{A}$, which \eg\ enables the interpretation of the exclusive dimuon ATLAS measurement~\cite{Burmasov:2021phy}, as pointed out earlier by Ref.~\cite{Baltz:2009jk}. We stress the difference with respect to Ref.~\cite{Burmasov:2021phy}, as we have extended the fluxes for the generic $w_\mathrm{A}\neq 0$ ion profile case, and also kept the higher-order terms in $e^{-n\frac{R_\mathrm{A}}{a_\mathrm{A}}}$ for $n>1$ in the ChFF $F_{\text{ch},A}(q)$ function.\\

Figure~\ref{fig:photonnumberdensity} shows the EDFF (red solid) and ChFF (blue dashed) photon number densities for Pb (top), Ar (middle), and p (bottom) ions at LHC energies, for two indicative low ($E_\gamma = 2$~GeV) and very high ($E_\gamma = 500$~GeV) photon energies. The fluxes have clearly different shapes at low impact parameters: a continuous powerlaw-like decrease (divergent for $b\to 0$) in the EDFF case, and a rising ChFF flux with impact parameter up to a few fm followed by a falloff that is very similar to the EDFF one. However, the $b_{1,2}>R_\mathrm{A,B}$ requirement (indicated by the vertical dashed lines in the plots) implemented in the EDFF two-photon integral, Eq.~(\ref{eq:twophotonyield}), renders such low-$b$ flux differences with the ChFF case less relevant in terms of actual photon-photon luminosities. At very high $\gamma$ energies, one can see that the ChFF fluxes for heavy ions show an oscillatory pattern, which is however unlikely to have any experimental impact given the large beam luminosities needed to reach such high $E_\gamma$ values.\\

\begin{figure}[!htpb]
\centering
\includegraphics[width=0.49\textwidth]{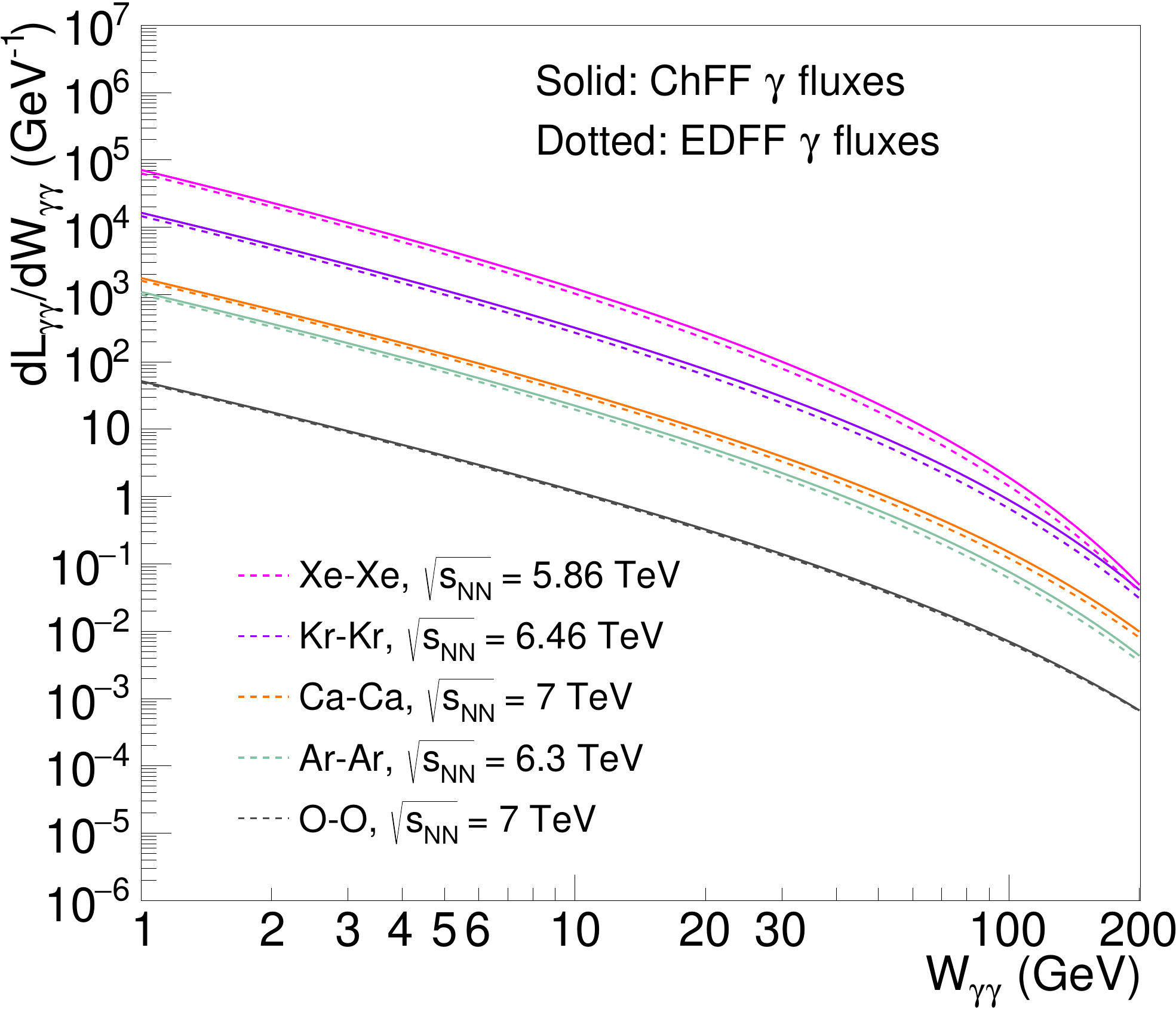}
\includegraphics[width=0.49\textwidth]{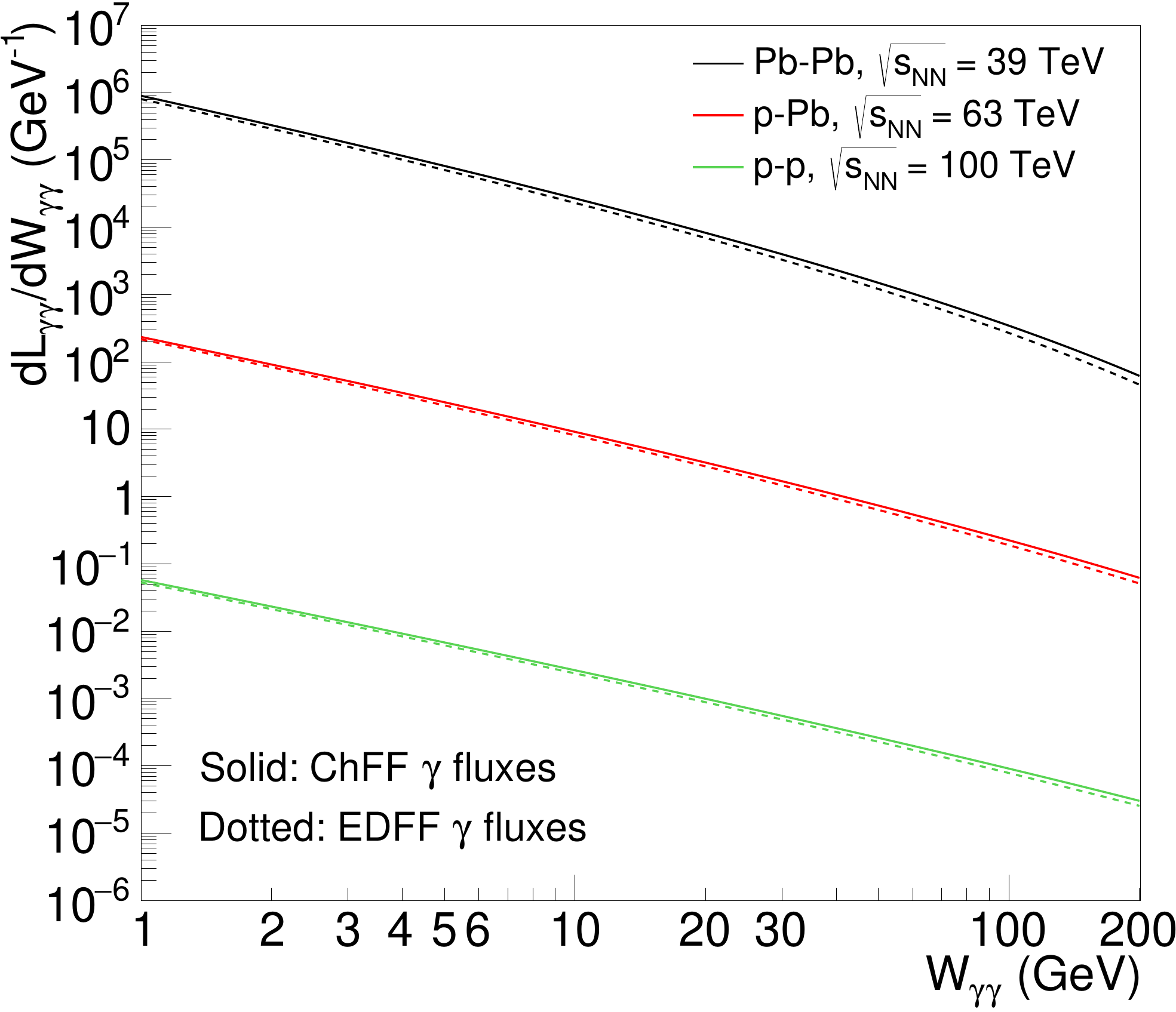}
\caption{Comparison of the effective photon-photon luminosities $\mathrm{d}{\Lumi}_{\gaga}/\mathrm{d}W_{\gaga}$ as a function of $W_{\gaga}$, Eq.~(\ref{eq:gagalumi}), for UPCs of various ion species at the LHC (left) and for \pp, \pPb, and \PbPb\ collisions at the FCC (right). Solid and dotted curves are obtained using ChFF and EDFF photon fluxes, respectively.
\label{fig:dLdWUPClightIons}}
\end{figure}

The effective photon-photon luminosities $\mathrm{d}{\Lumi}_{\gaga}/\mathrm{d}W_{\gaga}$ for \pp, \pPb, and \PbPb\ UPCs at the LHC, as obtained from Eq.~(\ref{eq:gagalumi}) using the EDFF (with and without the hadronic nonoverlap requirement) and ChFF functions, are shown in Fig.~\ref{fig:dLdWUPC}. 
In the lower insets of Fig.~\ref{fig:dLdWUPC}, the corresponding ratios over the EDFF $\gaga$ luminosity results are plotted. The first observation is that, as expected, the $P_{\text{no\,inel}}\neq 1$ requirement (dashed curves) reduces the photon-photon luminosities for increasing $W_{\gaga}$ values (\ie\ for lower impact parameters), in particular for \PbPb\ UPCs where the nonoverlap condition depletes the effective luminosity by 50\% above $W_{\gaga}\approx 50$~GeV, and by about a factor of three above 200~GeV (the impact of the nonoverlap requirement for the $\gaga$ luminosity of \pp\ collisions is much smaller, leading to a 1--5\% reduction over the considered mass range). The second observation is that the ChFF-based luminosities (solid curves) are overall larger than their EDFF counterparts by 10--30\% for \pp\ and \pPb\ UPCs, and by 15--50\% for \PbPb\ UPCs for small--large masses, respectively. As we will see in the next section, this implies that the ChFF cross sections for increasingly heavier final states are larger by about 10--20\% (for \pp\ and \pPb\ UPCs at the LHC) and 20--40\% (for \PbPb\ UPCs at the LHC) than the EDFF ones. In addition, Fig.~\ref{fig:dLdWUPClightIons} shows a comparison of the EDFF and ChFF effective photon-photon luminosities $\mathrm{d}{\Lumi}_{\gaga}/\mathrm{d}W_{\gaga}$ derived for UPCs with lighter heavy-ion systems at the LHC (Xe-Xe, Kr-Kr, Ar-Ar, Ca-Ca, and O-O; left), and for \pp, \pPb, and \PbPb\ UPCs at the FCC (right). All such colliding systems are incorporated by default in the \gammaUPC\ 
code. The theoretical precision of the EDFF- and ChFF-based predictions are being quantitatively estimated by varying all underlying \gammaUPC\ model input parameters within their uncertainties, and will be presented in an upcoming work~\cite{inpreparation}.

\section{Total photon-photon cross sections results}
\label{sec:results}

In this section we present predictions for total photon-fusion cross sections at LHC and FCC energies for a large variety of spin-even (scalar or tensor) resonances; for pairs of $\jpsi$ mesons, W bosons, Z bosons, and top quarks; and for axionlike particles and massive gravitons; all produced in \pp, \pA, and \AaAa\ UPCs. In all cases, results derived with EDFF and ChFF photon fluxes are presented. 

\subsection{\texorpdfstring{$C$}{C}-even resonances}

\begin{table}[htpb!]
\tabcolsep=3.5mm
\centering
\caption{List of all known $C$-even resonances above $m_X\approx 3$~GeV that can be produced via two-photon fusion. For each particle, we quote its $J^{PC}$ quantum numbers, mass $m_X$, 
and diphoton partial width $\Gamma_{\gaga}$ from measurements~\cite{Zyla:2020zbs} or theoretical predictions (for $\etabTwoS$, $\chibZero,\chibTwo$, $\tata$, and H, see text for details).
}
\vspace{0.2cm}
\label{tab:resonances}
\begin{tabular}{lccc}\toprule
Resonance &  $J^{PC}$ & $m_X$ (GeV) & 
$\Gamma_{\gaga}$ (MeV)\\ 
\midrule
$\etacOneS$ & $0^{-+}$ & \texttt{$2.9839 \pm 0.0005$} & 
\texttt{$(5.06 \pm 0.34)\cdot 10^{-3}$} 
\\
$\etacTwoS$ & $0^{-+}$ & \texttt{$3.6375 \pm 0.0011$} & 
\texttt{$(2.15 \pm 1.47)\cdot 10^{-3}$} 
\\
$\chicZero$ & $0^{++}$ & \texttt{$3.41471  \pm 0.00030$} & 
\texttt{$(2.203 \pm 0.097)\cdot 10^{-3}$} 
\\
$\chicTwo$ & $2^{++}$ & \texttt{$3.55617 \pm 0.00007$} & 
\texttt{$(5.614 \pm 0.197)\cdot 10^{-4}$} 
\\
$\tata$ & $0^{-+}$ & \texttt{$3.5537 \pm 0.0002$} 
&  \texttt{$1.83\cdot 10^{-8}$} 
\\
$\etabOneS$ & $0^{-+}$ & \texttt{$9.3987 \pm 0.0020$} & 
\texttt{$(4.8 ^{+2.5}_{-2.0})\cdot 10^{-4}$} 
\\
$\etabTwoS$ & $0^{-+}$ & \texttt{$9999 \pm 4$} 
& \texttt{$(2.4 ^{+1.2}_{-1.0})\cdot 10^{-4}$} 
\\
$\chibZero$ & $0^{++}$ &\texttt{$9.85944  \pm 0.00052$} 
& \texttt{$(0.15^{+0.05}_{-0.03})\cdot 10^{-3}$} 
\\
$\chibTwo$ & $2^{++}$ & \texttt{$9.91221  \pm 0.00040$} & 
\texttt{$(9.3^{+1.3}_{-6.2})\cdot 10^{-6}$} 
\\
H & $0^{++}$ & \texttt{$125.250  \pm 0.170$} & 
\texttt{$(9.3 \pm 0.2)\cdot 10^{-3}$} 
\\
\bottomrule
\end{tabular}
\end{table}

The cross section for the exclusive production of a $C$-even resonance $X$ through $\gaga$ fusion in an UPC is given by Eq.~(\ref{eq:sigma_AA_X}), and is completely determined from its spin $J=0,2$, two-photon width $\Gamma_{\gaga}(X)$, and the photon-photon effective luminosity of the colliding system at the particle mass. In Table~\ref{tab:resonances}, we list the relevant properties of all presently known\footnote{Any new exotic spin-0 multiquark hadron, such as the candidate $(cs\bar{u}\bar{d})$ tetraquark $X_0(2900)$ state~\cite{LHCb:2020bls,LHCb:2020pxc}, can be likely produced via photon fusion provided its diphoton width is not too small.} scalar and tensor resonances from $m_X\approx 3$~GeV up to the Higgs boson. Except for the Higgs and ditauonium cases, the rest of spin-even particles over this mass range are charmonium and bottomonium bound states. Masses are precisely determined for all the particles, although not all their two-photon widths have been experimentally measured~\cite{Zyla:2020zbs}. All charmonium resonances have diphoton widths known to within 3--6\% except for $\etacTwoS$, which is badly known and has a $\pm60\%$ uncertainty presently. The $\gaga$ decays of four $\bbbar$ resonances ($\etabOneS$, $\etabTwoS$, $\chibZero$, $\chibTwo$) remain unobserved so far. For the $\etabOneS$ and $\etabTwoS$ cases, predictions exist in nonrelativistic QCD (NRQCD) for their two-photon partial widths~\cite{Chung:2010vz,Penin:2004ay}. Due to the spin symmetry of heavy quarks, the two-photon $\etabOneS\to\gaga$ and leptonic $\etabOneS\to\ell^+\ell^-$ decay widths are proportional to the same wavefunction at NLO accuracy. This suggests that the decay ratio $\Gamma(n^3\mathrm{S}_1\to \epem)/\Gamma(n^1\mathrm{S}_0\to \gaga)$ is more appropriate to obtain reliable results, stable against the renormalization scale variations. The diphoton partial width of $\Gamma(\etabTwoS\to \gaga)$ is thus evaluated by rescaling $\Gamma(\etabOneS\to \gaga)$ with the wavefunctions at origin in the Buchm\"uller-Tye potential model~\cite{Eichten:1995ch}. The diphoton widths of $\chibZero$, $\chibTwo$ and the Higgs boson are from~\cite{Wang:2018rjg} and~\cite{LHCHiggsCrossSectionWorkingGroup:2016ypw}, respectively. The one from ditauonium ($\tata$) has been derived in~\cite{dEnterria:2022alo}.\\

Table~\ref{tab:xsecs2} lists the theoretical predictions for the total photon-fusion cross sections for ten scalar/tensor resonances produced in UPCs for various colliding systems at LHC and FCC \cm\ energies, derived using Eq.~(\ref{eq:sigma_AA_X})  and the properties listed in Table~\ref{tab:resonances}, for EDFF and ChFF $\gamma$ fluxes. Uncertainties in the cross sections (not quoted) are dominated by the propagated uncertainty of the corresponding $\Gamma_{\gaga}$ widths and vary between 5\% and 100\%. One can see first, as expected from Eq.~(\ref{eq:sigma_AA_X}), that all cross sections decrease rapidly with resonance mass due to the intrinsic $\propto\!m_X^{-2}$ dependence of the photon-fusion cross section as well as the steep decrease with $W_{\gaga}$ of the two-photon effective luminosities (Figs.~\ref{fig:dLdWUPC} and~\ref{fig:dLdWUPClightIons}). Second, one can also see that the cross sections obtained with EDFF are systematically lower by 15--25\% compared to the ChFF ones: heavier systems featuring larger differences, as indicated by the ratios of ChFF/EDFF two-photon luminosities shown in the bottom panels of Fig.~\ref{fig:dLdWUPC}. Lastly, for the \pp\ UPC case, the iWW cross sections derived neglecting hadronic overlaps overestimate the EDFF (ChFF) results by 15--30\% (8--15\%), whereas ignoring the survival factors ($S^2_{\gaga}=1$) leads to a relatively moderate rise in the cross sections (by 2--8\%, increasing with $m_X$) compared to the default EDFF values.\\

\begin{table}[htpb!]
\centering
\tabcolsep=1.2mm
\caption{Total photon-fusion cross sections for all known spin-even resonances with masses above $m_X\approx 3$~GeV (Table~\ref{tab:resonances}) in UPCs for various colliding systems at LHC and FCC \cm\ energies. Results derived with EDFF and ChFF are shown for all systems. Associated uncertainties (not quoted) are discussed in the text. In the \pp\ case, we list also the iWW results using the \mgshort\ default EPA flux, Eq.~(\ref{eq:flux_p}), as well as the EDFF cross sections assuming 100\% survival probability ($S^2_{\gaga}=1$).
\label{tab:xsecs2}}
\vspace{0.2cm}
\resizebox{\textwidth}{!}{
\begin{tabular}{l|c|cccccccccc} \hline
Colliding  & Form  & \multicolumn{10}{c}{\gammaUPC\;\;$\sigma(\gaga\to X)$}\\
system & factor & $\etacOneS$ & $\etacTwoS$ & $\chicZero$ &  $\chicTwo$ & $\etabOneS$ & $\etabTwoS$ & $\chibZero$ & $\chibTwo$ & $\tata$ & H \\ \hline
\multirow{3}{*}{\pp, 14~TeV} & iWW & 61 pb & 13 pb & 17 pb & 19 pb & 110 fb & 44 fb & 29 fb & 8.9 fb & 0.12 fb & 0.17 fb \\
& EDFF\,($S^2_{\gaga}=1$) & 51 pb & 11 pb & 14 pb & 15 pb & 88 fb & 35 fb & 23 fb & 7.1 fb & 0.10 fb & 0.12 fb \\
& EDFF & 50 pb & 11 pb & 14 pb & 15 pb & 86 fb & 35 fb & 23 fb & 7.0 fb & 0.10 fb & 0.11 fb \\
& ChFF & 56 pb & 12 pb & 15 pb & 17 pb & 99 fb & 40 fb & 26 fb & 8.0 fb & 0.11 fb & 0.14 fb \\\hline
\multirow{2}{*}{\pPb, 8.8~TeV} & EDFF & 0.16 $\mu$b & 33 nb & 43 nb & 46 nb & 0.23 nb & 92 pb & 60 pb & 18 pb & 0.31 pb & 0.11 pb\\
& ChFF & 0.18 $\mu$b & 38 nb & 49 nb & 53 nb & 0.27 nb & 106 pb & 70 pb & 21 pb & 0.35 pb & 0.14 pb \\\hline
\multirow{2}{*}{O-O, 7~TeV} & EDFF & 76 nb & 16 nb & 21 nb & 23 nb & 0.10 nb & 42 pb & 28 pb & 8.5 pb & 0.15 pb & 31 fb\\
& ChFF & 82 nb & 17 nb & 22 nb & 24 nb & 0.11 fb & 44 pb & 29 pb & 9.0 pb & 0.16 pb & 32 fb \\\hline
\multirow{2}{*}{Ca-Ca, 7~TeV} & EDFF & 2.5 $\mu$b & 0.50 $\mu$b & 0.63 $\mu$b & 0.70 $\mu$b & 3.1 nb & 1.2 nb & 0.81 nb & 0.25 nb & 4.6 pb & 0.48 pb\\
& ChFF & 2.7 $\mu$b & 0.58 $\mu$b & 0.74 $\mu$b & 0.81 $\mu$b & 3.5 nb & 1.4 nb & 0.91 nb & 0.29 nb & 5.2 pb & 0.62 pb \\\hline
\multirow{2}{*}{Ar-Ar, 6.3~TeV} & EDFF & 1.5 $\mu$b & 0.31 $\mu$b & 0.40 $\mu$b & 0.42 $\mu$b & 1.8 nb & 0.73 nb & 0.48 nb & 0.15 nb & 2.9 pb & 0.25 pb\\
& ChFF & 1.6 $\mu$b & 0.34 $\mu$b & 0.44 $\mu$b & 0.49 $\mu$b & 2.1 nb & 0.83 nb & 0.55 nb & 0.17 nb &  3.1 pb & 0.31 pb \\\hline
\multirow{2}{*}{Kr-Kr, 6.46~TeV} & EDFF & 22 $\mu$b & 4.4 $\mu$b & 5.9 $\mu$b & 6.3 $\mu$b & 25 nb & 10 nb & 6.7 nb & 1.9 nb & 41 pb & 2.5 pb\\
& ChFF & 25 $\mu$b & 5.1 $\mu$b & 6.4 $\mu$b & 7.0 $\mu$b & 31 nb & 12 nb & 7.9 nb & 2.3 nb & 46 pb & 3.4 pb \\\hline
\multirow{2}{*}{Xe-Xe, 5.86~TeV} & EDFF & 89 $\mu$b & 18 $\mu$b & 24 $\mu$b & 26 $\mu$b & 98 nb & 38 nb & 26 nb & 7.7 nb & 0.16 nb & 4.8 pb\\
& ChFF & 101 $\mu$b & 21 $\mu$b & 27 $\mu$b & 29 $\mu$b & 116 nb & 46 nb & 31 nb & 9.2 nb & 0.19 nb & 6.2 pb \\\hline
\multirow{2}{*}{\PbPb, 5.52~TeV} & EDFF & 0.39 mb & 79 $\mu$b & 0.10 mb & 0.11 mb & 0.40 $\mu$b & 0.15 $\mu$b & 0.10 $\mu$b & 31 nb & 0.71 nb & 9.3 pb\\
& ChFF & 0.46 mb & 95 $\mu$b & 0.12 mb & 0.13 mb & 0.50 $\mu$b & 0.19 $\mu$b & 0.13 $\mu$b & 38 nb & 0.86 nb & 13 pb \\\hline 
\multicolumn{12}{@{}c@{}}{\hrulefill} \\
\multirow{3}{*}{\pp, 100~TeV} & iWW & 0.13 nb & 28 pb & 35 pb & 39 pb & 0.26 pb & 104 fb & 69 fb & 21 fb & 0.26 fb & 0.65 fb \\
& EDFF\,($S^2_{\gaga}=1$) & 0.11 nb & 24 pb & 30 pb & 34 pb & 0.22 pb & 88 fb & 58 fb & 18 fb & 0.22 fb & 0.51 fb \\
& EDFF & 0.11 nb & 24 pb & 30 pb & 33 pb & 0.21 pb & 87 fb & 57 fb & 17 fb & 0.22 fb & 0.49 fb \\
& ChFF & 0.12 nb & 26 pb & 33 pb & 37 pb & 0.24 pb & 96 fb & 63 fb & 19 fb & 0.24 fb & 0.57 fb \\\hline
\multirow{2}{*}{\pPb, 62.8~TeV} & EDFF & 0.41 $\mu$b & 89 nb & 0.11 $\mu$b & 0.13 $\mu$b & 0.75 nb & 0.29 nb & 0.19 nb & 60 pb & 0.82 pb & 1.1 pb \\
& ChFF & 0.46 $\mu$b & 100 nb & $0.13$ $\mu$b & 0.14 $\mu$b & 0.83 nb & 0.33 nb & 0.22 nb & 67 pb & 0.91 pb & 1.4 pb \\\hline
\multirow{2}{*}{\PbPb, 39.4~TeV}  & EDFF & 1.3 mb & 0.29 mb & 0.37 mb & 0.41 mb & 2.1 $\mu$b & 0.85 $\mu$b & 0.57 $\mu$b & 0.17 $\mu$b & 2.7 nb & 1.5 nb \\
& ChFF & 1.6 mb & 0.33 mb & 0.43 mb & 0.47 mb & 2.5 $\mu$b & 1.0 $\mu$b & 0.66 $\mu$b & 0.19 $\mu$b & 3.1 nb & 1.9 nb \\\hline
\end{tabular}
}
\end{table}

If one would naively take the average of EDFF and ChFF cross sections as the central prediction, and half their difference as their associated uncertainty, one would assign theoretical uncertainties linked to the choice of the photon flux\footnote{Uncertainties linked to the calculation of survival probabilities propagated from the imprecise knowledge of hadron profiles, as well as of $\sigma^{\mathrm{NN}}_{\text{inel}}$ and of $b_0$ (for protons), via Eqs.~(\ref{eq:Psurv}), are smaller than that~\cite{inpreparation}.} varying over 12--25\% for \PbPb, 7--15\% for \pPb, and 6--12\% for \pp\ UPCs in $\gaga\to X$ processes at low ($m_X\approx10$~GeV) and high ($m_X\approx100$~GeV) masses. Such uncertainties can nonetheless be significantly reduced by taking ratios of two exclusive photon-photon cross sections (\eg\ by using exclusive dimuon production as a reference baseline process in the denominator) at the same $W_{\gaga}$. Such results are consistent with the $\mathcal{O}(10\%)$ theoretical uncertainties often quoted in UPC studies at the LHC.\\

Given the LHC integrated luminosities per system listed in Table~\ref{tab:1}, the cross sections of Table~\ref{tab:resonances} indicate that most quarkonium $C$-even resonances should be in principle measurable in UPCs at the LHC (at least, in their dominant (hadronic) decay modes). A caveat is needed for \pp\ collisions, because their production via central exclusive (gluon-induced) processes has much larger cross sections~\cite{Harland-Lang:2010ajr} than via photon fusion, although imposing low final-state acoplanarities in their decay final states would largely reduce the former. Given their comparatively low masses $\mathcal{O}(3$--4~GeV), charmonium scalar and tensor resonances (as well as ditauonium~\cite{dEnterria:2022ysg}) can only be likely triggered-on and reconstructed at ALICE~\cite{ALICE:2008ngc} and LHCb~\cite{LHCb:2008vvz} with the required precision; whereas bottomonium bound states are also accessible to ATLAS~\cite{ATLAS:2008xda} and CMS~\cite{CMS:2008xjf}. On the other hand, the $\gaga$ production of the Higgs boson seems out of reach at the LHC, and one would need a machine like the FCC to observe it~\cite{dEnterria:2019jty}.
The motivation to perform studies of the scalar and tensor quarkonia via UPCs at the LHC listed in Table~\ref{tab:xsecs2} is driven by the fact that several important parameters of the states either need to be measured for the first time, or have conflicting experimental results in need of resolution. Examples include the poorly known diphoton width of $\etacTwoS$, the masses and widths of $\eta_{b}$ states, the $\chi_{\mathrm{c,b};0}$ widths, evidence for $\eta_{b}(2\mathrm{S})$ (which is below the 5-standard-deviations threshold today), the transitions between $\chi_{b}$ states, etc.
Ultimately, the best way to produce $\eta_{b}$ states is at Belle~II via $\Upsilon(4\mathrm{S})$ decays, where about four million $\etabOneS$ are expected with the total integrated luminosity of 50~ab$^{-1}$~\cite{Kou:2018nap}, but our work here motivates to follow up an alternative unexplored pathway for their study via photon-fusion production in UPCs at the LHC. 


\subsection{Exclusive di-\texorpdfstring{$\jpsi$}{J/Psi} mesons}

The exclusive production of a pair of J$/\psi$ mesons, both in central production~\cite{Harland-Lang:2014efa} and $\gaga$ fusion~\cite{Kwiecinski:1999hg}, is an interesting process for the study of BFKL-Pomeron dynamics~\cite{Kwiecinski:1998sa,Kwiecinski:1999hg,Chernyak:2014wra,Goncalves:2015sfy}. Such a process has been observed by the LHCb Collaboration~\cite{LHCb:2014zwa} in \pp\ at $\sqrts = 7$ and 8~TeV where central exclusive production dominates. With the \gammaUPC\,$+$\,\helaconia\ setup, one can easily obtain a theoretical prediction for the $\gaga\to \jpsi\jpsi$ process in \pp, \pPb, and \PbPb\ UPCs at the LHC. The corresponding cross sections are listed in Table~\ref{tab:xsecsdipsi} at LO accuracy with about $+50\%,-20\%$ theoretical uncertainties derived by varying the default renormalization scale within a factor of two to estimate the impact of missing higher-order corrections. For the total integrated \PbPb\ luminosity of $\LumiInt = 10$~nb$^{-1}$, one should expect about 15 exclusive double-$\jpsi$ events produced in the combined dielectron and dimuon $\jpsi$ decay channels in ALICE (although the actual measurable yields should be smaller taking into account detector acceptance and efficiencies). 

\begin{table}[htpb!]
\centering
\tabcolsep=4.5mm
\caption{Total cross sections for $\gaga\to \jpsi\jpsi$ in UPCs at the LHC, computed with EDFF and ChFF $\gamma$ fluxes and their average. The quoted asymmetric uncertainty is derived from the renormalization scale variation.
\label{tab:xsecsdipsi}}
\vspace{0.2cm}
\begin{tabular}{l|ccc} \hline
Process: $\gaga\to \jpsi\jpsi$ & \multicolumn{3}{c}{\gammaUPC~$\sigma$} \\
Colliding system, \cm\ energy & EDFF & ChFF & average \\ \hline
\pp\ at 14~TeV & $20_{-6}^{+11}$ fb & $23_{-7}^{+13}$ fb & $22_{-7}^{+12} \pm 2$ fb \\
\pPb\ at 8.8~TeV & $55_{-16}^{+30}$ pb & $64_{-18}^{+35}$ pb & $60_{-17}^{+32} \pm 4$ pb \\
\PbPb\ at 5.52~GeV & $103_{-29}^{+57}$ nb & $128_{-36}^{+71}$ nb &  $115_{-32}^{+64}\pm 12$ nb \\
\hline
\end{tabular}
\end{table}

\subsection{\texorpdfstring{$\gaga\to \mathrm{W^+W^-}$}{gamma-gamma to W+W-}}

The production of a pair of W bosons via photon-photon scattering constitutes a neat final state for the study of quartic gauge couplings (QGC) in the SM and searches for BSM effects~\cite{Pierzchala:2008xc,Maniatis:2008zz,deFavereaudeJeneret:2009db,Chapon:2009hh,Baldenegro:2017aen,Bailey:2022wqy}. The latter can be encoded into two dimension-6 operators $c_{WWW},c_{\tilde{W}WW}$ of the extended Lagrangian,  as follows~\cite{Degrande:2012wf}
\begin{eqnarray}
\mathcal{L}&\supset& \frac{c_{WWW}}{\Lambda^2}\mathrm{Tr}\left[W_{\mu\nu}W^{\nu\rho}W_{\rho}^{\mu}\right]+\frac{c_{\tilde{W}WW}}{\Lambda^2}\mathrm{Tr}\left[\tilde{W}_{\mu\nu}W^{\nu\rho}W_{\rho}^{\mu}\right],\label{eq:EWdim6}
\end{eqnarray}
where $\Lambda$ represents the BSM scale and $W_{\mu\nu}$ ($\tilde{W}_{\mu\nu}$) is the (dual) field strength of SU(2)$_L$. The trace $\mathrm{Tr}$ applies in the isospin space of SU(2). The total $\gaga\to\,$WW cross section can then be generically written as
\begin{eqnarray}
\sigma&=&\sigma_\mathrm{SM}+\left(\frac{c_{WWW}}{\Lambda^2}\times 1~\mathrm{TeV}^2\right)\sigma_{WWW}+\left(\frac{c_{\tilde{W}WW}}{\Lambda^2}\times 1~\mathrm{TeV}^2\right)\sigma_{\tilde{W}WW}+\mathcal{O}(\Lambda^{-4}).
\end{eqnarray}
The second operator in Eq.~(\ref{eq:EWdim6}) is CP odd, and its interference with the SM amplitude translates into $\sigma_{\tilde{W}WW}=0$ in the total phase-space integrated cross section. However, if one looks at asymmetry observables~\cite{Degrande:2021zpv}, one is able to probe the CP-violating effect.
Table~\ref{tab:xsecsWW} lists the expected SM cross sections $\sigma_\mathrm{SM}$ and QGC $\sigma_{WWW}$ contributions for $m_\mathrm{W}=80.419$ GeV and $c_{WWW}/\Lambda^2=1~\mathrm{TeV}^{-2}$. 
\begin{table}[hbp!]
\centering
\tabcolsep=4.mm
\caption{Total SM cross sections, and QGC $\sigma_{WWW}$ contributions, for $\gaga\to \mathrm{W^+W^-}$ in UPCs at the LHC and the FCC-hh, computed with EDFF and ChFF $\gamma$ fluxes and their average.
\label{tab:xsecsWW}}
\vspace{0.2cm}
\begin{tabular}{l|cc|cc|cc} \hline
Process: $\gaga\to \mathrm{W^+W^-}$ & \multicolumn{2}{c|}{\gammaUPC\ EDFF} & \multicolumn{2}{c|}{\gammaUPC\ ChFF} & \multicolumn{2}{c}{\gammaUPC\ average} \\
Colliding system, \cm\ energy & $\sigma_\mathrm{SM}$ & $\sigma_{WWW}$ & $\sigma_\mathrm{SM}$ & $\sigma_{WWW}$ & $\sigma_\mathrm{SM}$ & $\sigma_{WWW}$ \\ \hline
\pp\ at 14~TeV & 52.4 fb & 44.7 ab & 73.6 fb & 60.6 ab & $63 \pm 11$ fb & $53 \pm8$ ab \\
\pPb\ at 8.8~TeV & 20.9 pb & 23.1 fb & 30.3 pb & 32.8 fb & $26\pm 5$ pb & $28\pm 5$ fb \\
\PbPb\ at 5.52~TeV & 233 pb & 330 fb & 321 pb & 458 fb & $277\pm 44$ pb & $394\pm 64$ fb \\\hline
\pp\ at 100~TeV & 460 fb & 291 ab &  572 fb & 351 ab & $516\pm56$ fb & $320\pm30$ ab \\
\pPb\ at 62.8~TeV & 650 pb & 516 fb & 814 pb & 634 fb & $730\pm 80$ pb & $575\pm60$ fb \\
\PbPb\ at 39.4~TeV & 351 nb & 368 pb & 485 nb & 504 pb & $420\pm65$ nb & $436\pm68$ pb \\
\hline
\end{tabular}
\end{table}
In this particular case, the impact of BSM effects on the total cross section is at the permille level, whereas differences due to the $\gamma$ photon flux (EDFF or ChFF) are at the $\mathcal{O}(30\%)$, calling for the need of differential observables more sensitive to aQGC.

\subsection{\texorpdfstring{$\gaga\to \mathrm{Z}\gamma$}{gamma gamma to Zgamma} and \texorpdfstring{$\gaga\to \mathrm{ZZ}$}{gamma gamma to ZZ}}

The UPC $\gaga\to \mathrm{Z}\gamma$ and $\gaga\to\mathrm{ZZ}$ processes are loop-induced in the SM and particularly sensitive to aQGC effects~\cite{Gounaris:1999ux,Pierzchala:2008xc,deFavereaudeJeneret:2009db,Chapon:2009hh,Baldenegro:2017aen}. In addition, they constitute a continuum background for any search for resonances decaying into the same final states. The SM cross sections, computed with \madgraph~v2.6.6~\cite{Alwall:2014hca,Hirschi:2015iia} and our \gammaUPC\ setup, are tiny as can be seen in Tables~\ref{tab:xsecsza} and~\ref{tab:xsecszz}, and would require FCC energies and luminosities for their observation. Obviously, the observation of any signal with the expected LHC luminosities would be an indication of a BSM-related enhancement.

\begin{table}[htpb!]
\centering
\tabcolsep=4.5mm
\caption{Total SM cross sections for $\gaga\to Z\gamma$ in UPCs at the LHC and the FCC-hh, computed with EDFF and ChFF $\gamma$ fluxes and their average.
\label{tab:xsecsza}}
\vspace{0.2cm}
\begin{tabular}{l|ccc} \hline
Process: $\gaga\to Z\gamma$ & \multicolumn{3}{c}{\gammaUPC~$\sigma$} \\
Colliding system, \cm\ energy  & EDFF & ChFF & average \\ \hline
\pp\ at 14~TeV & 36.2 ab & 44.7 ab & $40.5\pm 4.3$ ab \\
\pPb\ at 8.8~TeV & 10.3 fb & 15.6 fb & $13.0 \pm 2.6$ fb \\
\PbPb\ at 5.52~TeV & 109 fb & 152 fb & $130\pm 22$ fb \\\hline
\pp\ at 100~TeV & 350 ab & 440 ab & $400\pm 50$ ab \\
\pPb\ at 62.8~TeV & 437 fb & 540 fb & $490\pm 50$ fb \\
\PbPb\ at 39.4~TeV & 169 pb & 217 pb & $195\pm 25$ pb \\
\hline
\end{tabular}
\end{table}


\begin{table}[htpb!]
\centering
\tabcolsep=4.5mm
\caption{Total SM cross sections for $\gaga\to ZZ$ in UPCs at the LHC and the FCC-hh, computed with EDFF and ChFF $\gamma$ fluxes and their average.
\label{tab:xsecszz}}
\vspace{0.2cm}
\begin{tabular}{l|ccc} \hline
Process: $\gaga\to ZZ$ & \multicolumn{3}{c}{\gammaUPC~$\sigma$} \\
Colliding system, \cm\ energy & EDFF & ChFF & average \\ \hline
\pp\ at 14~TeV & 52.8 ab & 78.4 ab & $66 \pm 13$ ab \\
\pPb\ at 8.8~TeV & 12.3 fb & 18.8 fb & $ 15.5 \pm 3.2$ fb \\
\PbPb\ at 5.52~TeV & 46.8 fb & 63.2 fb & $55 \pm 8$ fb \\\hline
\pp\ at 100~TeV & 664 ab & 854 ab  & $760 \pm 90$ ab \\
\pPb\ at 62.8~TeV & 684 fb & 940 fb & $810\pm 130$ fb \\
\PbPb\ at 39.4~TeV & 217 pb & 296 pb & $260 \pm 40$ pb \\
\hline
\end{tabular}
\end{table}

\subsection{\texorpdfstring{$\gaga\to\ttbar$}{gamma-gamma to t-tbar}}

Table~\ref{tab:xsecsttbar} lists the SM cross sections for the photon-fusion production of a pair of top quarks in UPCs with protons and ions at LHC and FCC computed at LO and NLO pQCD accuracy with our setup. This process probes anomalous top-quark e.m.\ couplings~\cite{deFavereaudeJeneret:2009db,dEnterria:2009cwl}. The NLO corrections augment the theoretical cross sections by about 50\% and have only few percent uncertainties due to missing higher-order terms (evaluated here by varying the default renormalization scale within a factor of two). This result emphasizes the need to include NLO corrections for the accurate calculation of cross sections for any hadronic final state in UPCs. At the LHC, the cross sections are in the fb range and can only be observed in \pp\ collisions with forward proton tagging (for which the acceptance should be large, given the heavy mass of the central $\ttbar$ system).

\begin{table}[htpb!]
\centering
\tabcolsep=3.5mm
\caption{Total LO and NLO QCD cross sections for $\gaga\to \ttbar$ in UPCs at the LHC, computed with EDFF and ChFF $\gamma$ fluxes, and their average for the NLO case. The quoted asymmetric NLO uncertainty is derived from the renormalization scale variation.
\label{tab:xsecsttbar}}
\vspace{0.2cm}
\begin{tabular}{l|cc|ccc} \hline
Process: $\gaga\to \ttbar$ & \multicolumn{2}{c|}{\gammaUPC~$\sigma_\mathrm{LO}$} & \multicolumn{3}{c}{\gammaUPC~$\sigma_\mathrm{NLO}$} \\
Colliding system, \cm\ energy & EDFF & ChFF & EDFF & ChFF & average \\ \hline
\pp\ at 14~TeV & $0.164$ fb & $0.238$ fb & $0.198_{-0.003}^{+0.004}$ fb & $0.287_{-0.004}^{+0.005}$ fb & $0.242_{-0.004}^{+0.005}\pm 0.045$ fb \\
\pPb\ at 8.8~TeV & $28.3$ fb & $46.4$ fb & $36.5_{-0.7}^{+0.8}$ fb & 
$59.3_{-1.1}^{+1.3}$ fb & $48_{-0.9}^{+1.0} \pm 11$ fb \\
\PbPb\ at 5.52~TeV & $9.23$ fb & $13.6$ fb & $12.6_{-0.3}^{+0.4}$ fb & $18.8_{-0.4}^{+0.5}$ fb & $15.7_{-0.4}^{+0.5} \pm 3.1$ fb \\
\hline
\pp\ at 100~TeV & $1.86$ fb & $2.29$ fb & $2.19_{-0.03}^{+0.03}$ fb & $2.70_{-0.03}^{+0.04}$ fb & $2.45_{-0.03}^{+0.04} \pm 0.26$ fb \\
\pPb\ at 62.8~TeV & $2.38$ pb & $3.05$ pb & $2.86_{-0.04}^{+0.05}$ pb & $3.62_{-0.05}^{+0.06}$ pb & $3.24_{-0.05}^{+0.06} \pm 0.38$ pb \\
\PbPb\ at 39.4~TeV & $0.66$ nb & $0.95$ nb & $0.830_{-0.015}^{+0.018}$ nb & $1.19_{-0.02}^{+0.03}$ nb & $1.00_{-0.02}^{+0.03} \pm 0.18$ nb \\
\hline
\end{tabular}
\end{table}

\subsection{\texorpdfstring{$\gaga\to \mathrm{HH}$}{gamma-gamma to HH}}

Table~\ref{tab:xsecsHH} lists the SM cross sections for the photon-fusion production of a pair of Higgs bosons in UPCs with protons and ions at LHC and FCC, a process that probes the Higgs potential~\cite{Belusevic:2004pz} and the quartic $\gamma\gamma \mathrm{HH}$ coupling. The SM double-Higgs cross sections are in the sub-attobarn range and will likely remain unobservable in such a production mode. Even in the most favourable case of \pp\ collisions at FCC with forward proton taggers to remove backgrounds, one expects $N_\mathrm{HH}\approx 1~\mathrm{ab} \times 20~\mathrm{ab}^{-1}\times \mathcal{B}(H\to\bbbar)^2\approx 7$~events in the dominant 4 b-jets decay channel (on top of a much larger expected $\gaga\to 2(\bbbar)$ continuum background).

\begin{table}[htpb!]
\centering
\tabcolsep=3.5mm
\caption{Total cross sections for $\gaga\to \mathrm{HH}$ in UPCs at the LHC, computed with EDFF and ChFF $\gamma$ fluxes and their average. 
\label{tab:xsecsHH}}
\vspace{0.2cm}
\begin{tabular}{l|ccc} \hline
Process: $\gaga\to \mathrm{HH}$ & \multicolumn{3}{c}{\gammaUPC~$\sigma$} \\
Colliding system, \cm\ energy & EDFF & ChFF & average \\ \hline
\pp\ at 14~TeV & $0.080$ ab & $0.12$ ab & $0.10\pm 0.02$ ab \\
\pPb\ at 8.8~TeV & $18.2$ ab & $28.6$ ab & $23.4\pm 5.2$ ab \\
\PbPb\ at 5.52~TeV & $21.6$ ab & $29.0$ ab & $25.3\pm 3.7$ ab \\
\hline
\pp\ at 100~TeV & $0.88$ ab & $1.09$ ab & $1.0\pm0.1$ ab \\
\pPb\ at 62.8~TeV & $1.14$ fb & $1.46$ fb & $1.3\pm0.2$ fb \\
\PbPb\ at 39.4~TeV & $0.38$ pb & $0.54$ pb & $0.46\pm 0.08$ pb \\
\hline
\end{tabular}
\end{table}

\subsection{Axion-like particles}

\begin{figure}[htbp!]
\centering
\includegraphics[width=0.8\textwidth]{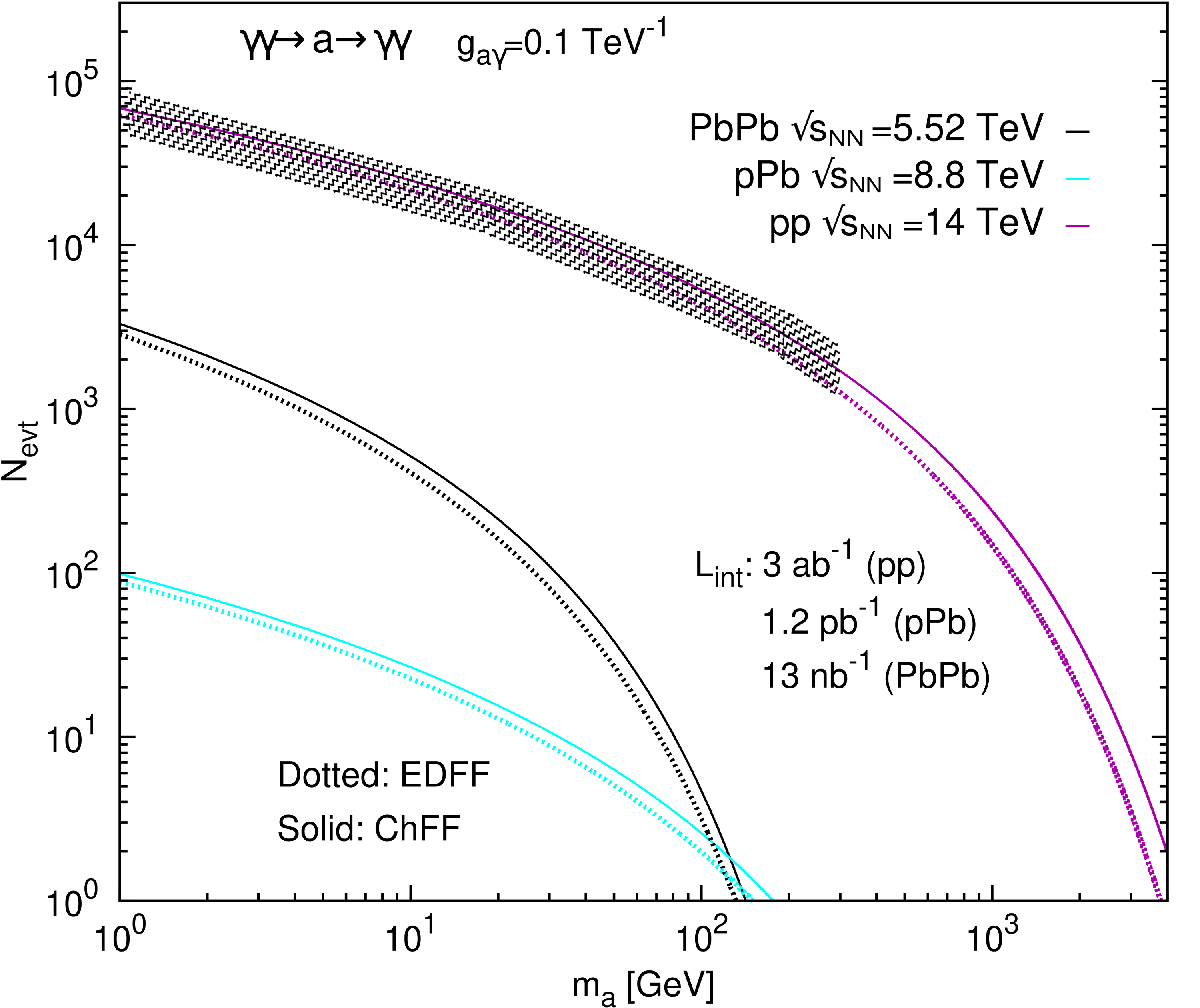}
\caption{Total number of ALPs events expected via $\gaga \to a \to \gaga$ in \pp, \pPb, and \PbPb\ UPCs at the LHC as a function of ALP mass, for fixed ALP-photon coupling $g_{a\gamma}=0.1~\text{TeV}^{-1}$ (approximately corresponding to the current experimental limits over this mass range~\cite{dEnterria:2021ljz}) computed with EDFF (dotted) and ChFF (solid) $\gamma$ fluxes. The hatched area around the \pp\ curve indicates the range of masses below $m_a \approx 300$~GeV where the detection is hindered due to pileup and lack of proton tagging acceptance.
\label{fig:axionxs}
}
\end{figure}

The photon-fusion production of axion-like particles in UPCs decaying back into two photons, provides arguably the most competitive search channel over the ALP mass range $m_a \approx 1$--100~GeV at present and future hadron colliders~\cite{Knapen:2016moh,dEnterria:2021ljz}. The effective Lagrangian for an ALP of mass $m_a$ preferentially coupling to photons reads 
\begin{eqnarray}
\mathcal{L}&\supset& \frac{1}{2}\partial_\mu a \partial^\mu a-\frac{m_a^2}{2}a^2-\frac{g_{a\gamma}}{4}a F^{\mu \nu}\tilde{F}_{\mu\nu}
\end{eqnarray}
where $a$ is the ALP field, $F^{\mu \nu}\,(\tilde{F}_{\mu\nu})$ is the photon field strength (dual) tensor, and the dimensionful ALP-photon coupling strength $g_{a\gamma} \propto 1/\Lambda$ is inversely proportional to the high-energy scale $\Lambda$ associated with the spontaneous breaking of a new global U$(1)$ approximate symmetry. This Lagrangian determines the ALP photon-fusion production cross section and its corresponding diphoton decay width, which is $\Gamma_{a\to \gaga} = g_{a\gamma}^2m_a^3/(64\pi)$.
Exclusive searches in \PbPb\ UPCs provide today the best exclusion limits for ALP masses $m_a \approx 5$--100 GeV for axion-photon couplings down to $g_{a\gamma} \approx 0.1$~TeV$^{-1}$~\cite{Knapen:2016moh,CMS:2018erd,ATLAS:2020hii}. For such a value of $g_{a\gamma}$, Fig.~\ref{fig:axionxs} shows the expected $\gaga \to a \to \gaga$ cross sections in \pp, \pPb, and \PbPb\ UPCs at the LHC, as a function of ALP mass, for the EDFF and ChFF $\gamma$ fluxes. The hatched area around the \pp\ luminosities indicate that for the range of masses below $m_a \approx 300$~GeV, ALP detection is hindered in \pp\ UPCs due to pileup and lack of proton tagging acceptance. The plot confirms that \PbPb\ UPCs provide the most competitive means to search for ALPs in the region $m_a \approx 1$--100~GeV, but that \pp\ UPCs will rapidly take over beyond this mass with the full LHC integrated luminosity and forward proton taggers to remove pileup background~\cite{Baldenegro:2018hng}, probing ALP masses above a few TeV.

\subsection{Massive gravitons}

\begin{figure}[htbp!]
\centering
\includegraphics[width=0.8\textwidth]{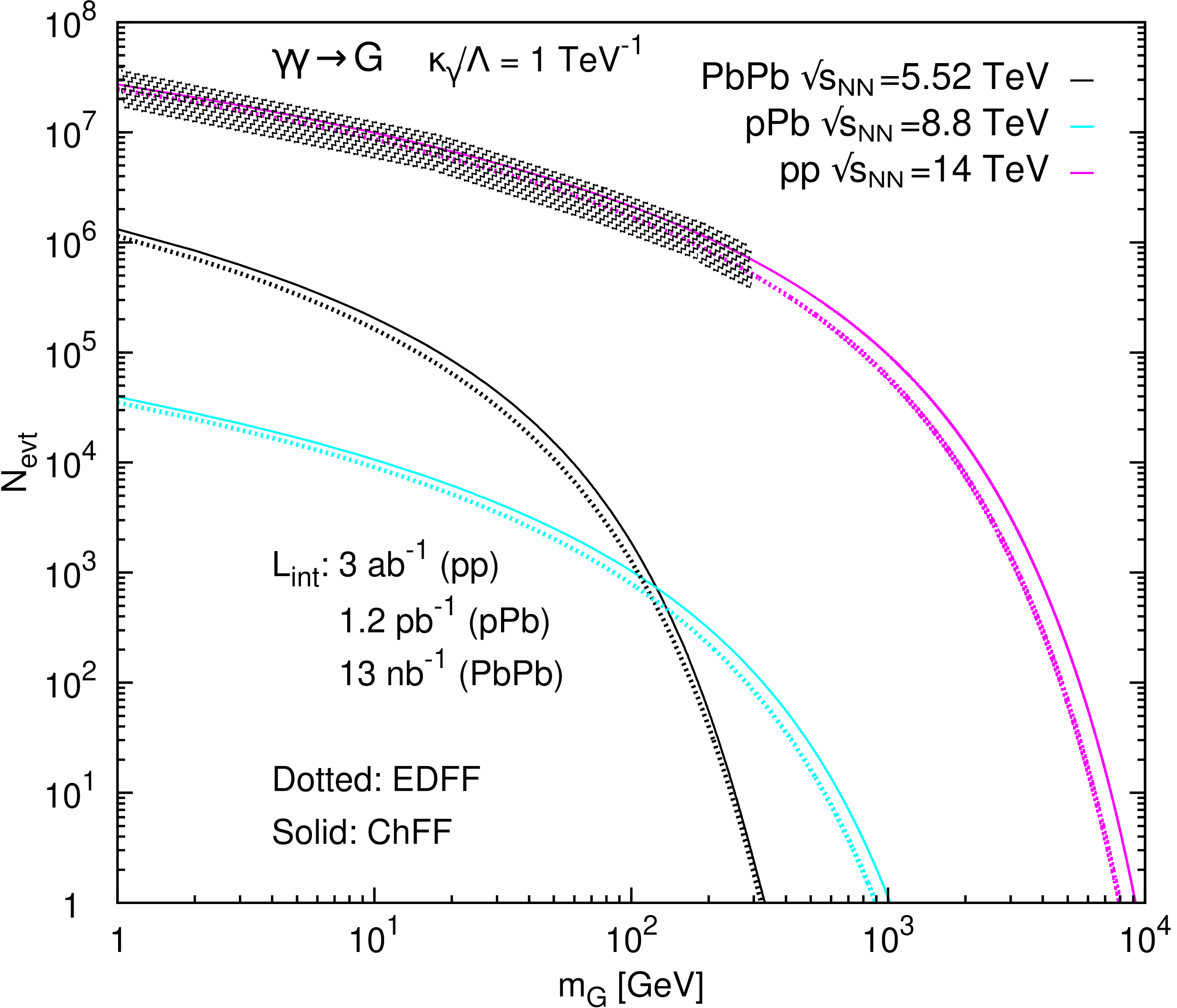}
\caption{Total number graviton events expected via $\gaga \to G$ in \pp, \pPb, and \PbPb\ UPCs at the LHC, as a function of graviton mass, for fixed $\kappa_\gamma/\Lambda=1~\mathrm{TeV}^{-1}$ graviton-photon coupling computed with EDFF (dotted) and ChFF (solid) $\gamma$ fluxes. The hatched area around the \pp\ curve indicates the range of masses below $m_G \approx 300$~GeV where the detection is hindered due to pileup and lack of proton tagging acceptance.
\label{fig:gravitonnevt}
}
\end{figure}

The production of spin-2 massive gravitons in UPCs can be also computed with our setup. We consider the effective field theory of a massive graviton $G$ interacting with the photon field, where the kinetic term of $G$ is the well-known Fierz--Pauli Lagrangian with the positive-energy condition $\partial_\mu G^{\mu\nu}=0$. The interaction between the $G$ and $\gamma$ is then described by the Lagrangian~\cite{Das:2016pbk}
\begin{eqnarray}
    \mathcal{L}&\supset&-\frac{\kappa_\gamma}{\Lambda}T_{\mu\nu}^\gamma G^{\mu\nu},
\end{eqnarray}
where $T_{\mu\nu}^\gamma$ is the energy-momentum tensor of the photon, and $\kappa_\gamma/\Lambda$ the effective graviton-photon coupling. The LO decay width is given by $\Gamma_{G\to \gaga}=\kappa_\gamma^2 m_{G}^3/(80\pi \Lambda^2)$.
The number of total events of $\gaga\to G$ at the LHC are displayed in Fig.~\ref{fig:gravitonnevt} for a choice of coupling $\kappa_\gamma/\Lambda=1~\mathrm{TeV}^{-1}$ in \pp, \pPb, and \PbPb\ UPCs, as a function of $G$ mass, for the EDFF and ChFF $\gamma$ fluxes. The hatched area around the \pp\ luminosities indicate that for the range of masses below $m_G \approx 300$~GeV, graviton detection is hindered in \pp\ UPCs due to pileup and lack of proton tagging acceptance. As for ALPs, the plot confirms that \PbPb\ UPCs provide the most competitive means to search for massive gravitons in the region $m_G \approx 1$--100~GeV, but that searches with \pp\ UPCs can eventually reach $m_G$ values in the multi-TeV scale, with the full LHC integrated luminosity and forward proton taggers to remove pileup background.

\section{Differential photon-photon cross section results: Data vs.\ \texorpdfstring{\gammaUPC}{gamma-UPC}}
\label{sec:diff_results}

In this section we present differential cross sections for exclusive dileptons, $\gaga\to\ell^+\ell^-$ and light-by-light scattering, $\gaga\to\gaga$, in \PbPb\ UPCs at $\sqrtsnn = 5.02$~TeV where our calculations can be compared to existing LHC data and to alternative predictions from the UPC-dedicated \starlight\ and \superchic\ MC generators. In all cases, \gammaUPC\ results derived with EDFF and ChFF photon fluxes are presented.

\subsection{Exclusive dielectrons in \PbPb\ UPCs  \texorpdfstring{$\sqrtsnn= 5.02$}{sqrt(s)= 5.02}~TeV}

The exclusive production of electron-positron pairs in photon-photon collisions, $\gaga\to\epem$, known as the Breit--Wheeler (B--W) process~\cite{Breit:1934zz}, is the simplest elementary process in two-photon physics. In addition, the B--W continuum constitutes a background for the measurement of multiple dielectron resonances (in particular vector meson ones produced via exclusive photon-hadron collisions), which needs to be properly understood and subtracted. The simplicity and large cross section of the B--W process has facilitated its measurement in hadronic UPCs multiple times (by the WA93~\cite{Vane:1992ms}, CERES/NA45~\cite{CERESNA45:1994cpb}, STAR~\cite{STAR:2004bzo,STAR:2019wlg}, PHENIX~\cite{PHENIX:2009xtn}, CDF~\cite{CDF:2006apx}, ALICE~\cite{ALICE:2013wjo}, CMS~\cite{CMS:2012cve,CMS:2018erd,CMS:2018uvs}, and ATLAS~\cite{ATLAS:2015wnx,ATLAS:2017fur,ATLAS:2020mve} experiments), and has become a clean final state to test the theoretical ingredients of UPC cross section calculations.

\begin{table}[htpb!]
\centering
\tabcolsep=2.mm
\caption{Fiducial exclusive dielectron cross sections measured in \PbPb\ UPCs at $\sqrtsnn=5.02$~TeV ($\ET^{e}>2$~GeV , $|y^{e}|<2.4$, $m_{\epem}>5$~GeV, $p_{\mathrm{T},\epem}<1$~GeV), compared to the theoretical \gammaUPC\ results obtained with EDFF and ChFF $\gamma$ fluxes (and their average), and to the \starlight\ and \superchic\ MC predictions.\label{tab:xsecsdielec}}
\vspace{0.2cm}
\begin{tabular}{l|c|ccc|c|c} \hline
Process, system & Scaled CMS data~\cite{CMS:2018erd} & \multicolumn{3}{c|}{\gammaUPC~$\sigma$} & \starlight~$\sigma$ & \superchic~$\sigma$ \\
 & & EDFF & ChFF & average & &\\
 $\gaga\to\epem$, \PbPb\ at 5.02 TeV & $ 275 \pm 55$~$\mu$b & 272~$\mu$b & 326~$\mu$b & $298\pm 28$~$\mu$b & 285~$\mu$b & 318~$\mu$b \\\hline
\end{tabular}
\end{table}

\begin{figure}[htbp!]
\centering
\includegraphics[width=0.49\textwidth]{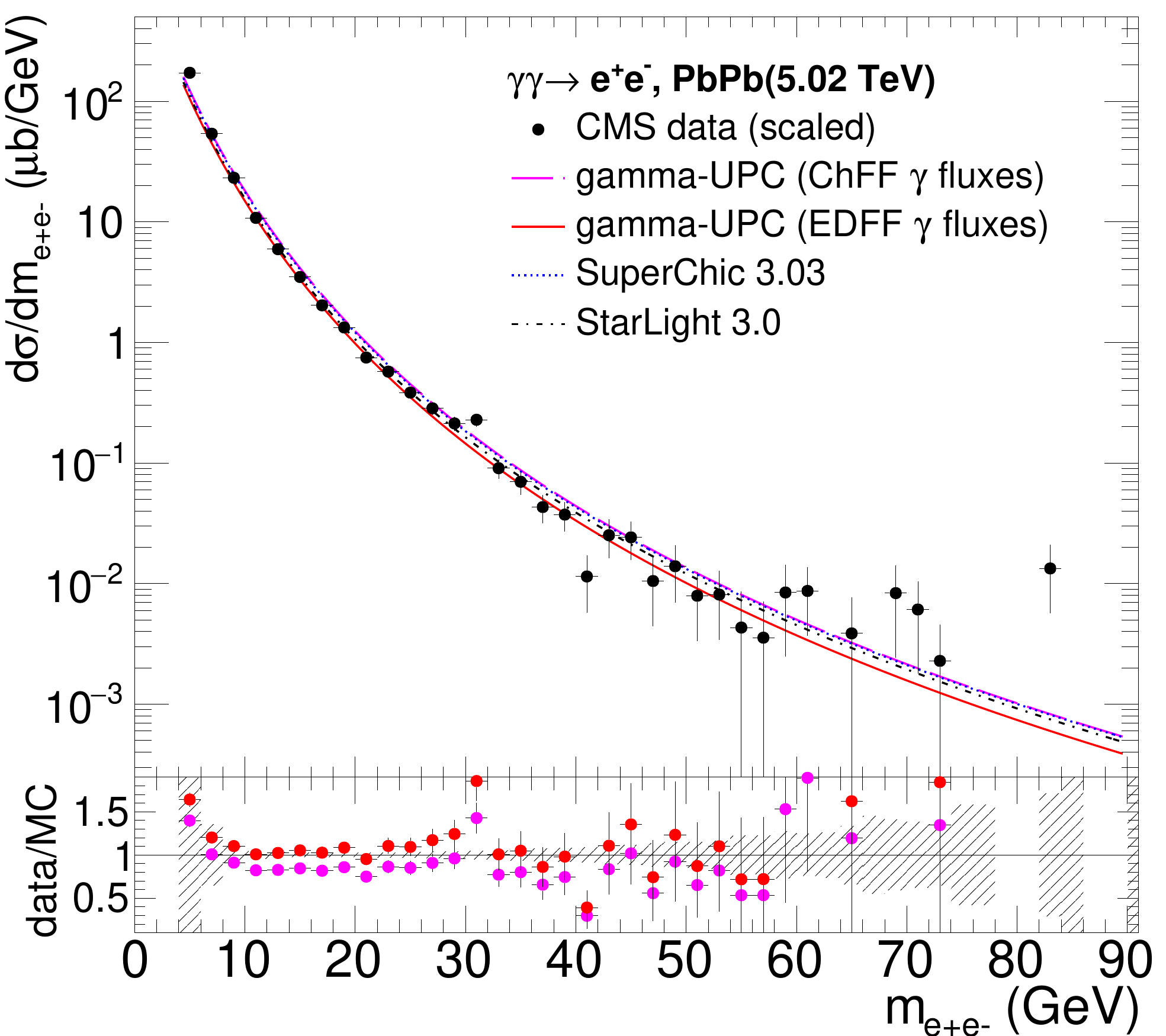}
\includegraphics[width=0.49\textwidth]{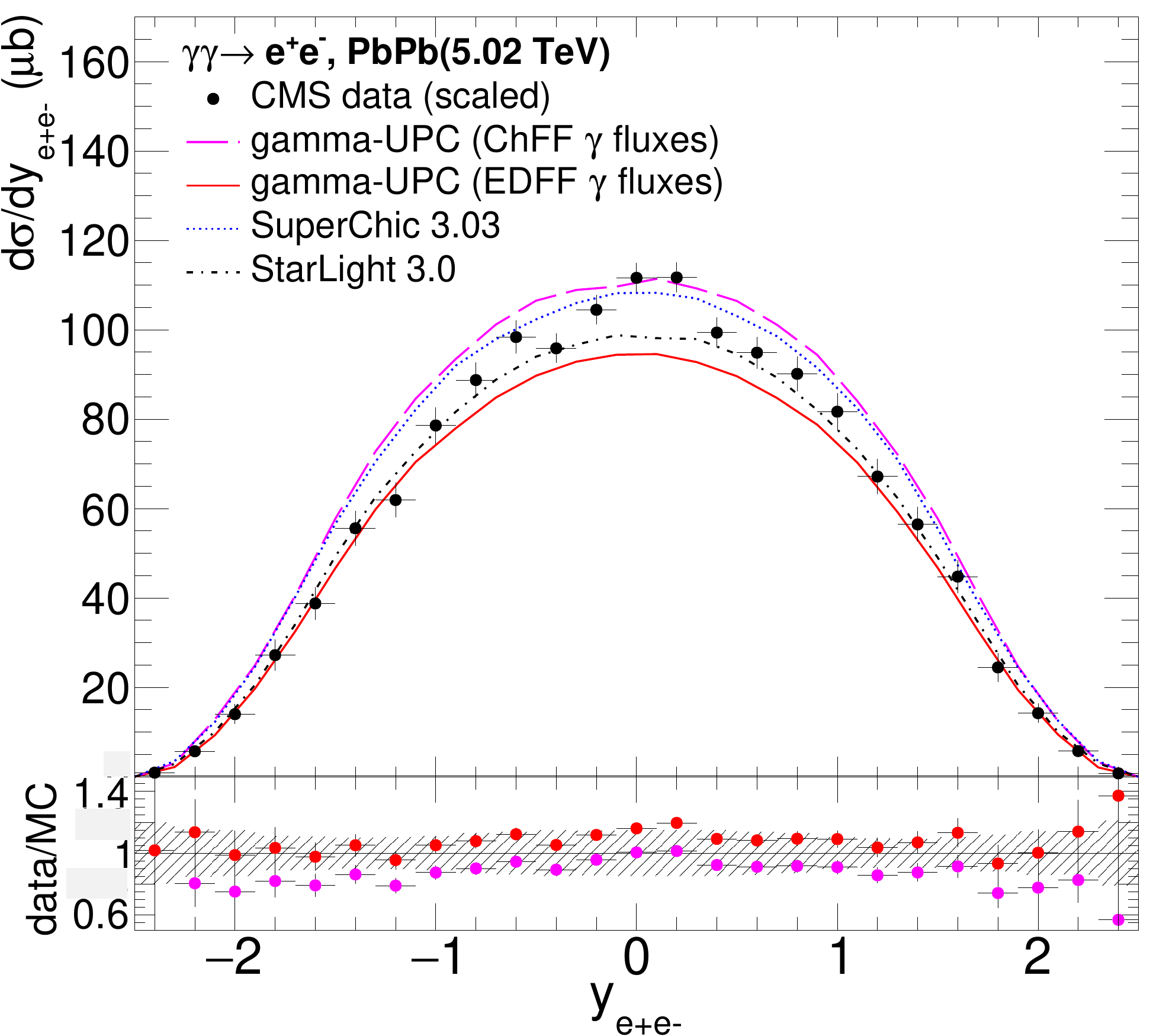}
\caption{Comparison of the differential fiducial cross sections for exclusive $\epem$ production in \PbPb\ UPCs at $\sqrtsnn= 5.02$~TeV as a function of pair invariant mass (left) and rapidity (left) predicted by \gammaUPC\ (EDFF and ChFF $\gamma$ fluxes), \starlight, and \superchic. The data points show the CMS results~\cite{CMS:2018erd} scaled as explained in the text. The bottom insets show the ratio of the CMS results (with associated systematic uncertainties indicated by hashed boxes) to the EDFF (red) and ChFF (purple) \gammaUPC\ predictions.
\label{fig:excl_ee}}
\end{figure}

Table~\ref{tab:xsecsdielec} lists the integrated fiducial cross sections, measured by CMS in \PbPb\ UPCs at $\sqrtsnn= 5.02$~TeV~\cite{CMS:2018erd} compared to our \gammaUPC\ calculations with the two form factors (and their average), as well as to the \starlight~3.0~\cite{Klein:2016yzr} and \superchic~3.03~\cite{Harland-Lang:2018iur} predictions. For comparison purposes, the original CMS experimental uncorrected yields have been scaled to a fully corrected cross section by using their known ratio to the corresponding reconstructed \starlight\ result over the measured fiducial phase space ($\ET^{e}>2$~GeV , $|y^{e}|<2.4$, $m_{\epem}>5$~GeV, $p_{\mathrm{T},\epem}<1$~GeV)~\cite{CMS:2018erd}.  The first observation is that the EDFF and \starlight\ (as well as ChFF and \superchic) results are very similar, and the data seem to fall in between all predictions. In Fig.~\ref{fig:excl_ee}, we plot the B--W differential distributions as a function of dielectron invariant mass (left) and rapidity (right) compared to all theoretical predictions. Within the current experimental uncertainties, all calculations are consistent with the measurement, calling for upcoming higher-precision B--W measurements (\eg\ in the higher $m_{\epem}\gtrsim8$~GeV mass region which features smaller systematic uncertainties) to be able to better discriminate among the different model ingredients.

\subsection{Exclusive dimuons in \PbPb\ UPCs at \texorpdfstring{$\sqrtsnn= 5.02$}{sqrt(s)= 5.02}~TeV}

Like its dielectron counterpart, the exclusive dimuon production in UPCs is also a clean standard-candle process that can be used to calibrate our theoretical understanding of EPA fluxes, survival probabilities, higher-order QED corrections, etc. At the LHC, the process has been measured with proton~\cite{CMS:2011vma,ATLAS:2015wnx,ATLAS:2017sfe,CMS:2018uvs,ATLAS:2020mve} and nuclear beams~\cite{CMS:2020skx,ATLAS:2020epq}, and a detailed discussion of the \superchic\ and \starlight\ predictions confronted to the differential ATLAS data has been presented in~\cite{Harland-Lang:2021ysd}. In Table~\ref{tab:xsecsdimuo}, we compare the integrated fiducial cross section measured in \PbPb\ UPCs at $\sqrtsnn = 5.02$~TeV to the \gammaUPC, \starlight, and \superchic\ predictions.
The results with ChFF flux (and \superchic) seem to overshoot the total fiducial cross section of the ATLAS measurement by 18\%, while the EDFF (and \starlight) cross section undershots it by 6\%. The ChFF and EDFF average agrees perfectly with the data.

\begin{table}[htpb!]
\centering
\tabcolsep=2.5mm
\caption{Fiducial exclusive dimuon cross sections measured in \PbPb\ UPCs at $\sqrtsnn=5.02$~TeV (with $\pT^{\mu} > 4$~GeV, $|\eta^\mu| < 2.4$, $m_{\mumu} > 10$~GeV, $p_{\mathrm{T},\mumu} < 2$~GeV), compared to the theoretical \gammaUPC\ results obtained with EDFF and ChFF $\gamma$ fluxes (and their average), as well as with the \starlight\ and \superchic\ MC predictions.\label{tab:xsecsdimuo}}
\vspace{0.2cm}
\begin{tabular}{l|c|ccc|c|c} \hline
Process, system & ATLAS data~\cite{ATLAS:2020epq} & \multicolumn{3}{c|}{\gammaUPC~$\sigma$} & \starlight~$\sigma$ & \superchic~$\sigma$ \\
 & & EDFF & ChFF  & average & &\\
 $\gaga\to\mumu$, \PbPb\ at 5.02 TeV & $34.1\pm 0.8$
 ~$\mu$b & $32.1$~$\mu$b & $40.4$~$\mu$b & $36.2 \pm 4.2$~$\mu$b & $32.1$~$\mu$b & $38.9$~$\mu$b \\\hline
\end{tabular}
\end{table}

\begin{figure}[htbp!]
\centering
\includegraphics[width=0.33\textwidth]{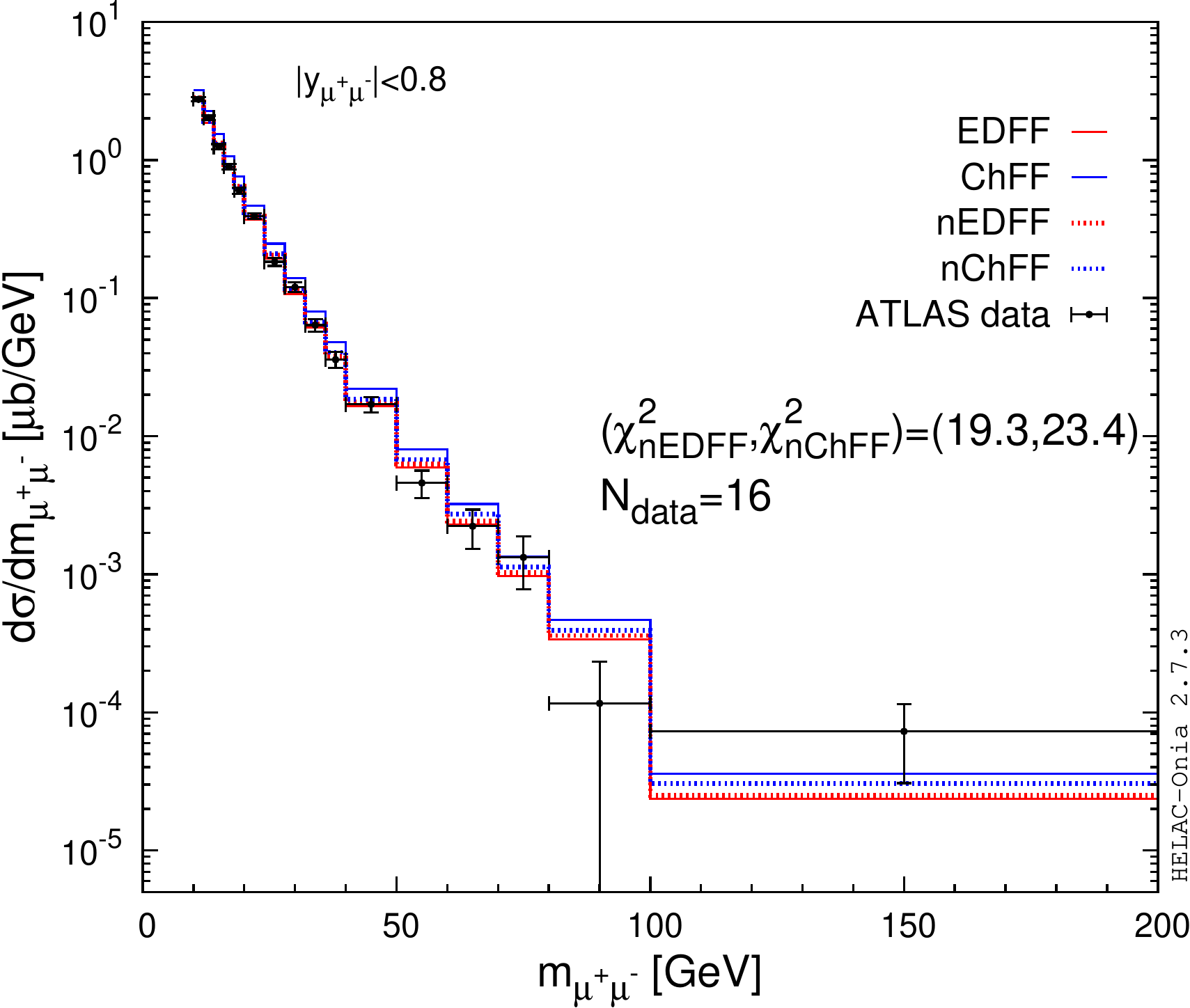}
\includegraphics[width=0.33\textwidth]{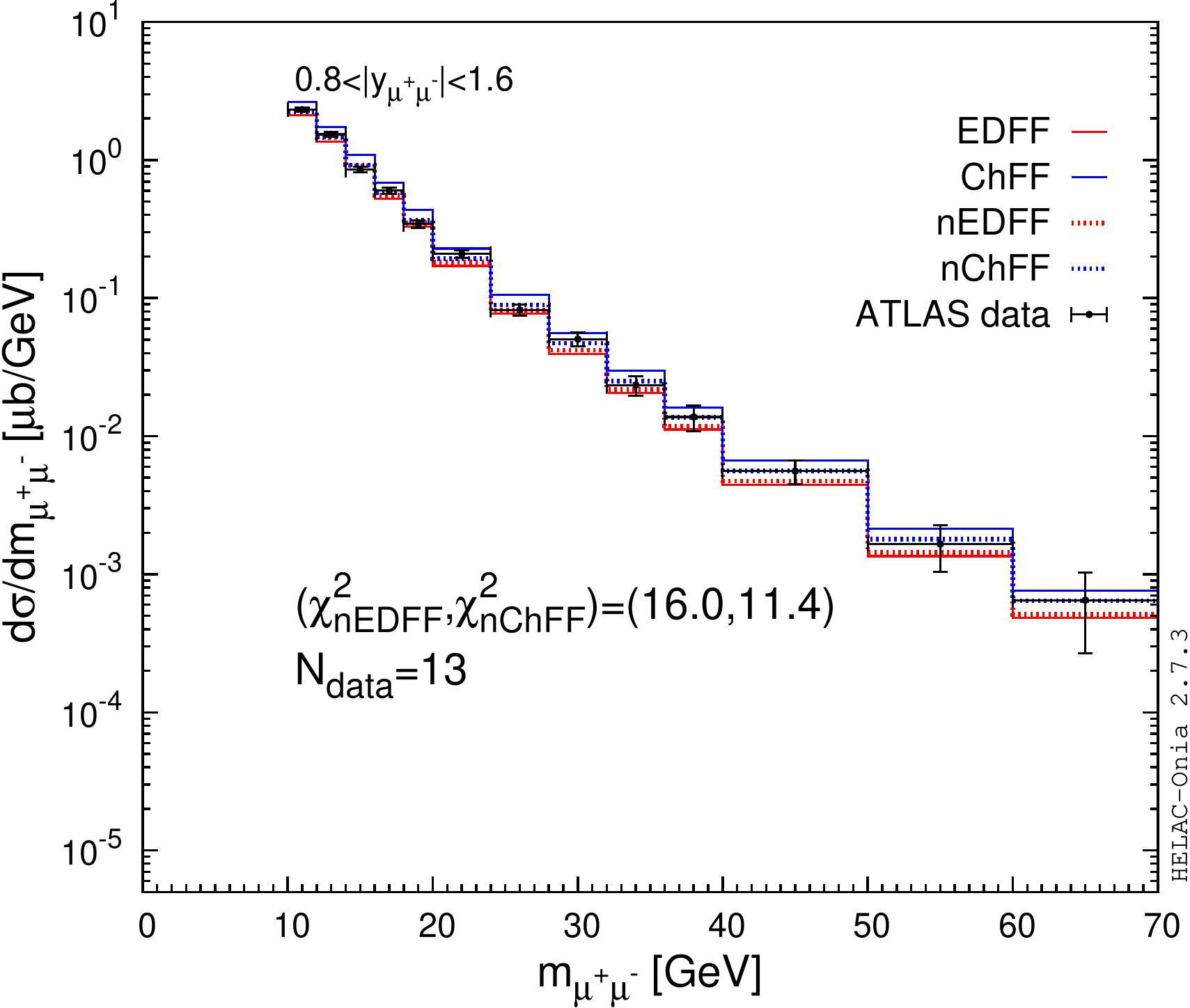}
\includegraphics[width=0.33\textwidth]{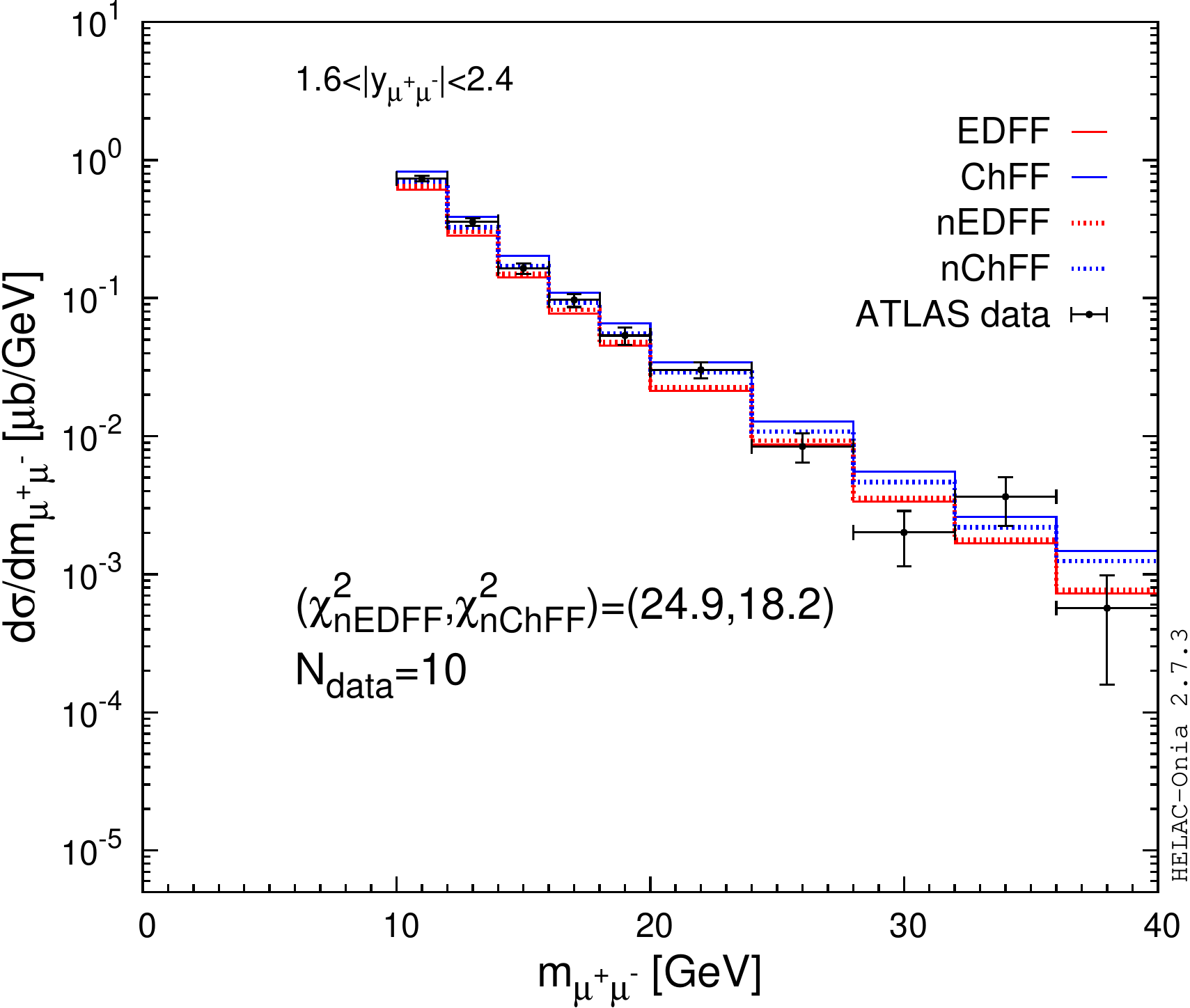}\\
\includegraphics[width=0.33\textwidth]{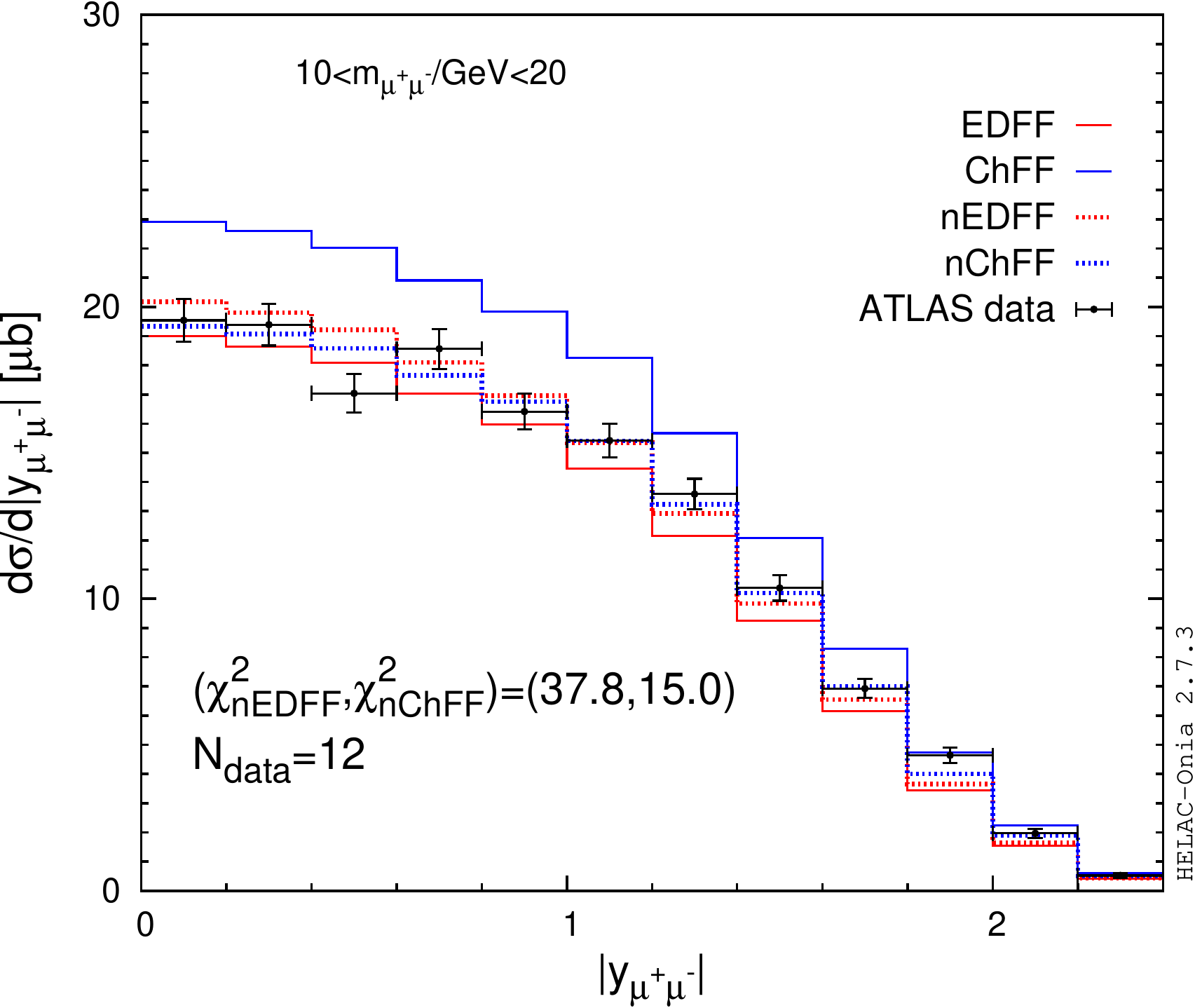}
\includegraphics[width=0.33\textwidth]{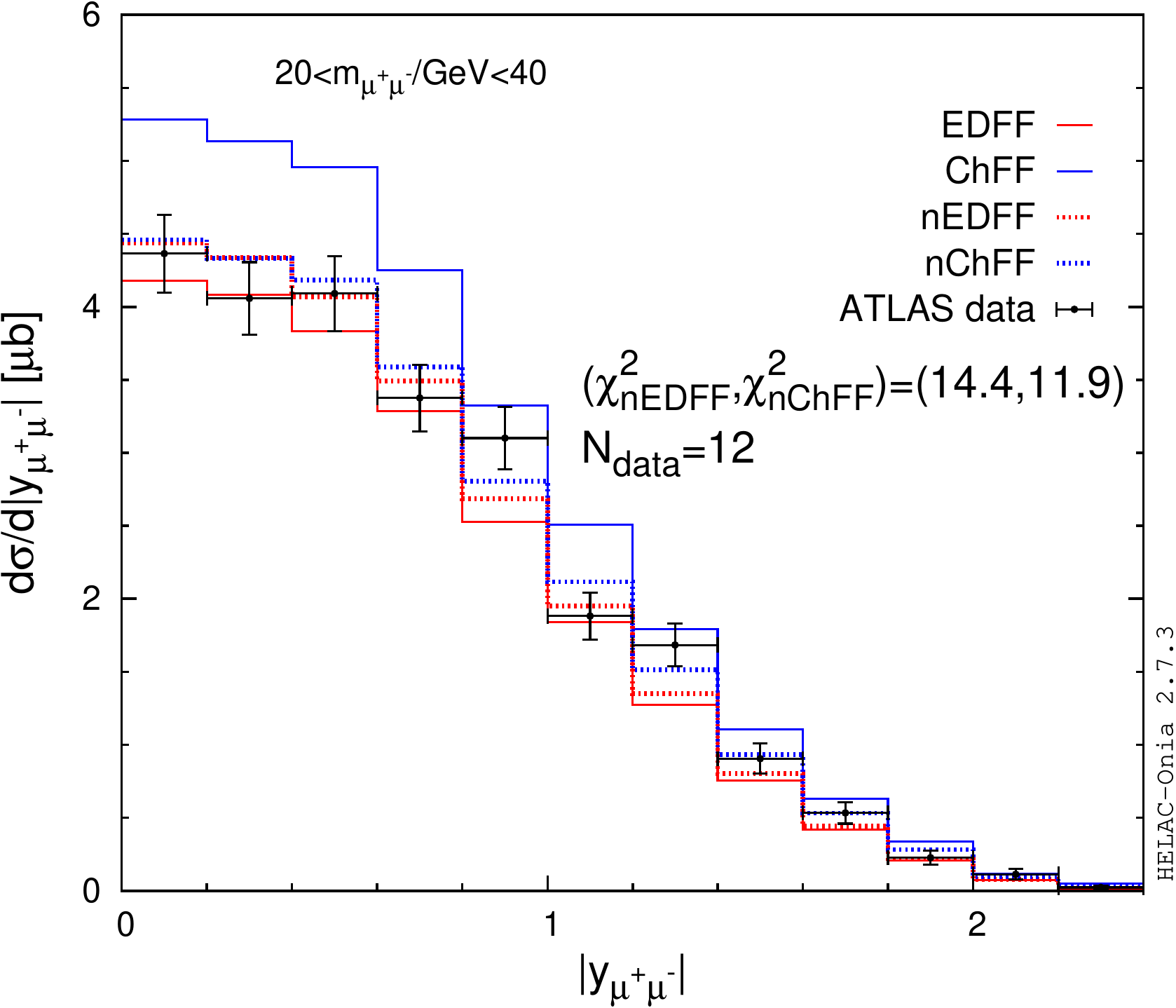}
\includegraphics[width=0.33\textwidth]{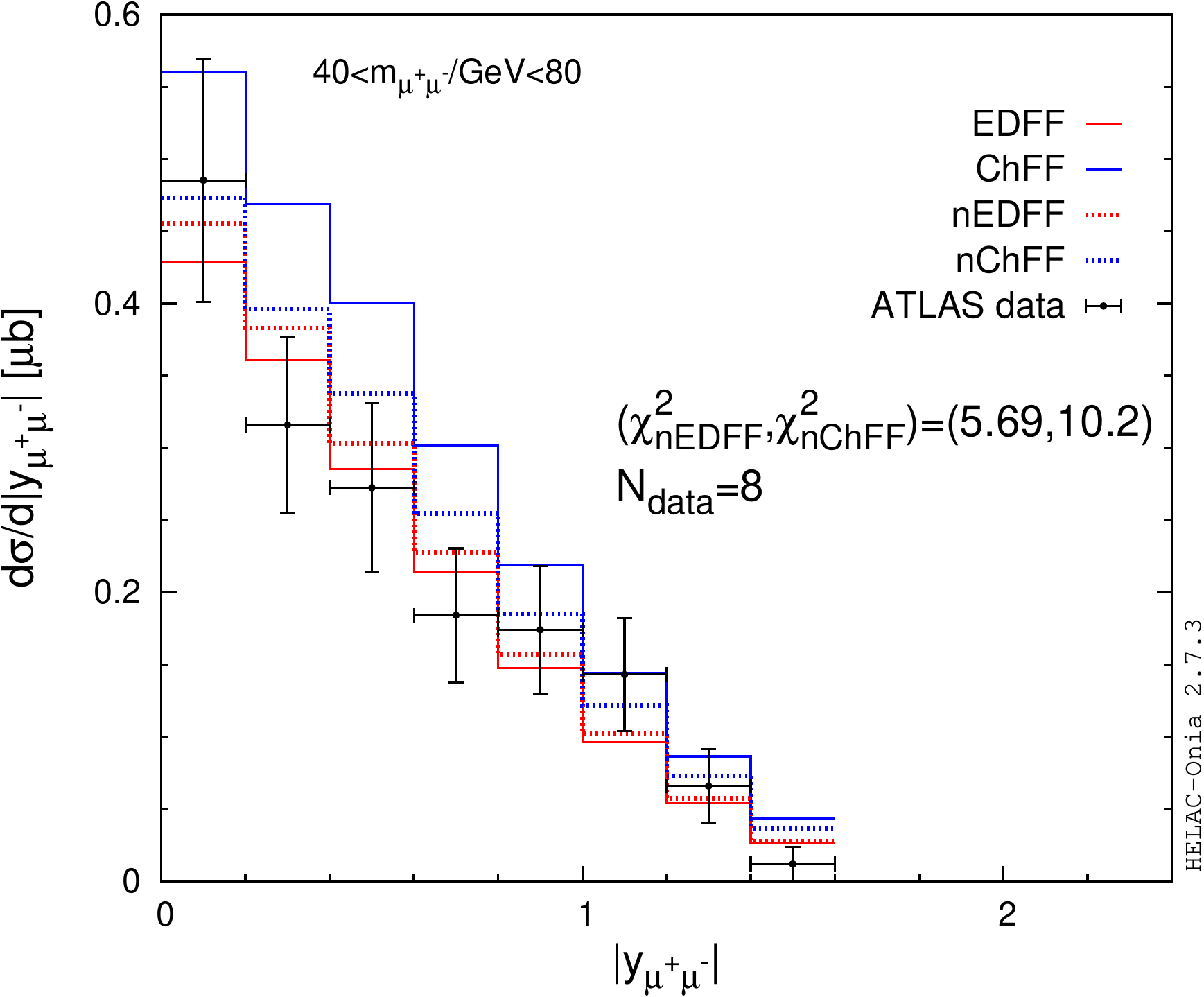}\\
\includegraphics[width=0.33\textwidth]{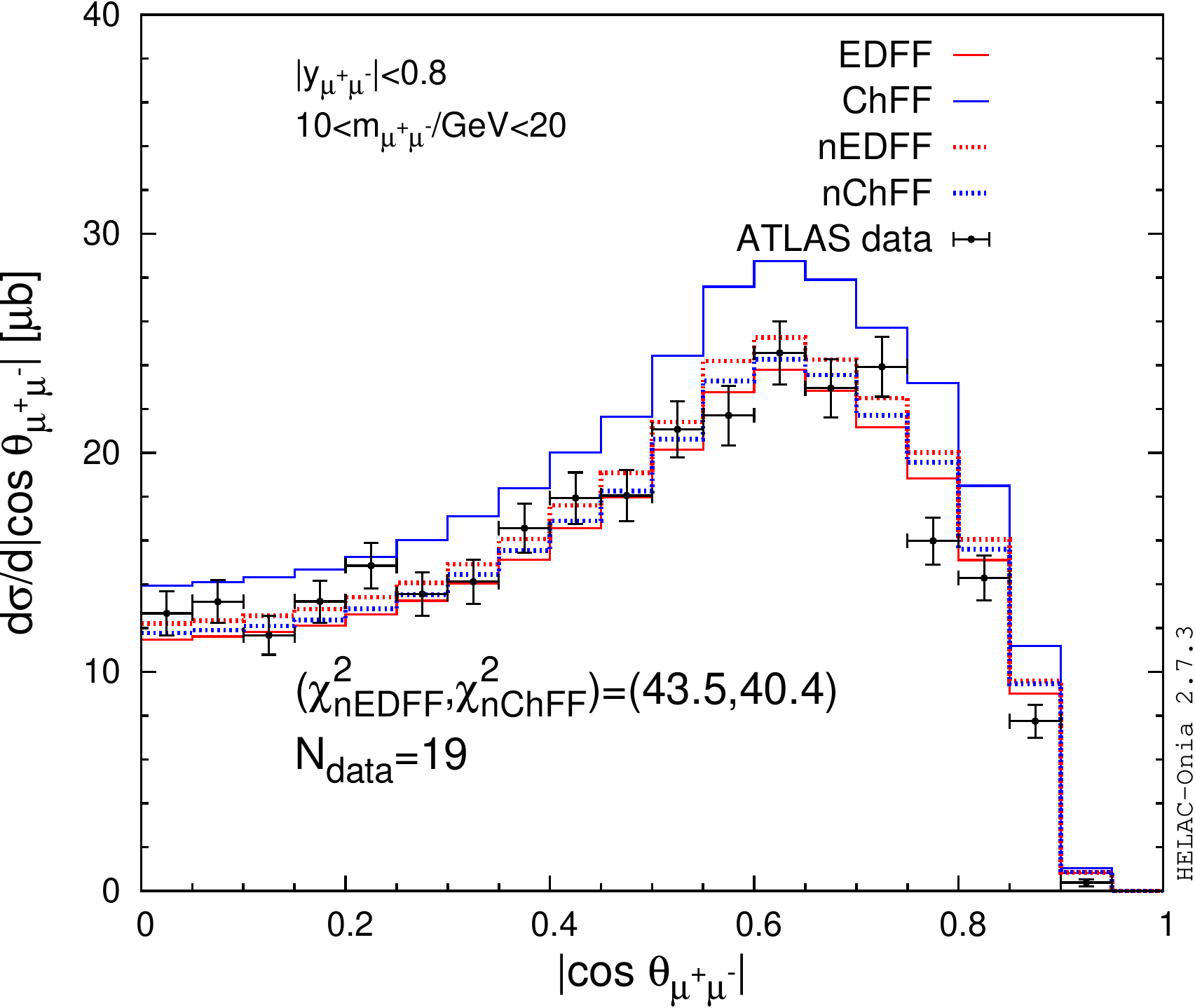}
\includegraphics[width=0.33\textwidth]{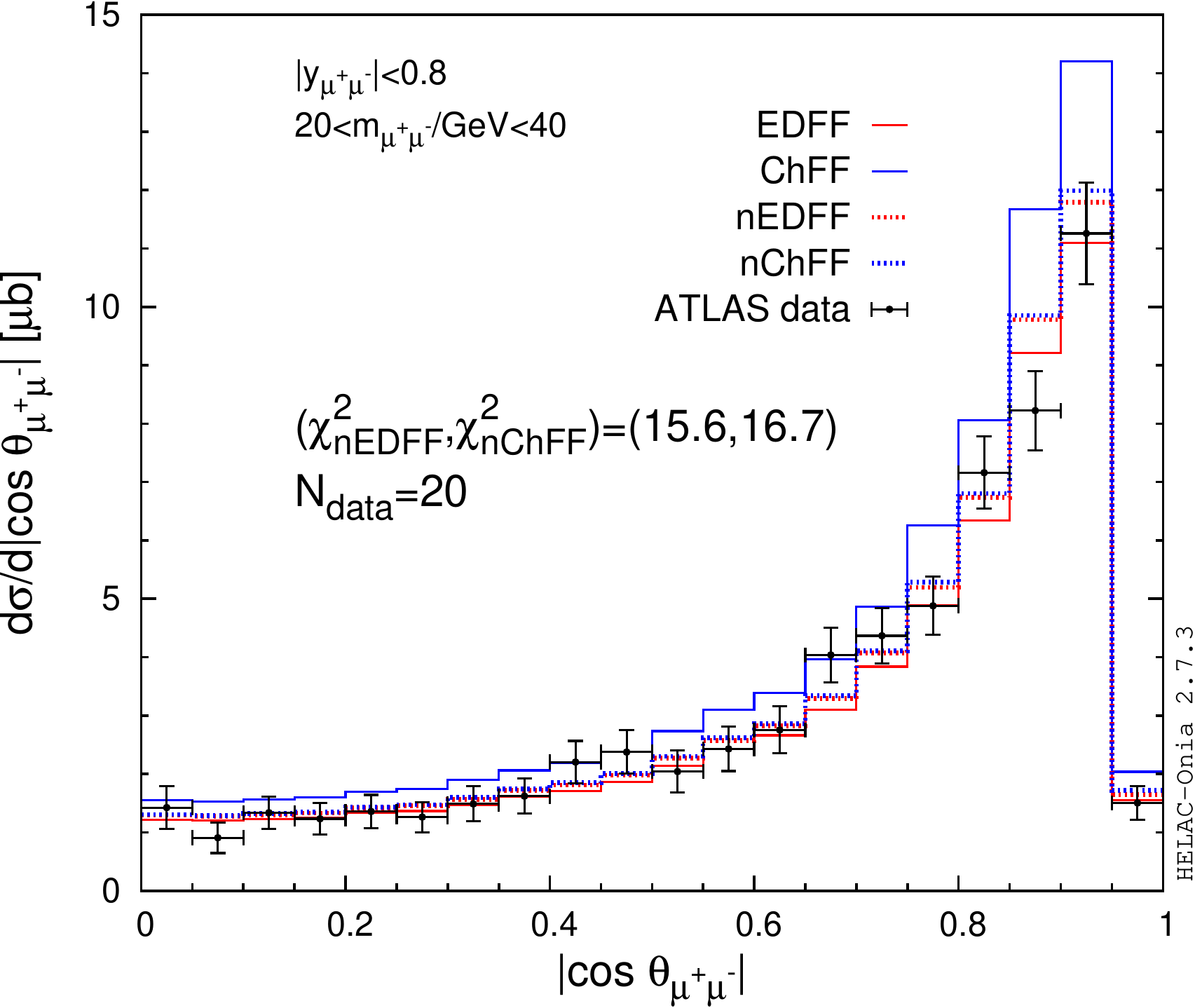}
\includegraphics[width=0.33\textwidth]{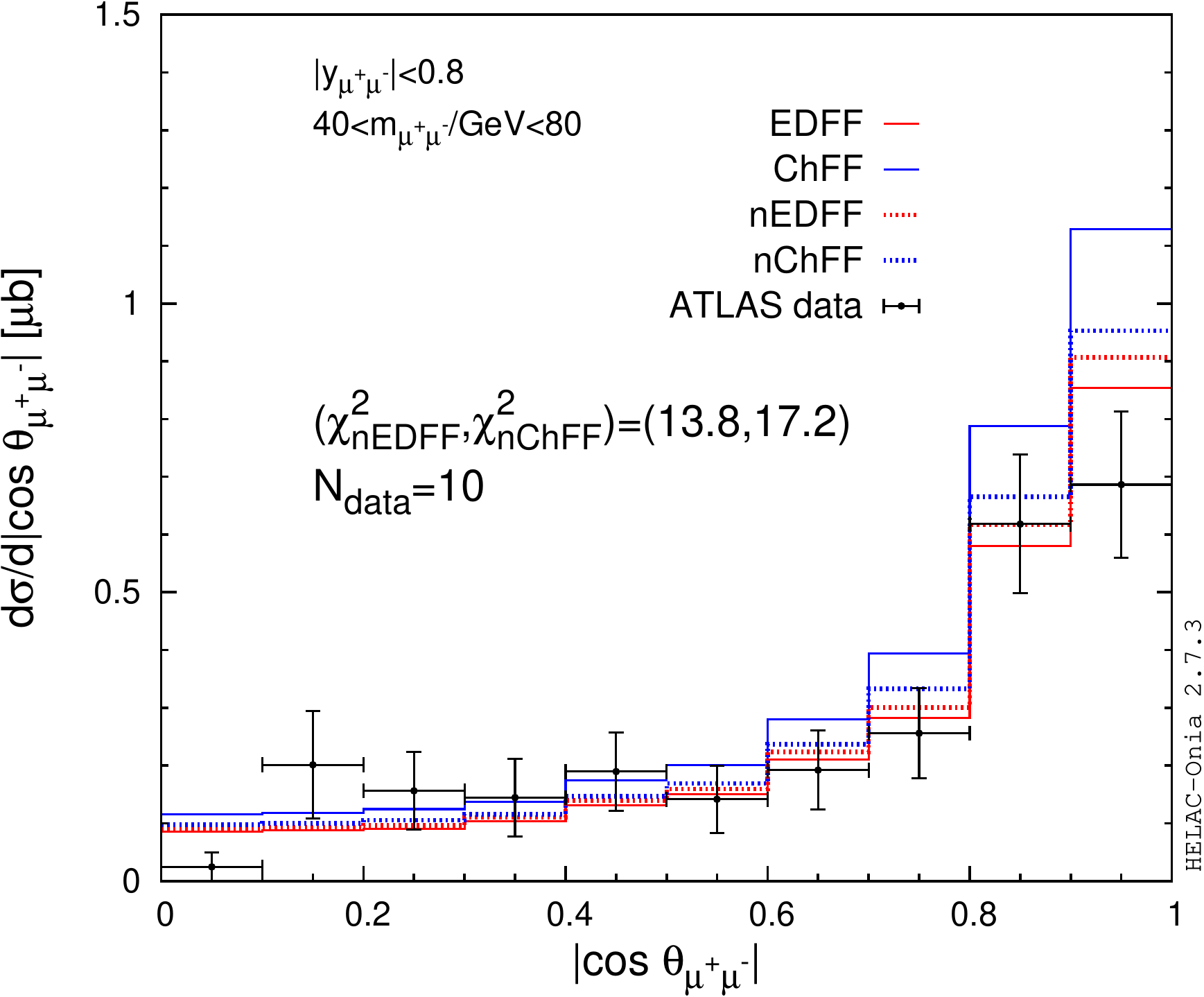}\\
\includegraphics[width=0.33\textwidth]{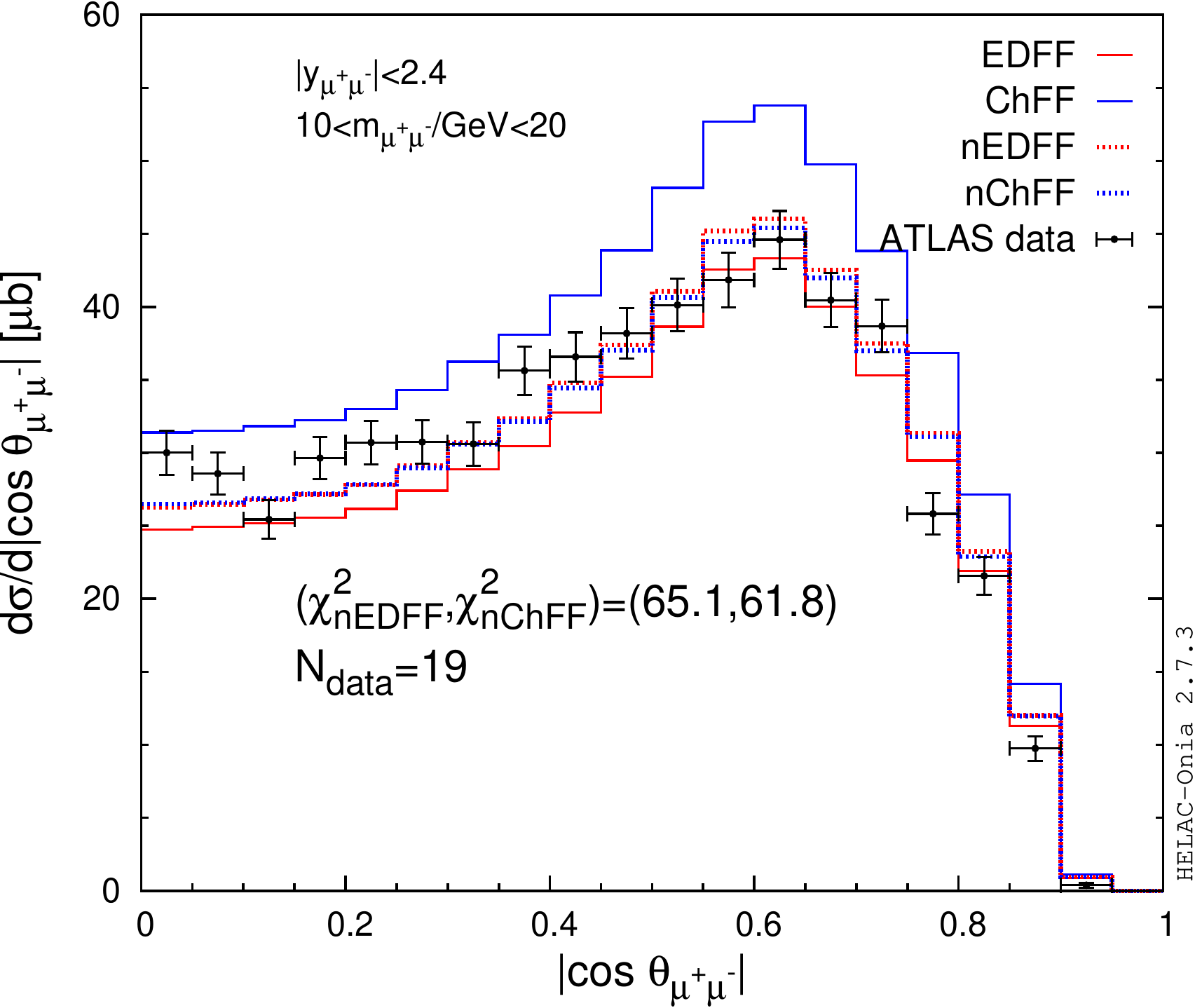}
\includegraphics[width=0.33\textwidth]{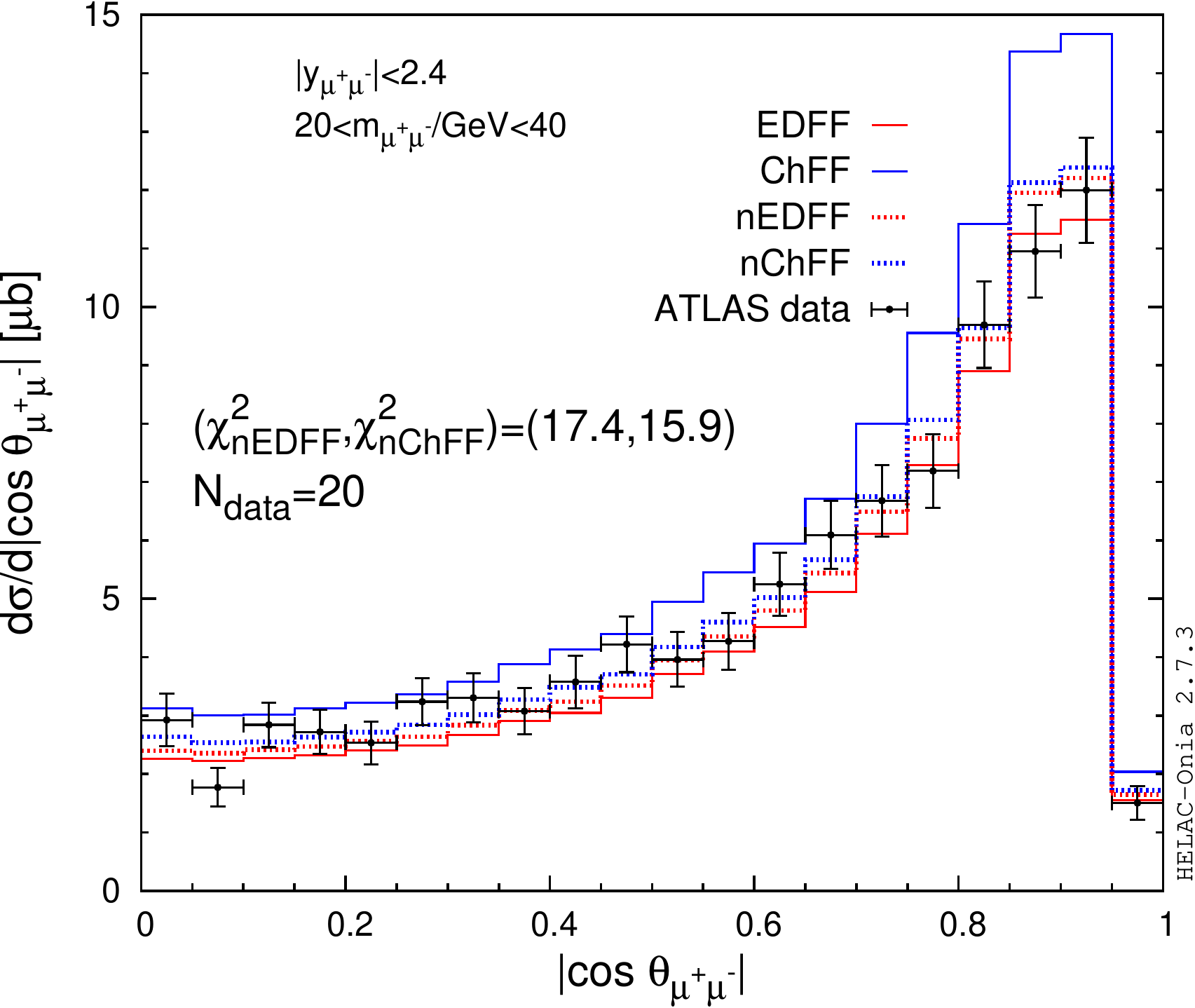}
\includegraphics[width=0.33\textwidth]{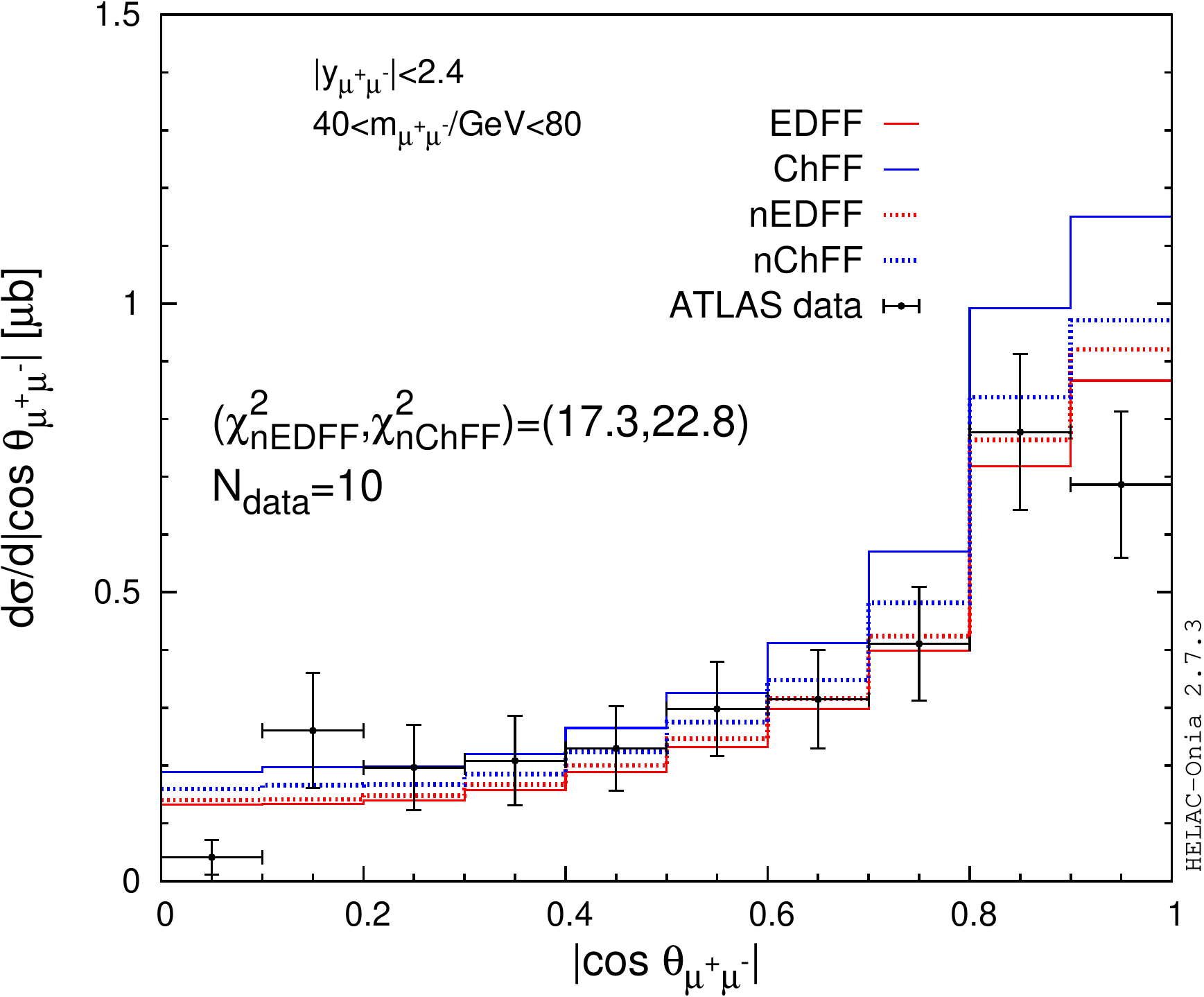}
\caption{Differential cross section of exclusive dimuon production in \PbPb\ UPCs $\sqrtsnn=5.02$~TeV as a function of various kinematic variables in different regions of phase space. The data (black points)~\cite{ATLAS:2020epq} are compared to \gammaUPC\ predictions (histograms) with EDFF and ChFF fluxes. The dotted histograms, nEDFF and nChFF, have been obtained normalizing the EDFF and ChFF predictions,  respectively, to match the measured total fiducial cross section. Data-theory $\chi^2$ values are quoted for nEDFF and nChFF fluxes.
\label{fig:dimuondM}}
\end{figure}

\begin{figure}[htbp!]
\centering
\includegraphics[width=0.49\textwidth]{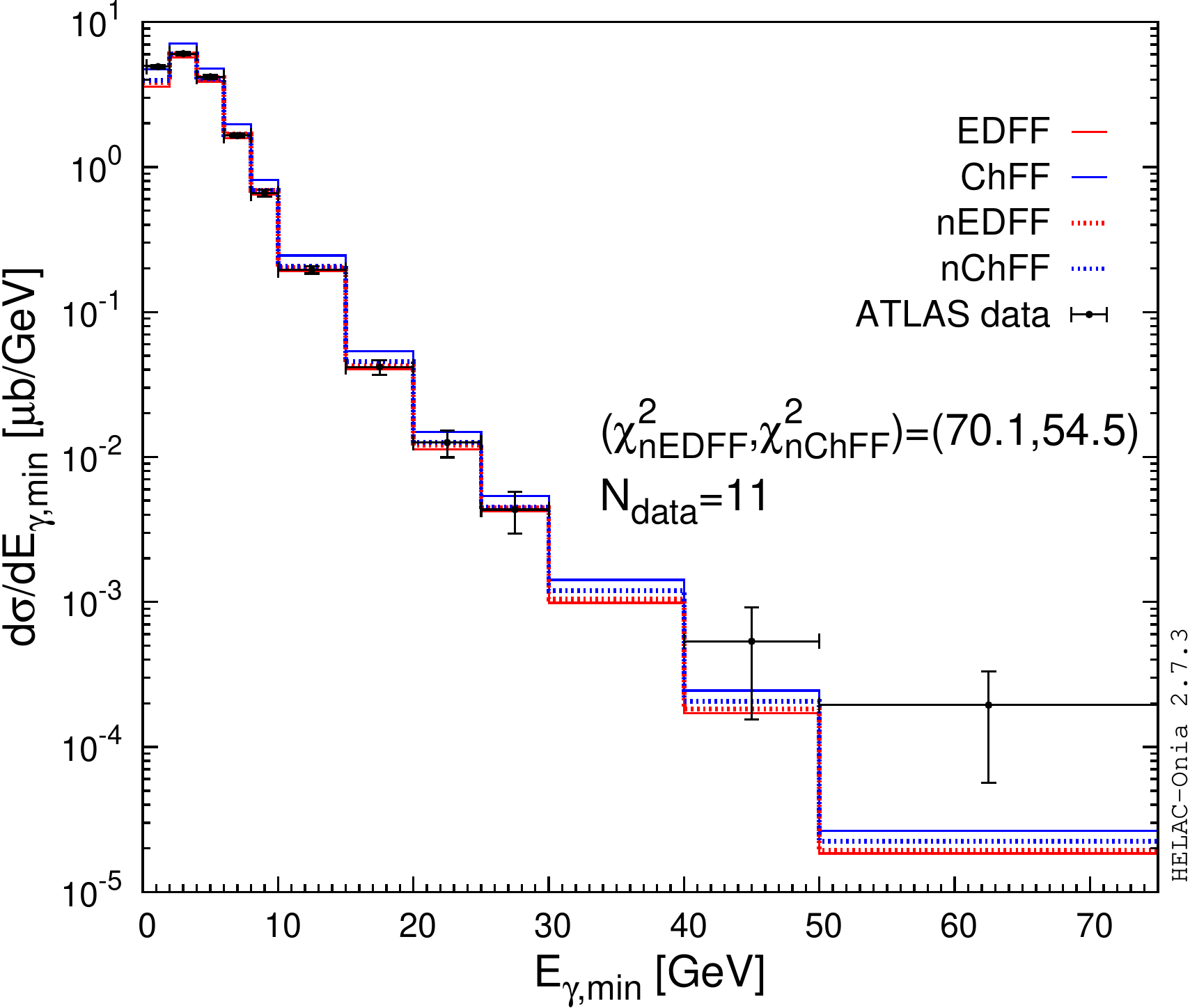}
\includegraphics[width=0.49\textwidth]{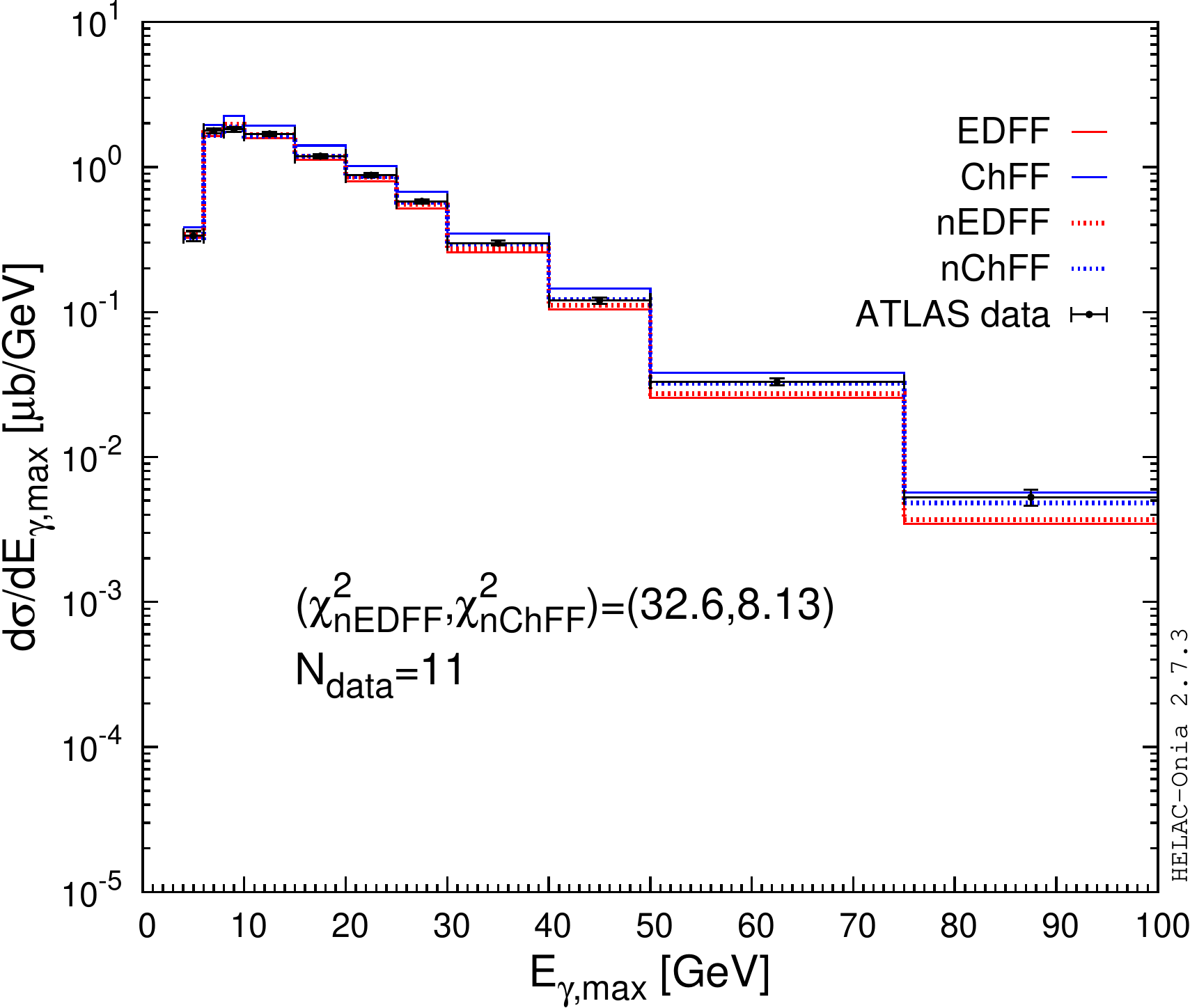}
\caption{Differential cross section in terms of minimum (left) and maximum (right) initial photon energies in exclusive dimuon production in \PbPb\ UPCs at $\sqrtsnn=5.02$~TeV. The data (black points)~\cite{ATLAS:2020epq} are compared to the histograms with the \gammaUPC\ predictions normalized (nEDFF and nChFF) and not (EDFF and ChFF) to the measured fiducial cross sections. Data-theory $\chi^2$ values are quoted for nEDFF and nChFF fluxes\label{fig:dimuondE}}
\end{figure}

In Fig.~\ref{fig:dimuondM}, the differential cross sections of exclusive dimuons measured by ATLAS as a function of invariant mass (top), pair rapidity (second row), and cosine of the pair polar angle (third and bottom rows) are plotted in different regions of phase space (from left to right) compared to the corresponding \gammaUPC\ results with EDFF and ChFF $\gamma$ fluxes, and to these same predictions but normalized (nEDFF and nChFF) to match the measured fiducial cross section. The $\chi^2$ goodness-of-fit, determined considering only experimental uncertainties, and number of data points for the rescaled theoretical predictions with respect to the experimental data are listed in each panel. The total $\chi^2$ for the overall predictions with nEDFF and nChFF fluxes are respectively $393$ and $327$ for $191$ data points. Namely, the data-theory comparison is slightly better with nChFF than nEDFF fluxes, indicating that the ChFF spectrum provides a better shape agreement with the data. Figure~\ref{fig:dimuondE} shows the differential exclusive-dimuon cross section as a function of mininum (left) and maximum (right) initial photon energy in data and theory.

\subsection{Light-by-light scattering in \PbPb\ UPCs at \texorpdfstring{$\sqrtsnn= 5.02$}{sqrt(s)= 5.02}~TeV}

The loop-induced LbL signal is generated with \gammaUPC\ plus \madgraph~v2.6.6~\cite{Alwall:2014hca,Hirschi:2015iia} with the virtual box contributions computed at leading order. Table~\ref{tab:xsecsLbL} compares the integrated fiducial cross sections measured by ATLAS~\cite{ATLAS:2020hii} with the \gammaUPC\ using EDFF and ChFF $\gamma$ fluxes and the \superchic\ predictions. The measured cross section is about 2 standard deviations above the \gammaUPC\ and \superchic\ predictions. 

\begin{table}[htpb!]
\centering
\tabcolsep=4.mm
\caption{Fiducial light-by-light cross sections measured in \PbPb\ UPCs at $\sqrtsnn=5.02$~TeV (with $\ET^{\gamma}>2.5$~GeV , $|\eta^{\gamma}|<2.4$, $m_{\gaga}>5$~GeV, $p_{\mathrm{T},\gaga}<1$~GeV), compared to the theoretical \gammaUPC\ results obtained with EDFF and ChFF $\gamma$ fluxes (and their average), as well as with the \superchic\ MC prediction.
\label{tab:xsecsLbL}}
\vspace{0.2cm}
\begin{tabular}{l|c|ccc|c} \hline
Process, system & ATLAS data~\cite{ATLAS:2020hii} & \multicolumn{3}{c|}{\gammaUPC~$\sigma$} 
& \superchic~$\sigma$ \\
 & & EDFF & ChFF & average & \\
 $\gaga\to\gaga$, \PbPb\ at 5.02 TeV & 
 $120\pm 22$~nb 
 & $63$~nb & $76$~nb &  $70 \pm 7$~nb & $78 \pm 8$~nb \\\hline
\end{tabular}
\end{table}

\begin{figure}[!htbp]
\centering
\includegraphics[width=0.49\textwidth]{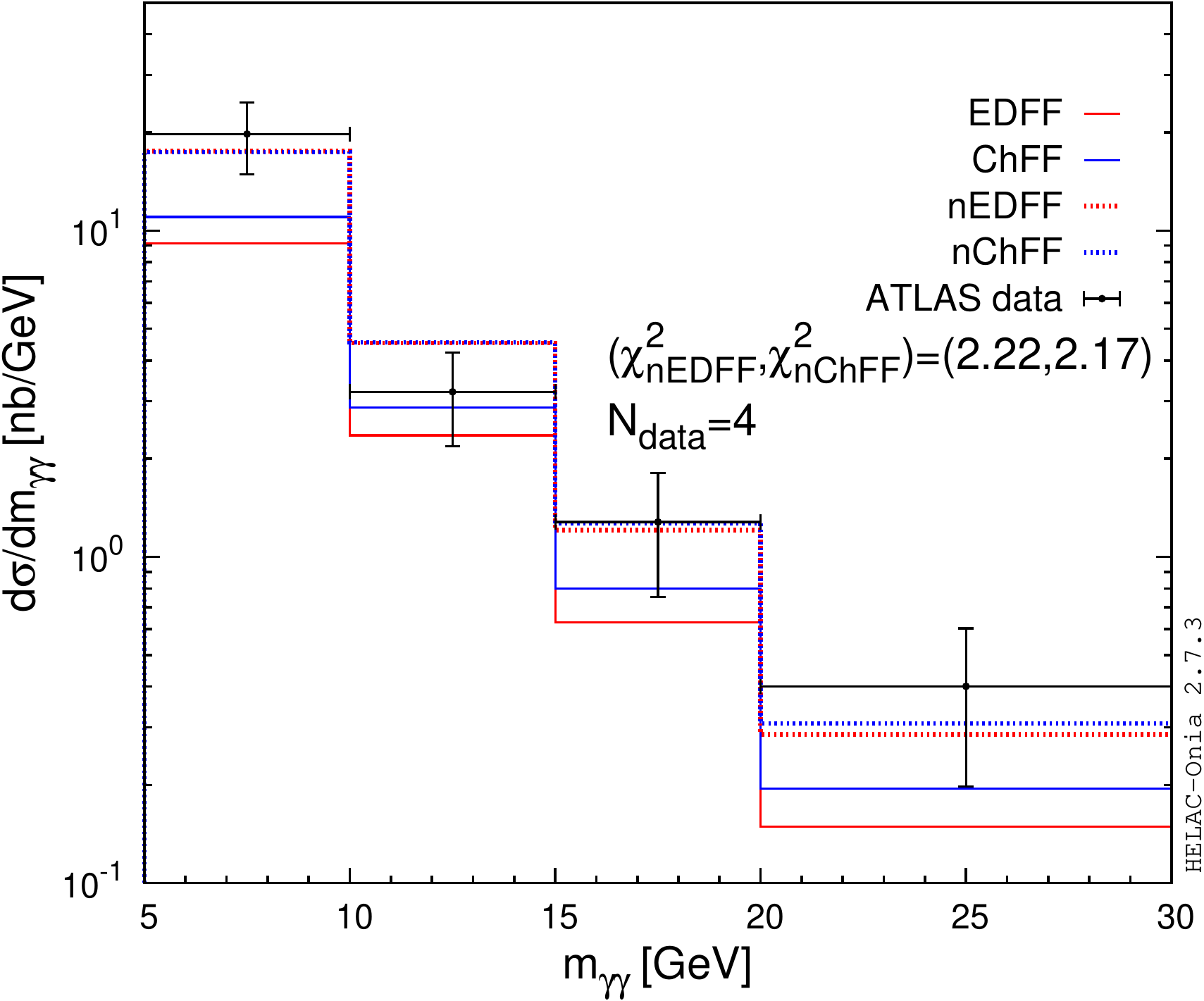}
\includegraphics[width=0.49\textwidth]{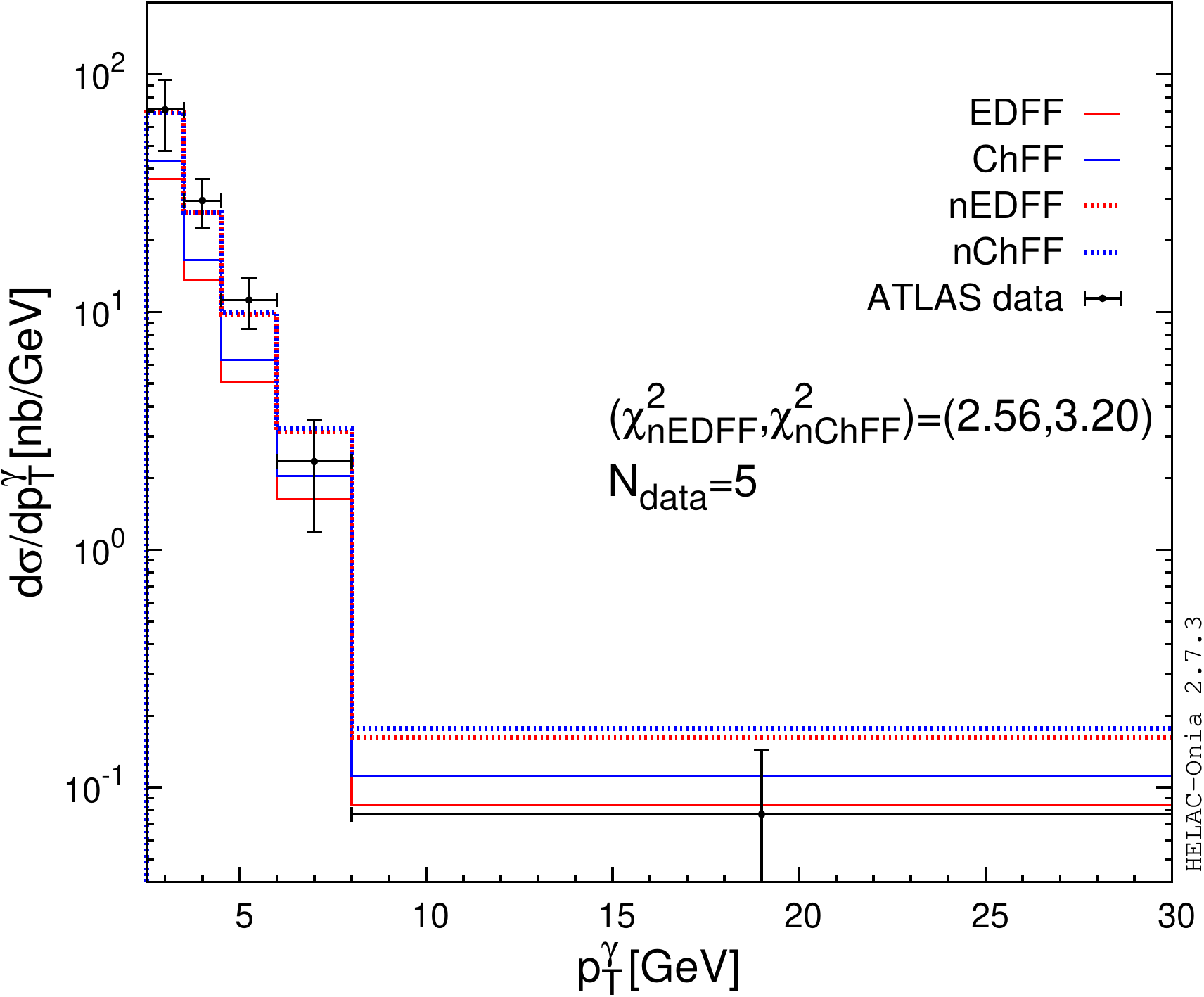}\\
\includegraphics[width=0.49\textwidth]{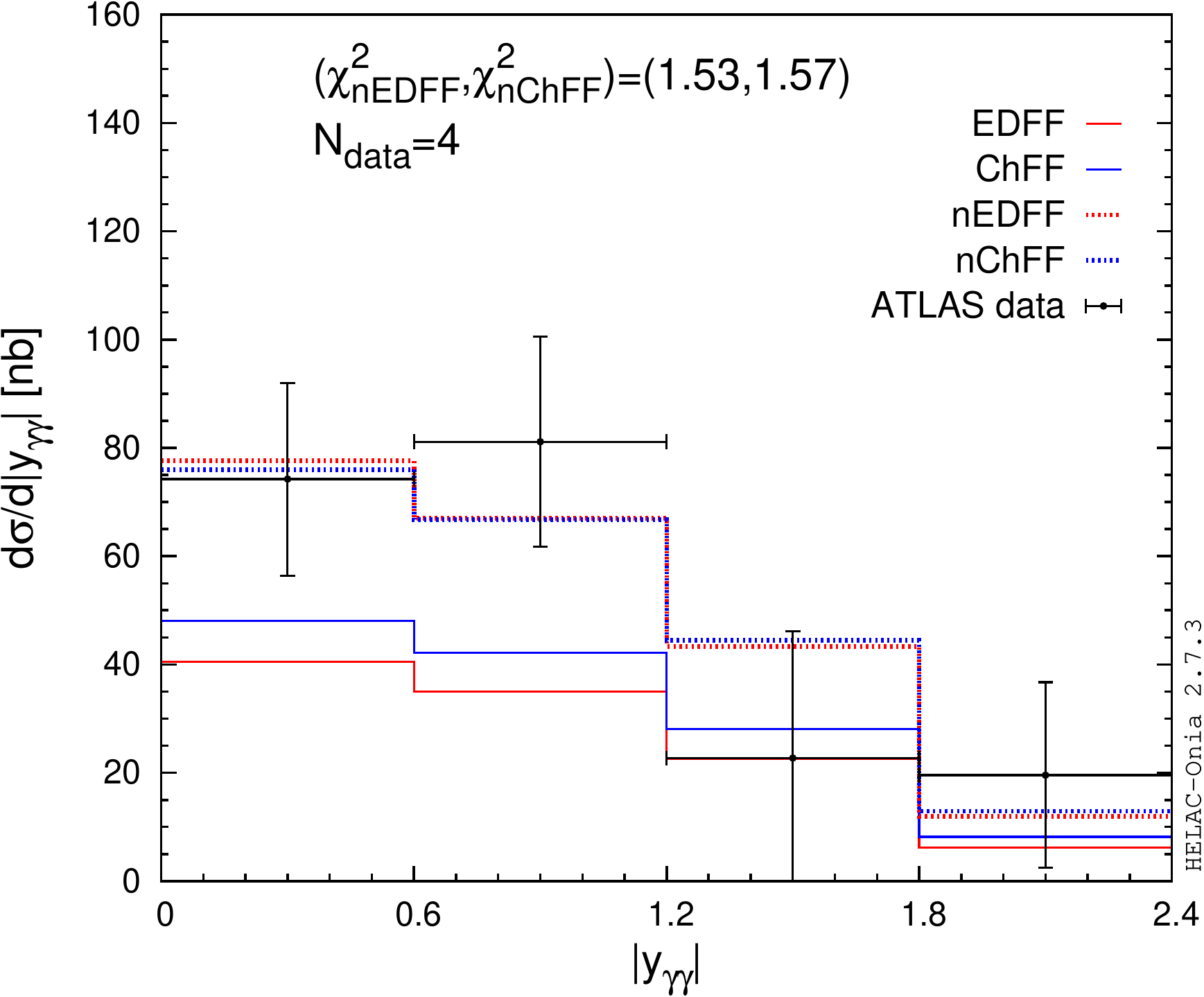}
\includegraphics[width=0.49\textwidth]{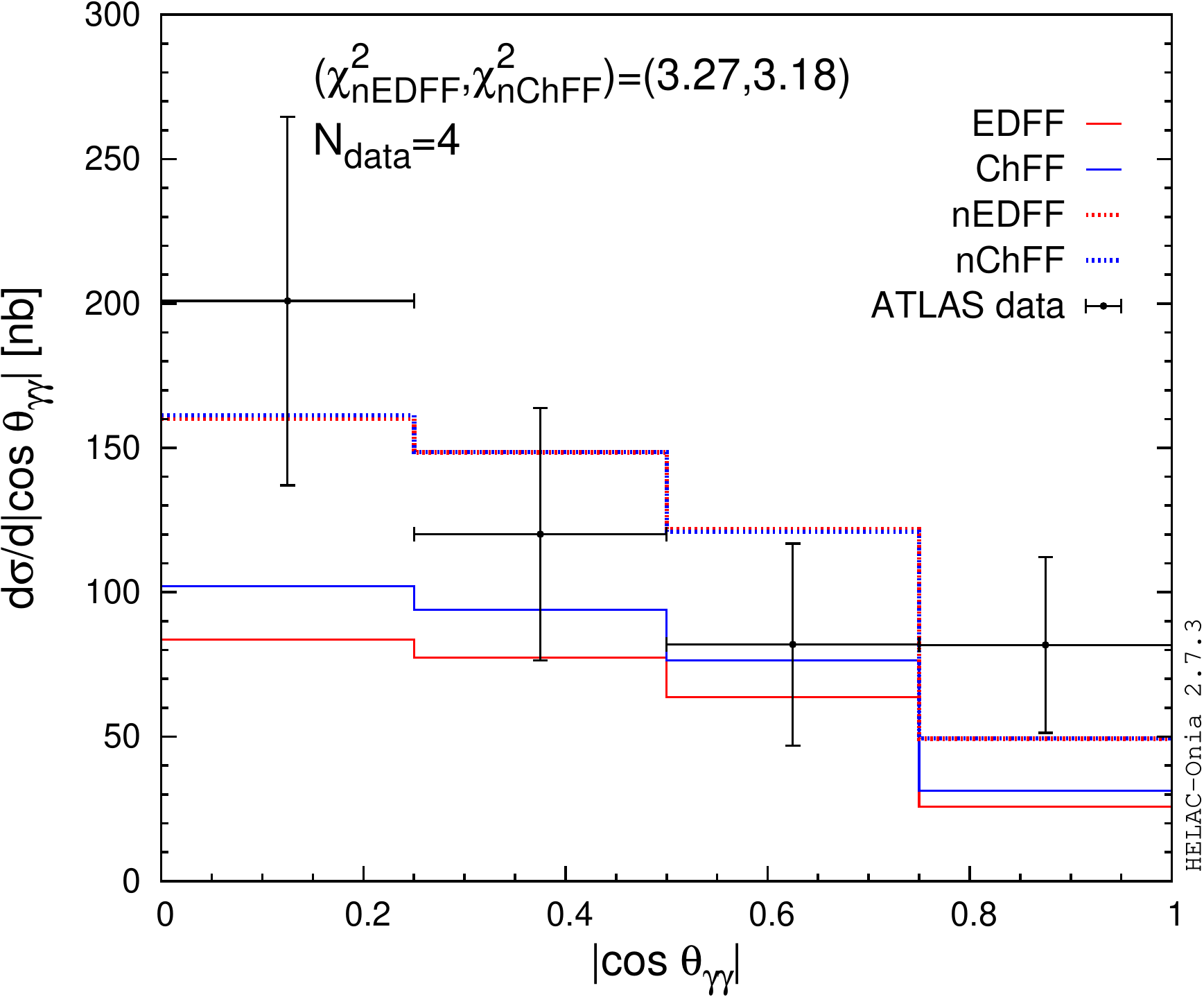}
\caption{Differential cross sections for light-by-light scattering as a function of various diphoton variables measured in \PbPb\ UPCs at $\sqrtsnn=5.02$~TeV (black data points)~\cite{ATLAS:2020hii} compared to our theoretical predictions (red and blue solid-line histograms for EDFF and ChFF, respectively). The dotted histograms (labeled as normalized EDFF and ChFF, nEDFF and nChFF, respectively) are the same predictions rescaled to match the experimental value of the fiducial cross section.
\label{fig:LbL}
}
\end{figure}

In Fig.~\ref{fig:LbL}, the differential LbL cross sections measured by ATLAS as a function of invariant mass (top left), single photon $\pT$ (top right), pair rapidity (bottom left), and cosine of the pair polar angle (bottom right) are compared to the corresponding \gammaUPC\ results with absolute (EDFF and ChFF) and normalized (nEDFF and nChFF) $\gamma$ fluxes. The overall $\chi^2$ is $9.58$ and $10.1$ for nEDFF and nChFF fluxes, respectively, with $17$ data points. The data-theory $\chi^2$ comparisons are very similar for nChFF and nEDFF fluxes ($\chi^2/N_\text{data}\approx 0.6$) indicating that both reproduce well the shapes of the LbL distributions measured in data within the relatively large experimental uncertainties. More accurate and precise LbL data are needed in order to understand if the moderate ``excess'' apparent in the first mass bin ($m_{\gaga} = 5$--10~GeV) with respect to the predictions is real.

\section{\texorpdfstring{\gammaUPC}{gamma-UPC} output and upcoming improvements}
\label{sec:upcoming}

The first release of the \gammaUPC\ code contains all the theoretical ingredients described previously in Sections~\ref{sec:sigma_gaga} and \ref{sec:gaga_lumis} that lead to the results presented in Sections~\ref{sec:results} and~\ref{sec:diff_results}. Such a code provides the baseline framework to compute the production cross section and event generation of any UPC final state of interest at the LHC and other hadron colliders (RHIC, FCC,...). We provide next a few more details on the \gammaUPC\ event generation output and ongoing/future developments.\\

The output of \gammaUPC\ is not just the photon-fusion fiducial or differential cross sections (in pb units) for the chosen process, but also unweighted MC events are generated in LHE format~\cite{Alwall:2006yp} using the default machinery of the \madgraph\ and \helaconia\ codes. The produced LHE output file contains the standard input kinematics and cross section of the generated process in the \texttt{<init>} block, as well as the four-momenta of all produced \textit{central} particles for each \texttt{<event>}.
In the \pp\ case, since most of the exclusive photon-photon processes are measured at the LHC employing \textbf{forward proton tagging} to get rid of the large pileup background, the \gammaUPC\ code can also provide as output the 4-momenta kinematics of the two outgoing protons (in the form of a second ancillary LHE file). Such information can then be used to transport the protons, through the beamline magnetic lattice, from the interaction point up to the down- and up-stream taggers in order to determine the experimental acceptance and efficiency of the latter for the physics process in question~\cite{CMS:2021ncv,Tasevsky:2015xya}.\\

Further improvements and extensions of \gammaUPC\ are ongoing or under consideration and will likely be part of a second release of the code, among which: 
\begin{enumerate}
    \item \textbf{Nonzero photon transverse momentum} $\pmb{\kt}$: Although our ChFF flux, Eq.~(\ref{eq:ChFF}), contains an explicit dependence on the photon $\kt$ (related to the photon virtuality via $Q^2 =  \kt^2 + E_{\gamma}^2/\gamma_\mathrm{L}^2$), our cross sections are fully integrated over the $Q$ and $\kt$ of the colliding photons, and therefore the centrally produced system $\gaga\to X$ is produced at rest, $\pT^{X}=0$. As the photon density follows a $1/\kt^2$ dependence and the $\kt$ values are many orders-of-magnitude smaller than the longitudinal photon energy, the approximation that both photons are real ($Q\approx 0$) has no actual numerical impact on the computed cross sections. In addition, the assumption that the colliding photons have zero $\kt$, \ie\ that the central system is produced exactly at rest, has no real experimental implication either because the detector resolution smears out the energies of the decay products of the central system leading to $\pT^{X}$ values that, though nonzero, are still well below $\pT^{X}\approx 1$~GeV (the usual upper limit imposed in the experimental analyses to remove nonexclusive backgrounds). Nonetheless, as discussed in the introduction, in reality the colliding photons can have very small but nonzero virtualities up to about $Q^{2} < 1/R^{2}\approx~0.08$~GeV$^2$ for protons and $Q^2<10^{-3}$~GeV$^2$ for Pb nuclei, and the next release of the code will include the impact of this small (few tens or hundred MeV) extra photon $\kt$ in the MC event generator output.
    \item \textbf{Semiexclusive photon-photon processes}: Our calculations use the elastic $\gamma$ fluxes for both hadrons, but $\gaga$ collisions can also occur in semielastic processes where one of the photons is emitted from the constituents (partons or nucleons in \pp\ or \AaAa\ UPCs, respectively) of one of the hadrons leading to its breakup. Although the cross sections for such semiexclusive collisions are suppressed compared to the fully coherent cases (\eg\ they scale at most as $Z^3$ compared to the $Z^4$ dependence of the \AaAa\ UPCs case), they can constitute a background to the elastic cross sections in the absence of detectors at very forward angles (Roman Pots and Zero Degree Calorimeters for \pp\ and \AaAa\ UPCs, respectively) that can be used to veto activity from the hadronic breakup. Our setup can be easily extended to incorporate semiexclusive collisions of inelastic photons from the hadron constituents, on the one hand, with elastic photons from the other intervening hadron, on the other.
    \item \textbf{NLO QED and weak corrections}: The availability of full NLO corrections accounting for  virtual and real QED and weak emissions is a requirement for accurate and precise calculations of photon-photon cross sections. In particular when comparing the data to theory to extract precision SM parameters (such as \eg\ the $g-2$ of the tau lepton via $\gaga\to\tautau$~\cite{delAguila:1991rm,Atag:2010ja,Beresford:2019gww,Dyndal:2020yen}) and/or to search for absolute or differential cross section deviations from the SM prediction due to new physics contributions. Theoretical developments in this direction are already part of \madgraph~\cite{Frederix:2018nkq} and need to be properly interfaced with the \gammaUPC\ setup to account for the particularities of photon-photon collisions.
    \item \textbf{Electroweak boson fusion processes with elastic photons}: Photon-photon collisions are actually a fraction of the multiple combinations of fusion processes among electroweak vector bosons (W, Z, and $\gamma$). Interesting possibilities exist if one considers semiexclusive photon-V collisions where the photon is radiated coherently from one hadron, and the weak boson V = W or Z is emitted from the constituent partons of the other\footnote{The coherent emission of a weak boson from the proton or nucleus as a whole is very much suppressed given the very short range of the weak interaction.}. 
    Such ``hybrid'' photon-W collisions at hadron colliders have been considered in the literature~\cite{Alva:2014gxa} and can be also in principle incorporated into our \gammaUPC\ setup by combining the equivalent W flux (the effective W/Z fluxes from leptons have been implemented in \mgshort\ recently~\cite{Ruiz:2021tdt}) or Z flux (for loop-induced $\gamma-Z$ fusion) from one hadron with the coherent photon of the other hadron.
    \item \textbf{UPCs in electron-proton,nucleus collisions}: The photon flux of an electron has larger virtualities than that of a hadron beam, but photon-photon collisions have been studied at electron-proton colliders for a long time~\cite{Vermaseren:1982cz,Schuler:1997ex}. The planned Electron-Ion-Collider (EIC)~\cite{AbdulKhalek:2021gbh} will allow for the first time to study $\gaga$ collisions issuing from the fusion of $e^\pm$ and heavy-ion photon fluxes, providing novel opportunities for studies of interest~\cite{Chwastowski:2022fzk,Davoudiasl:2021mjy}. The extension of \gammaUPC\ to handle and combine the incoming fluxes of photons from electrons and protons or heavy-ions is also under consideration to facilitate the preparation of EIC feasibility studies.
    \item \textbf{Forward neutron emission}: The exclusive photon-photon fusion cross sections calculated with \gammaUPC\ are fully inclusive with respect to any additional potential electromagnetic soft excitation(s) of the colliding nuclei (which in principle completely factorize from the photon-photon fusion process itself), and which may lead to later-time nuclear deexcitations with very forward neutron emission. For this reason, the data--theory comparisons shown in Figs.~\ref{fig:excl_ee}-- \ref{fig:LbL} are fully inclusive in forward neutron topology. However, one of the main advantages of generating $\gaga$ collisions with the dedicated \starlight\ MC code is the possibility of calculating cross sections for UPCs with ions including or vetoing the concurrent emission of $Xn$ (with $X=0,1,\cdots$) forward neutrons from one or both interacting ions. Events with neutron multiplicity indicate the presence of mutual e.m.\ excitation of the passing-by ions, or their nuclear breakup. Experimentally, such neutrons are usually detected in Zero Degree Calorimeters (ZDCs)~\cite{ALICE:1999edx,Adler:2000bd,Grachov:2006ke,White:2010zzd} and their veto helps to reduce nonexclusive backgrounds. A dedicated  stand-alone MC code exists, called  $\textbf{nOOn}$, for the calculation of forward neutron emission in UPCs with heavy ions~\cite{Broz:2019kpl} that can be eventually combined with the \gammaUPC\ setup.
\end{enumerate}
These upcoming expected improvements will be reported in the \gammaUPC\ code version information at the \href{http://cern.ch/hshao/gammaupc.html}{http://cern.ch/hshao/gammaupc.html} webpage.

\section{Summary}
\label{sec:summ}

We have presented a new phenomenological code development that is able of automatically generating arbitrary photon-photon collision events in ultraperipheral collisions (UPCs) of protons and heavy ions, \ABgagaX, at high energies. Two types of elastic photon fluxes, as well as associated survival probabilities of the photon-emitting hadrons, have been implemented into the \madgraph\ and \helaconia\ codes, based on the electric-dipole (EDFF) and charge (ChFF) form factors for proton and light and heavy nuclei. This setup, named \gammaUPC\ (downloadable from \href{http://cern.ch/hshao/gammaupc.html}{http://cern.ch/hshao/gammaupc.html}), can compute the cross sections and generate any exclusive final state of interest producing SM (in particular quarkonia) and BSM particles in UPCs at high energies, including higher-order real corrections for processes with extra photons and/or gluons emitted. From the differences found between the EDFF- and ChFF-based results, theoretical uncertainties in the cross sections linked to the elastic $\gamma$ spectrum and hadron survival probabilities for $\gaga\to X$ processes at low ($m_X\approx10$~GeV) and high ($m_X\approx100$~GeV) masses are estimated to vary over 12--25\% for \PbPb, 7--15\% for \pPb, and 6--12\% for \pp\ UPCs. Such uncertainties can nonetheless be significantly reduced by taking ratios of two exclusive $\gaga$ cross sections (\eg\ by using exclusive dimuon production as a reference baseline process in the denominator) at the same photon-photon \cm\ energy $W_{\gamma\gamma}$.\\

Illustrative examples of $\gaga$ cross sections computed with this setup have been shown for proton-proton, proton-nucleus, and nucleus-nucleus UPCs at the Large Hadron Collider (LHC) and Future Circular Collider (FCC). Total photon-fusion cross sections for the exclusive production of spin-0,\,2 resonances (four charmonium states, four bottomonium states, paraditauonium, and the Higgs boson), as well as for pairs of SM particles ($\jpsi\jpsi$, WW, ZZ, Z$\gamma$, $\ttbar$, HH) and for BSM particles (axionlike and massive gravitons) have been presented. All such processes provide valuable novel SM tests ($\tau$ and top-quark electromagnetic moments, quartic gauge couplings, properties of QCD and QED bound states, etc.) and unique BSM searches. Differential cross sections for the production of exclusive dielectrons, dimuons, and  light-by-light scattering have been compared to existing LHC~\PbPb\ data as well as to predictions from other UPC-dedicated MC models such as \starlight\ and \superchic. These more detailed comparisons indicate that, for the processes implemented in the two latter MC codes, the \gammaUPC\ EDFF and ChFF results are, respectively, very consistent with the \starlight\ and \superchic\ ones (and can be, therefore, used as ``proxies'' of the latter whenever the physics process is not available in them).\\

Ongoing and upcoming developments that will extend the \gammaUPC\ features (semiexclusive collisions, weak-boson fusion processes, UPCs in e-p,A, etc.) have been also outlined. This code provides a novel useful tool for carrying out studies of any arbitrary final state produced in photon-photon collision at hadron colliders, providing not only the cross section calculation and automatic generation of events for any SM/BSM signal of interest, but also of any potential associated backgrounds. The upcoming incorporation of full electroweak corrections at next-to-leading-order accuracy and beyond in \madgraph\ will allow for a reduction of theoretical uncertainties and the possibility of carrying out more precise SM tests, and BSM searches, with exclusive photon-photon processes employing our setup.\\ 


\paragraph*{Acknowledgments.---} 
Support from the European Union's Horizon 2020 research and innovation program (grant agreement No.824093, STRONG-2020, EU Virtual Access ``NLOAccess''), the French ANR (grant ANR-20-CE31-0015, ``PrecisOnium''), and the CNRS IEA (grant No.205210, ``GlueGraph"), are acknowledged.

\clearpage
\appendix
\section{Basic code instructions}
\label{sec:app}

The \gammaUPC\ code is written in {\sc\small Fortran90}. A brief set of instructions on how to compile and run \gammaUPC\ stand-alone, or with \madgraph\ or \helaconia\ are provided below. More technical details can be found at \href{http://cern.ch/hshao/gammaupc.html}{http://cern.ch/hshao/gammaupc.html}, where the code can be downloaded.

\subsection{Standalone usage}

The \gammaUPC\ can be run stand-alone. This package 
contains a module, {\tt test.f90}, which acts as the driver when working 
in this mode. The code is compiled with the usual shell command
\begin{lstlisting}
> make test
\end{lstlisting}
We assume a {\tt gfortran} compiler. The test program embedded in
{\tt test.f90} can be run by executing:
\begin{lstlisting}
> ./test
\end{lstlisting}
If one just compiles the code via
\begin{lstlisting}
> make
\end{lstlisting}
a static library {\tt libgammaUPC.a} will be generated. The \gammaUPC\ subroutines can be accessed by including the {\sc\small Fortran90} module via
\begin{lstlisting}
USE ElasticPhotonPhotonFlux
\end{lstlisting}
The common parameters of defining the two beams can be found in {\tt run90.inc} via
\begin{lstlisting}
INCLUDE `run90.inc`
\end{lstlisting}
The energies per nucleon of the two beams are {\tt ebeam(1)} and {\tt ebeam(2)} in units of GeV, while the nuclear mass and charge numbers of the first (second) beam are defined via the integers {\tt nuclearA\_beam1} ({\tt nuclearA\_beam2}) and {\tt nuclearZ\_beam1} ({\tt nuclearZ\_beam2}), respectively. The value of $\alpha$ can be changed from its default of $1/137$ by assigning {\tt alphaem\_elasticphoton} a new value. The bool flag {\tt USE\_CHARGEFORMFACTOR4PHOTON} is used to select EDFF ({\tt .FALSE.}) or ChFF ({\tt .TRUE.}) $\gamma$ fluxes. After the above preparation, 
one can call the function {\tt dLgammagammadW\_UPC} to obtain the effective two-photon luminosity at a given resonance mass {\tt m}, i.e. $\frac{d{\Lumi}^{(\mathrm{A\,B})}_{\gaga}}{dW_{\gaga}}\big|_{W_{\gaga}=m}$, as follows:
\begin{lstlisting}
dLdW=dLgammagammadW_UPC(m,icoll,1)
\end{lstlisting}
where the {\tt icoll}$=1,2,3$ argument applies to \pp, \pA, \AaBa\ collisions, respectively. The two-photon differential distribution normalized by $(x_1x_2)$, i.e., $\frac{1}{x_1x_2}\frac{\mathrm{d}^2N^{(\mathrm{AB})}_{\gamma_1/\mathrm{Z}_1,\gamma_2/\mathrm{Z}_2}}{\mathrm{d}E_{\gamma_1}\mathrm{d}E_{\gamma_2}}$ can be accessed via
\begin{lstlisting}
dNpp=PhotonPhotonFlux_pp(x1,x2)
dNpA=PhotonPhotonFlux_pA_WoodsSaxon(x1,x2)
dNAB=PhotonPhotonFlux_AB_WoodsSaxon(x1,x2)
\end{lstlisting}
for \pp, \pA, \AaBa\ collisions respectively, where {\tt x1} and {\tt x2} are the fractions $x_1$ and $x_2$ of the hadron energy carried out by the photons, for the two incoming beams. The initialization for generating grids in the first call may take a few minutes. However, the numerical evaluations should be fast enough and suitable for the numerical phase space integrations as long as the grids have been successfully produced.

\subsection{Usage of \gammaUPC\ in \helaconia}

The program \gammaUPC\ has been integrated into \helaconia~\cite{Shao:2012iz,Shao:2015vga} for the exclusive two-photon production of quarkonia bound states, and easily extendable to any spin-even resonance by introducing a ``fake'' $\qqbar$ state with any arbitrary mass and diphoton width, as \eg\ done for ditauonium~\cite{dEnterria:2022ysg}. 
A few parameters need to be specified before launching the jobs, as follows:
\begin{lstlisting}
HO> set colpar = 14
HO> set nuclearA_beam1 = <an integer>
HO> set nuclearA_beam2 = <an integer>
HO> set nuclearZ_beam1 = <an integer>
HO> set nuclearZ_beam2 = <an integer>
HO> set UPC_photon_flux_type = <an integer between 1 to 6>
\end{lstlisting}
where the {\tt nuclearA\_beam1} ({\tt nuclearA\_beam2}) and {\tt nuclearZ\_beam1} ({\tt nuclearZ\_beam2}) integers are nuclear mass and atomic numbers for the first (second) beam, respectively. The parameter {\tt UPC\_photon\_flux\_type} determines the usage of the UPC photon-photon fluxes as explained in {\tt input/default.inp}. Namely, setting {\tt UPC\_photon\_flux\_type}=$1,6$ selects EDFF and ChFF fluxes, respectively, with their corresponding hadronic-nonoverlap requirement. In such a case, the two initial particles must be photons. The parameters {\tt energy\_beam1} and {\tt energy\_beam2} (in GeV/nucleon) are interpreted as the energy of the beams per nucleon.

\subsection{Usage of \gammaUPC\ in \madgraph}

One can also directly call \gammaUPC\ within \madgraph
~\cite{Alwall:2014hca} for the exclusive two-photon production of any SM or BSM final state. The two initial particles of the generated process must be two photons. In order to call \gammaUPC, one needs to specify the following parameters in {\tt run\_card.dat}, taking here \pPb\ UPCs at $\sqrt{s_{\rm NN}}=8.16$ TeV as an example:\\

\begin{minipage}{\textwidth}
\fontsize{10pt}{10pt}\selectfont
\begin{verbatim} 
#*********************************************************************
# Collider type and energy                                           *
# lpp: 0=No PDF, 1=proton, -1=antiproton,                            *
#                2=elastic photon of proton/ion beam                 *
#             +/-3=PDF of electron/positron beam                     *
#             +/-4=PDF of muon/antimuon beam                         *
#*********************************************************************
  2     = lpp1 ! beam 1 type
  2     = lpp2 ! beam 2 type
  7000.0        = ebeam1 ! beam 1 total energy in GeV
  574080.0      = ebeam2 ! beam 2 total energy in GeV
  
#*********************************************************************
# PDF CHOICE: this automatically fixes alpha_s and its evol.         *
# pdlabel: lhapdf=LHAPDF (installation needed) [1412.7420]           *
#          iww=Improved Weizsaecker-Williams Approx.[hep-ph/9310350] *
#          eva=Effective W/Z/A Approx.       [2111.02442]            *
#          edff=EDFF in gamma-UPC            [2207.03012]            *
#          chff=ChFF in gamma-UPC            [2207.03012]            *
#          none=No PDF, same as lhapdf with lppx=0                   *
#*********************************************************************
  edff  = pdlabel ! PDF set
  
#*********************************************************************
# Heavy ion PDF / rescaling of PDF                                   *
#*********************************************************************
  1    = nb_proton1 # number of protons for the first beam
  0    = nb_neutron1 # number of neutrons for the first beam
  82   = nb_proton2 # number of protons for the second beam
  126  = nb_neutron2 # number of neutrons for the second beam
\end{verbatim}  
\end{minipage}

\vskip 0.5truecm
\noindent Note that unlike the previous two cases (running stand-alone and with \helaconia), the energy of the ion beam is its total energy (namely, $A\times E_\mathrm{beam}$, where $A=Z+N$ is the sum of the number of protons and neutrons, \ie\ the total number of nucleons) instead of the energy per nucleon. The two beam types ({\tt lpp1},{\tt lpp2}) must be chosen as $2$, and {\tt pdlabel} can be either {\tt iww} [cf. Eq.~(\ref{eq:flux_p})], {\tt edff} (EDFF), or {\tt chff} (ChFF) elastic photon fluxes. Note that the {\tt iww} choice is not applicable for ion beams, but only for protons. The parameters {\tt nb\_proton}$i$ and {\tt nb\_neutron}$i$ set the numbers of protons and neutrons, respectively, in the $i$th beam. These are hidden parameters in {\tt run\_card.dat}, which can be explicitly shown by using the prompt command `{\tt update ion\_pdf}' when editing the cards.

\clearpage
\bibliographystyle{myutphys}
\bibliography{gammagamma_colls_gammaUPC.bib}

\end{document}